\documentclass[aps,reprint,superscriptaddress,longbibliography,prb]{revtex4-2}
\usepackage{graphicx}
\usepackage{amsmath}
\usepackage{amsfonts}
\usepackage{bbm}
\usepackage{bm}
\usepackage{comment}
\usepackage{esint}
\usepackage{cancel}
\usepackage[dvipsnames]{xcolor}
\usepackage[unicode=true, colorlinks=true, citecolor={blue!80!black}, urlcolor={blue!50!black}, linkcolor = {blue!80!black}]{hyperref}
\usepackage{microtype}
\usepackage[normalem]{ulem}
\usepackage{mathrsfs}
\usepackage{multirow}
\usepackage{array}
\usepackage{lipsum}
\newcolumntype{C}[1]{>{\centering\arraybackslash}p{#1}}

\newcommand{\abs}[1]{\left\vert#1\right\vert}
\newcommand{\ket}[1]{\left\vert#1\right\rangle}

\graphicspath{{../figures/}}

\begin{document}
\title{Quantum phase slips in a resonant Josephson junction}

\author{Tereza Vakhtel}
\affiliation{Instituut-Lorentz, Universiteit Leiden, P.O. Box 9506, 2300 RA Leiden, The Netherlands}
\author{Bernard van Heck}
\affiliation{Leiden Institute of Physics, Universiteit Leiden, Niels Bohrweg 2, 2333 CA Leiden, The Netherlands}
\affiliation{Dipartimento di Fisica, Università di Roma “La Sapienza”, P.le Aldo Moro 5, 00185 Roma, Italy}

\date{\today}
\begin{abstract}
We investigate the consequences of resonant tunneling of Cooper pairs on the quantum phase slips occurring in a Josephson junction.
The amplitude for quantum tunneling under the Josephson potential barrier is modified by the Landau-Zener amplitude of adiabatic passage through an Andreev level crossing, resulting in the suppression of $2\pi$ phase slips.
As a consequence, close to resonance, $4\pi$ phase slips become the dominant tunneling process.
We illustrate this crossover by determining the energy spectrum of a transmon circuit, showing that a residual charge dispersion persists even at perfect transparency.
\end{abstract}

\maketitle

%\tableofcontents

\section{Introduction}

The phase difference across a Josephson junction can be driven by quantum fluctuations to change, or ``slip", by integer multiples of $2\pi$ \cite{haviland2010}.
Such quantum phase slips often determine the low-frequency behavior of microwave superconducting circuits \cite{chow1998,lau2001,mooij2006,pop2010,astafiev2012,manucharyan2012}.
In a long chain or loop of Josephson junctions, or in thin superconducting wires or rings, quantum phase slips compromise the spatial stiffness of the phase and can suppress superconductivity \cite{zaikin1997,hekking1997,fazio2001,golubev2001,matveev2002,buchler2004,refael2007,halperin2010}.
In general, quantum phase slips affect the energy levels of a coherent superconducting circuit~\cite{averin1985} and can therefore be measured with spectroscopic methods.

For instance, in a Cooper-pair-box circuit~\cite{bouchiat1998,nakamura1999,vion2002} in the transmon limit~\cite{koch2007}, quantum phase slips determine the charge dispersion of the energy levels~\cite{koch2007}, i.e.~the magnitude of their oscillation as a function of the charge induced on the superconducting island [see figure~\ref{fig:intro}(a-b)].
The charge dispersion of the fundamental frequency of the circuit is particularly important since it controls the dephasing time of superconducting qubits~\cite{koch2007}.
This fact motivated the development of the transmon~\cite{koch2007}, where the quantum phase slip amplitude is suppressed by a large ratio of the Josephson energy $E_J$ and the charging energy $E_c$, resulting in an exponential suppression of the charge dispersion~\cite{schreier2008}.

Setting aside qubit applications, devices with an appreciable charge dispersion remain of fundamental interest: thanks to their sensitivity to charge parity, they can be used to study quasiparticle poisoning and dynamics~\cite{riste2013,serniak2018,serniak2019,uilhoorn2021,kurter2022,erlandsson2022}, and, in a possible future, to measure fermion parity in topological Majorana qubits~\cite{hassler2011,aasen2016}.
These ongoing developments welcome further theoretical study of quantum phase slips, particularly given the emergence of hybrid semiconducting-superconducting qubit devices~\cite{aguado2020} and novel designs of noise-protected superconducting qubits~\cite{gyenis2021}.

In this paper, we compute in detail the amplitude of quantum phase slips in a Josephson junction with a resonant energy level.
We describe and pay particular attention to the competition between coherent $2\pi$ and $4\pi$ quantum phase slips that occurs in such a junction.
The competition is controlled by two independent parameters: the energy of the resonant level and the asymmetry between the tunneling rates to the superconducting leads.
The $4\pi$ phase slips become dominant close to resonance, and we argue that even though they were too small to be detected in recent experiments~\cite{kringhoj2020,bargerbos2020}, they can be observed in devices with a larger charging energy.
Towards the end, possible implications for qubit designs are also discussed.
The next section motivates our calculations, placing them in the context of previous theoretical and experimental research.

\section{$2\pi$ and $4\pi$ quantum phase slips}

\begin{figure*}
    \centering
    \includegraphics[width=\textwidth]{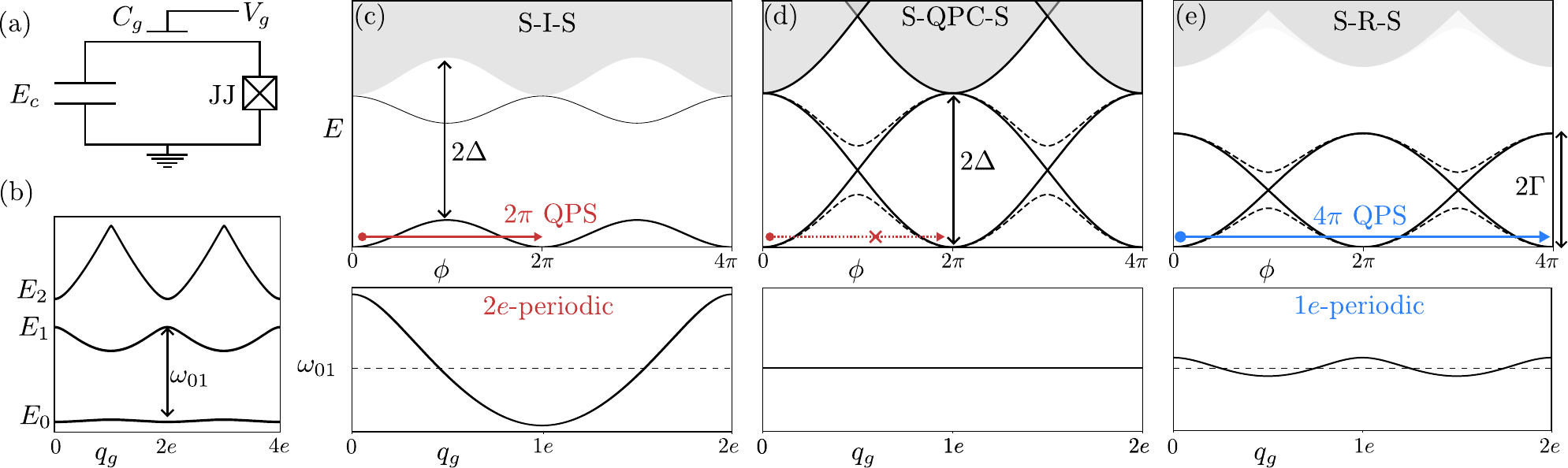}
    \caption{\emph{(a)}: A Cooper pair box consists of a superconducting island connected to ground by a capacitor and a Josephson junction. A gate voltage $V_g$ controls the charge induced on the island, $q_g = C_gV_g$. In the transmon limit of the Cooper pair box, the charging energy $E_c$ is much smaller than the Josephson tunneling strength. \emph{(b)} The energy levels $E_n$ of the Cooper pair box oscillate with $n_g$. The resulting charge dispersion can be determined by measuring the fundamental frequency $\omega_{01}=E_1-E_0$ as a function of $n_g$, for instance via microwave spectroscopy. \emph{(c-e)} Schematic energy spectrum of three different types of Josephson weak links (top row) and corresponding charge dispersion oscillations in the Cooper-pair box (bottom row). \emph{(c)}: $2e$-periodic dispersion due to $2\pi$ quantum phase slips in a tunnel junction. \emph{(d)}: Absence of charge dispersion in a quantum point contact at perfect transparency. \emph{(e)}: $1e$-periodic dispersion due to $4\pi$ quantum phase slips in a junction with a resonant energy level (e). Dashed lines in (d) and (e) show the Josephson potential away from perfect transparency, in which case $2\pi$ phase slips are weakly restored.}
    \label{fig:intro}
\end{figure*}

The amplitude of coherent quantum phase slips in a weak link is given by the tunneling amplitude between neighboring minima of the Josephson potential energy.
This amplitude can be qualitatively affected by the type of weak link where the phase slip occurs.
Figure~\ref{fig:intro}(c-e) compares three simple but paradigmatic scenarios: a low-transparency tunnel junction (S-I-S); a highly transparent single-channel quantum point contact (S-QPC-S); and finally a junction with a resonant level (S-R-S).
As we argue below, so far the S-R-S scenario has not been fully understood and described, despite its experimental relevance.

Figure~\ref{fig:intro}(c) illustrates the familiar setting of a tunnel junction, such as a quantum point contact close to pinch-off or an Al oxide junction, for which the potential energy is $\approx E_J(1-\cos\phi)$\footnote{A low-transparency QPC differs from an oxide junction because in the former the entire phase dispersion of the ground state originates from a single transport channel, and thus a single Andreev bound state, while in the latter from hundreds or even thousands of transport channels. The two junctions have equivalent ground state properties, but different densities of states close to the gap edge; the sketch in figure~\ref{fig:intro}c schematically depicts the first case.}.
Quantum phase slips connect the neighboring minima of the cosine potential, distant by $2\pi$ and, 
when $E_J\gg E_c$, they are suppressed exponentially with $\sqrt{E_J/E_c}$ \cite{koch2007}.
This classic result can be obtained using the WKB method or an instanton approach to the cosine potential \cite{holstein1988,altland}.
The charge dispersion of the energy levels is $2e$-periodic and, while exponentially small, remains finite at any value of $E_J$ due to the presence of back-scattering at the tunnel junction.

By contrast, figure~\ref{fig:intro}(d) shows the case of a quantum point contact at perfect transparency.
Its distinctive feature is the presence of a level crossing that disconnects the neighboring minima of the Josephson potential.
In fact, since each potential branch touches the continuum states at $E=2\Delta$, the Josephson potential is a-periodic~\cite{averin1999}.
As a consequence, quantum phase slips are forbidden altogether and the charge dispersion vanishes~\cite{ivanov1998,averin1999,averin1999b}.
Away from perfect transparency, the level crossing becomes a narrowly avoided crossing.
Quantum phase slips may then occur again, but only if the phase slips adiabatically though the crossing.
Hence, they are suppressed by the associated Landau-Zener transition amplitude and, near perfect transparency, it remains much smaller than in a S-I-S junction with comparable Josephson energy.

This enhanced suppression of the charge dispersion has been recently observed in spectroscopic measurements of transmon qubits realized with hybrid InAs/Al nanowire Josephson junctions~\cite{bargerbos2020,kringhoj2020}.
However, in these experiments the condition of almost perfect transparency was achieved by fine-tuning the nanowire junction to a resonance.
As shown in figure~\ref{fig:intro}(e), this  scenario differs qualitatively from that of a quantum point contact.

The normal-state transmission probability of a quantum point contact does not depend on energy on scales compared to the gap $\Delta$, while in the presence of a resonance it is a peaked function of energy, with a characteristic width $\Gamma$ that can be much smaller than $\Delta$.
As a consequence, the Andreev levels in the resonant case are detached from the continuum of energy levels even at zero phase difference~\cite{carlo1992,devyatov1997}, while they always touch the gap edge for a quantum point contact~\cite{carlo1991}.

This difference has important consequences for quantum phase slips:
if perfect transmission is achieved resonantly, the Josephson potential consists of two $4\pi$-periodic branches \cite{kurilovich2021}.
Thus, one expects $4\pi$ phase slips to occur even when $2\pi$ phase slips are forbidden.
As a result, one predicts a finite charge dispersion at resonance, but with a modified periodicity of $1e$ rather than $2e$.
In this respect, the situation is similar to that of a topological Josephson junction with coupled Majorana zero modes \cite{kitaev2001,fu2009}, with the crucial difference that in the resonant junction the two branches of the potential have the same fermion parity.

Given this scenario, it is appropriate to revisit quantum phase slips in the presence of a resonance, using as a starting point the existing knowledge on resonant Josephson tunneling~\cite{beenakker1992three,carlo1992,devyatov1997,martinrodero2011}, which has seen a revival~\cite{kurilovich2021} in view of experimental progress on microwave measurements of Andreev bound states~\cite{janvier2015,hays2018,hays2021,arno2022}.

\section{Model}

\begin{figure}[t!]
    \centering
    \includegraphics[width=\columnwidth]{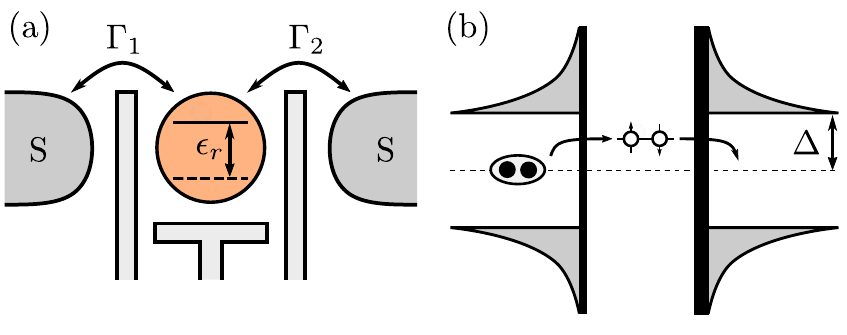}
    \caption{Illustration of the model of Eqs.~\eqref{eq:hamiltonian} and~\eqref{eq:potential_energy}. \emph{(a)}: A Josephson junction consisting of quantum dot (orange) with a single energy level. The detuning $\epsilon_r$ of the energy level from the Fermi level of the leads and the tunneling rates $\Gamma_1$, $\Gamma_2$ can be controlled via gate electrodes. \emph{(b)}: Transport of Cooper pairs across the two insulating barriers is mediated by the spin-degenerate resonant level.}
    \label{fig:layout}
\end{figure}

We consider a minimal model for a resonant Josephson junction in which the current between two superconducting electrodes is mediated via a single spin-degenerate energy level (see figure~\ref{fig:layout}).
The parameters of the model are the two tunneling rates $\Gamma_1$ and $\Gamma_2$ between the leads and the resonant level, and the energy $\epsilon_r$ of the resonant level, measured with respect to the Fermi level in the leads. In what follows, we will refer to $\epsilon_r$ as the detuning.

We consider the case in which $\Gamma_{1,2}\ll \Delta$, the superconducting gap in the leads. In this limit, it is possible to integrate out the fermionic degrees of freedom of the superconductors and obtain a simple effective Hamiltonian for the coupled dynamics of the superconducting phase difference $\phi$ and of the resonant level. The effective Hamiltonian is
\begin{equation}\label{eq:hamiltonian}
H = 4E_c(i\partial_\phi + n_g)^2 + V(\phi)
\end{equation}
Here, $E_c$ is the charging energy between the two electrodes, and $n_g=q_g/(2e)$ the charge induced by the electrostatic gates coupled to them, measured in units of $2e$.
The operator $-i\partial_\phi$ counts the number of Cooper pairs transferred between the two superconductors.
The matrix-valued potential energy $V(\phi)$ is \cite{meng2009,recher2010,oriekhov2021,kurilovich2021}
\begin{equation}\label{eq:potential_energy}
V =  -\epsilon_r\,\tau_z - \Gamma \cos(\phi/2)\,\tau_x - \delta\Gamma \sin(\phi/2)\,\tau_y\,,
\end{equation}
where we have introduced the total tunneling rate
\begin{equation}
\Gamma = \Gamma_1+\Gamma_2\,,
\end{equation}
and the asymmetry parameter
\begin{equation}
\delta\Gamma = \Gamma_1-\Gamma_2\,.
\end{equation}
The Pauli matrices $\tau_{x,y,z}$ encode the dynamics of the two-level system in which the resonant level is either empty ($\tau_z=+1$) or occupied by a Cooper pair ($\tau_z=-1$). 

The adiabatic eigenvalues $\pm E_A$ of the potential in Eq.~\eqref{eq:potential_energy} reproduce the well-known formula for the Andreev levels in a single-channel junction:
\begin{align}\label{eq:EA}
E_A(\phi) &= \sqrt{\epsilon_r^2 + \Gamma^2\cos^2(\phi/2) + \delta\Gamma^2\,\sin^2(\phi/2)} \\
&\equiv \Gamma_A\,\sqrt{1-T \sin^2\phi/2}\,,
\end{align}
with $\Gamma_A^2=\Gamma^2+\epsilon_r^2$ and $T=1-\abs{r}^2$
the transparency of the junction, controlled by the reflection coefficient
\begin{equation}\label{[eq:reflection_coefficient]}
r = \frac{\epsilon_r + i\delta\Gamma}{\Gamma_A}\,.
\end{equation}

The salient features of the Andreev spectrum are the following. First, at perfect transparency, which is achieved when $\epsilon_r=\delta\Gamma=0$ so that $r=0$, the spectrum evolves into two decoupled, $4\pi$-periodic branches with energy $\pm \Gamma \cos(\phi/2)$, with a zero-energy level crossing at $\phi=\pi$.
Second, , as long as $\Gamma_A\ll \Delta$, the Andreev bound state energy is well detached from the continuum spectrum for all values of $\phi$, including $\phi=0$ [see figure~\ref{fig:intro}(e)].
This fact, in particular, justifies neglecting excited states in the continuum when considering the adiabatic dynamics of the phase difference.

The derivation of the effective Hamiltonian of Eq.~\eqref{eq:hamiltonian}, which is carried out in appendix~\ref{app:effective_hamiltonian}, also yields the appropriate boundary condition for the spinor wave functions
\begin{equation}\label{eq:boundary_conditions}
\Psi(\phi+2\pi)=\tau_z \Psi(\phi).
\end{equation}
This twisted boundary condition incorporates a constraint on the dynamics that comes from charge conservation: if a Cooper pair occupies the resonant level, it must be subtracted from one of the two superconductors.
In other words, the tunneling of a Cooper pair between one of the two superconductors and the dot counts as \emph{half} of a Cooper pair transfer between the two superconductors.
This is the humble origin of the $4\pi$-periodicity of the tunneling terms in the effective Hamiltonian.

We also point out that, despite the complete similarity at the level of the Andreev spectrum, Eq.~\eqref{eq:EA}, the effective two-level Hamiltonian of Eq.~\eqref{eq:hamiltonian} is not the same as the corresponding two-level Hamiltonian for a quantum point contact~\cite{ivanov1999,zazunov2003}.
Besides the aforementioned fact that the Andreev levels are fully detached from the continuum, the main physical difference is that in the limit $T\to0$ a sub-gap state is present in the resonant level model (provided that $\epsilon_r$ is small enough), while no sub-gap state remains for the quantum point contact.

\begin{figure}[t!]
    \centering
    \includegraphics[width=\columnwidth]{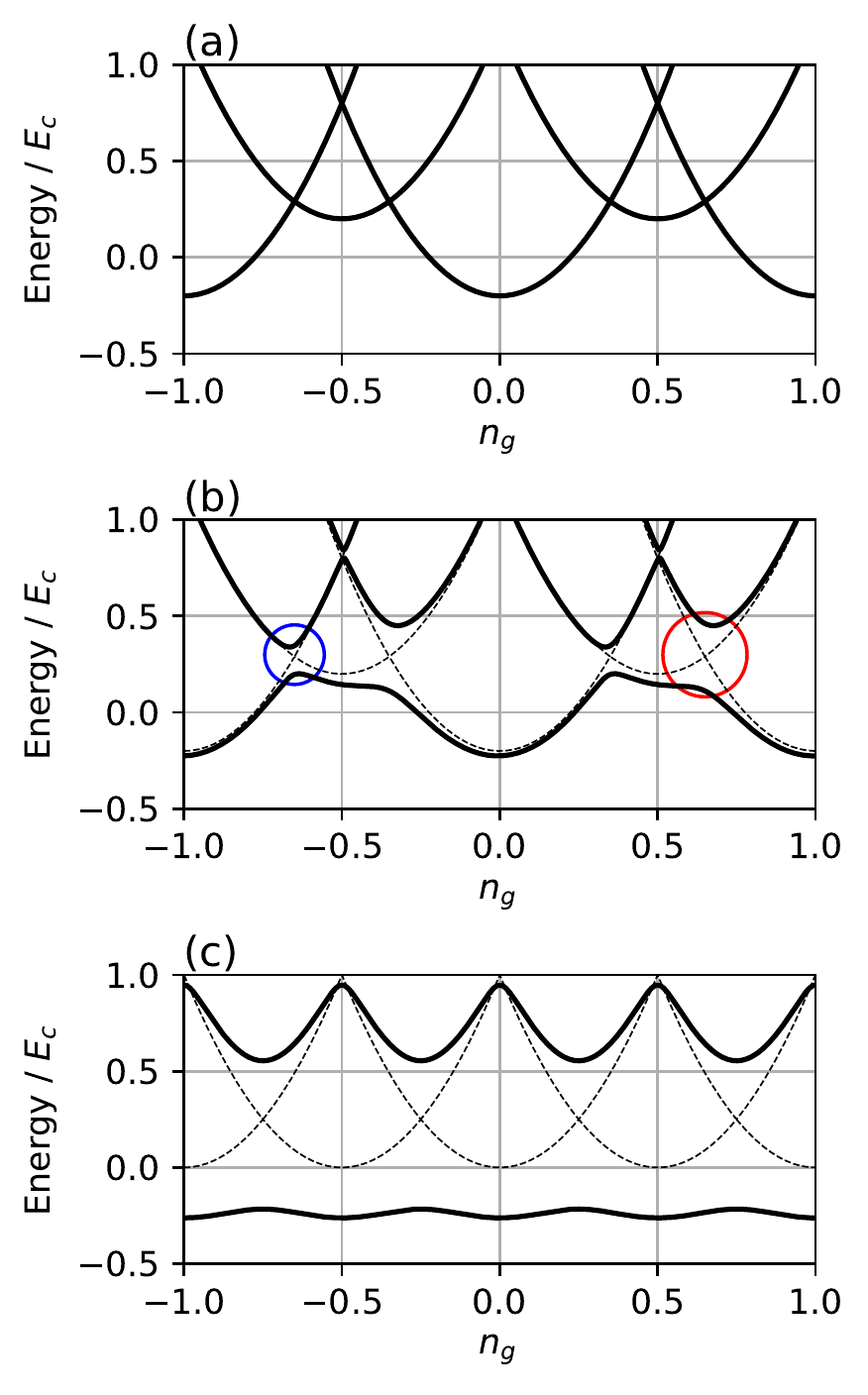}
    \caption{Energy spectrum of the model in the weak tunneling limit. We recall that $n_g$ is the charge induced on the island in units of $2e$. (\emph{a}): Energy levels of the model of Eq.~\eqref{eq:hamiltonian} with $\epsilon_r/E_c=0.2$, $\Gamma_1=\Gamma_2=0$. Note that charge parabolas with the dot empty (occupied) are centered around integer (half-integer) values of $n_g$. (\emph{b}): Energy levels with $\Gamma/E_c=0.12$ and $\delta\Gamma/E_c=0.06$. Blue and red circles identify avoided crossings opened by a finite $\Gamma_1$ and $\Gamma_2$, respectively. (\emph{c}): Energy levels for $\epsilon_r=0$, $\delta\Gamma=0$ and $\Gamma/E_c=0.8$. In panels (b) and (c) the dashed lines represent the charge parabolas for $\Gamma_1=\Gamma_2=0$.}
    \label{fig:parabolas}
\end{figure}

These circumstances can be elucidated by inspecting the energy spectrum in the absence of tunneling, at $\Gamma_1=\Gamma_2=0$, see figure~\ref{fig:parabolas}(a).
It consists of familiar parabolas with energy $E=4E_{c}\,(n-n_g)^2$, each corresponding to a charge $q=2e n$ transferred between the superconductors.
If the resonant level is empty, $n$ is integer, leading to a set of parabolas centered around integer values of $n_g$.
On the other hand, if the resonant level is occupied, $n$ is half-integer, leading to a second set of parabolas centered around half-integer values of $n_g$.
The resulting energy spectrum is always at least $2e$-periodic as a function of $n_g$, and it becomes $1e$-periodic if $\epsilon_r=0$.
If $\abs{\epsilon_r}<E_c$, as in figure~\ref{fig:parabolas}(a), there are two degeneracy points per period at which parabolas cross, otherwise only a single degeneracy point per period remains.

The effect of small but finite tunneling rates on the energy spectrum is shown in~figure~\ref{fig:parabolas}(b).
A small $\Gamma_1$ hybridizes the resonant level with the left superconductor, and thus opens avoided crossings at the degeneracy points between energy levels corresponding to $n$ and $n+\tfrac{1}{2}$ (with $n$ integer).
Conversely, a small $\Gamma_2$ hybridizes the resonant level with the right superconductor, and thus opens avoided crossings at the degeneracy points between energy levels corresponding to $n$ and $n-\tfrac{1}{2}$ (again, with $n$ integer).
If the tunneling rates are different, namely if $\delta\Gamma\neq 0$, the avoided crossing have different magnitudes.

These simple arguments indicate that the energy spectrum will be $2e$-periodic away from the resonant condition in which \emph{both} $\epsilon_r=0$ and $\delta\Gamma=0$.
At resonance, the energy spectrum is $1e$-periodic in $n_g$, as illustrated in figure~\ref{fig:parabolas}(c), since all the charge parabolas are aligned and the hybridization of the resonant level is balanced across the two leads.

Our discussion so far has been perturbative in nature, and it applies directly to the weak tunneling regime $T\Gamma_A \lesssim E_c$ of figure~\ref{fig:parabolas}.
However, the conclusions regarding the periodicity of the energy spectrum remain valid in the strong tunneling regime,
where they can be understood in terms of the relative strength of $2\pi$ and $4\pi$ phase slip amplitudes.
This will be the focus of the next section.

\section{WKB analysis}
\label{sec:wkb}

In this section we are going to derive approximate solutions for the energy levels of the Hamiltonian of Eq.~\eqref{eq:hamiltonian} under the boundary condition~\eqref{eq:boundary_conditions} using the WKB approximation.
The latter applies to the strong tunneling regime, defined as the parameter regime where the bandwidth of the Josephson potential is much larger than the charging energy: $T\,\Gamma_A \gg E_c$.
In this limit, the low-lying energy levels near the bottom of the potential are almost harmonic, with exponentially small corrections dictated by the tunneling under the potential barrier.
The calculation of the latter requires particular care near perfect transparency, $\abs{r}\ll 1$.

After moving the induced charge $n_g$ from the Hamiltonian to the boundary condition via a gauge transformation $\Psi\to e^{i\phi n_g}\Psi$, the problem to be solved is the stationary Schr\"{o}dinger equation
\begin{equation}\label{eq:sse}
- 4 E_c \Psi'' + V \Psi = (-\Gamma_A+E)\, \Psi\,.
\end{equation}
We have shifted the zero of the energy $E$ to the bottom of the Josephson potential, which is at energy $-\Gamma_A$, so that the eigenvalues are all positive.
We are interested in solutions near the bottom of the potential, $E\ll T\Gamma_A$.
In the WKB approximation, the solution $\Psi$ is taken to be a wave with a locally-varying wave vector
\begin{equation}\label{eq:kWKB}
k_\pm(\phi) = \sqrt{\frac{E-\Gamma_A\mp E_A(\phi)}{4E_c}}\,.
\end{equation}
where the $\pm$ index labels the two branches of the potential with energy $\pm E_A$.
The wave vector is real (imaginary) when $E$ is above (below) the potential energy.

The periodic boundary conditions~\eqref{eq:boundary_conditions} ensure that we need to solve Eq.~\eqref{eq:sse} in a $2\pi$ interval, say $[-\pi, \pi]$.
In this interval, the $-$ branch has a classically available region between the two turning points at $\pm \phi_c$, which are defined by the condition
\begin{equation}
E -\Gamma_A + E_A(\phi_c) = 0.
\end{equation}
On the other hand, the $+$ branch is classically forbidden in the entire interval, and thus for this branch the WKB ansatz consists of evanescent waves everywhere.

The WKB ansatz fails at the classical turning points, where the WKB momentum vanishes, and also, for small $r$, at $\phi=\pi$, because the adiabatic eigenstates (i.e. the spinors $\chi_s$ such that $V\chi_s=s E_A\chi_s$) rotate rapidly with the phase.
In both cases, it is possible to linearize the potential $V(\phi)$ at the problematic boundary and, from the solutions of the resulting differential equations, use the method of matching asymptotes to derive connection formulas for the WKB solutions on the two sides of the boundary.
At $\phi=\pm \phi_c$, the linearization involves only the $\sigma=-1$ energy branch and, as is well known, it leads to the Airy differential equation for the solutions close to the turning point~\cite{landau2013}.
In the case of the level crossing at $\phi=\pi$, the linearization involves both branches.
It leads to the $2 \times 2$ system of equations of the Landau-Zener problem with imaginary time~\cite{averin1999}, mathematically equivalent to a Weber differential equation whose solutions are parabolic cylinder functions \cite{gradshteyn2014}.

The result of these calculations, which are reproduced in detail in appendix~\ref{app:WKB}, is a bound state equation for the energy which takes the form:
\begin{equation}\label{eq:bound_state_equation}
\cos \sigma = w\, e^{-\tau}\cos(2\pi n_g + \delta) + e^{-\rho}e^{-\tau}\cos(4\pi n_g)
\end{equation}
On the left hand side, $\sigma$ is the integral of $k_-$ over the classically available region,
\begin{equation}\label{eq:sigma}
\sigma(E) = \int_{-\phi_c}^{\phi_c}\sqrt{\frac{E-\Gamma_A+E_A(\phi)}{4E_c}}\,d\phi\,.
\end{equation}
On the right hand side, $\tau$ and $\rho$ are WKB tunneling integrals, respectively under the smaller barrier of the $-$ branch and the larger barrier of the $+$ branch:
\begin{align}\label{eq:tau}
\tau(E) &= \int_{\phi_c}^{2\pi-\phi_c} \sqrt{\frac{\Gamma_A-E-E_A(\phi)}{4E_c}}\,d\phi\,,\\\label{eq:rho}
\rho(E) &= \int_{-\pi}^{\pi} \sqrt{\frac{\Gamma_A-E+E_A(\phi)}{4E_c}}\,d\phi\,.
\end{align}
Furhermore, on the right hand side of Eq.~\eqref{eq:bound_state_equation}, $w$ represents the amplitude for the wave function to remain on the lower branch when evolving through the avoided crossing. It is given by
\begin{equation}\label{eq:w}
w = \sqrt{\frac{2\pi}{\lambda}} \frac{e^{-\lambda}\,\lambda^\lambda}{\Gamma(\lambda)}\,,
\end{equation}
with 
\begin{equation}\label{eq:lambda}
\lambda = \frac{\abs{r}^2}{4}\,\frac{\Gamma_A}{\Gamma}\,\sqrt{\frac{\Gamma_A}{E_c}}
\end{equation}
the parameter controlling adiabaticity: $w$ tends to one for $\lambda \gg 1$ (adiabatic limit), while $w\sim \sqrt{2\pi\lambda}$ for $\lambda \ll 1$.
Note, in particular, that $w$ vanishes when $r=0$ (diabatic limit).
Finally, in Eq.~\eqref{eq:bound_state_equation}, $-\delta$ is the phase of the complex reflection coefficient $r$.

Before proceeding to solve the bound state equation, it is useful to discuss its structure.
The first and second term on the right hand side of Eq.~\eqref{eq:bound_state_equation} originate from $2\pi$ and $4\pi$ phase slips respectively, as revealed by their different periodicity with respect to the induced charge $n_g$.
The latter can be understood in terms of the Aharonov-Casher effect: in a $4\pi$ phase slip, the phase variable wraps around the circle twice, and so the wave function picks up a phase factor of $4\pi n_g$.
The comparison of the two terms also tells us that $2\pi$ phase slips dominate $4\pi$ phase slips when $w e^{\rho}\gg 1$, while in the opposite limit $w e^\rho \ll 1$ the $4\pi$-periodic component dominates.
Finally, we note that the appearance of the phase shift $\delta$ is a consequence of the twisted boundary conditions~\eqref{eq:boundary_conditions}.

Neglecting the occurrence of quantum phase slips means setting to zero the exponentially small tunneling amplitudes $e^{-\tau}$ and $e^{-\rho}$ on the right hand side of Eq.~\eqref{eq:bound_state_equation}.
In this case the left hand side yields a Bohr-Sommerfeld quantization condition for the energy levels $E_{n}$ in the Josephson potential,
\begin{equation}
\sigma(E_{n}) = \pi \left(n+\tfrac{1}{2}\right)\,,
\end{equation}
with $n=0, 1, 2\dots$ The effect of quantum phase slips can then be introduced as a small correction $\delta_n$ to the eigenvalues $E_{n}$ obtained via the Bohr-Sommerfeld condition.
This correction is the charge dispersion of the $n$-th energy level due to quantum phase slips.
Expanding the left hand side of Eq.~\eqref{eq:bound_state_equation} as described in appendix~\ref{app:WKB} leads to the expression
\begin{align}\label{eq:delta_n}\nonumber
\delta_n &= \frac{(-1)^{n+1}}{\sigma'_n}\,w e^{-\tau_n} \cos(2\pi n_g +\delta)\\\nonumber
& + \frac{(-1)^{n+1}}{\sigma'_n}\,e^{-\rho_n}e^{-\tau_n}\cos(4\pi n_g)\\&-\frac{\tau_n'}{2(\sigma_n')^2}w^2 e^{-2\tau_n}\,\cos(4\pi n_g + 2\delta)\,.
\end{align}
We adopted a shortened notation for the tunneling integrals evaluated at the eigenergies, e.g. $\tau_n \equiv \tau(E_{n})$.

\begin{figure*}[t!]
    \centering
    \includegraphics[width=0.8\textwidth]{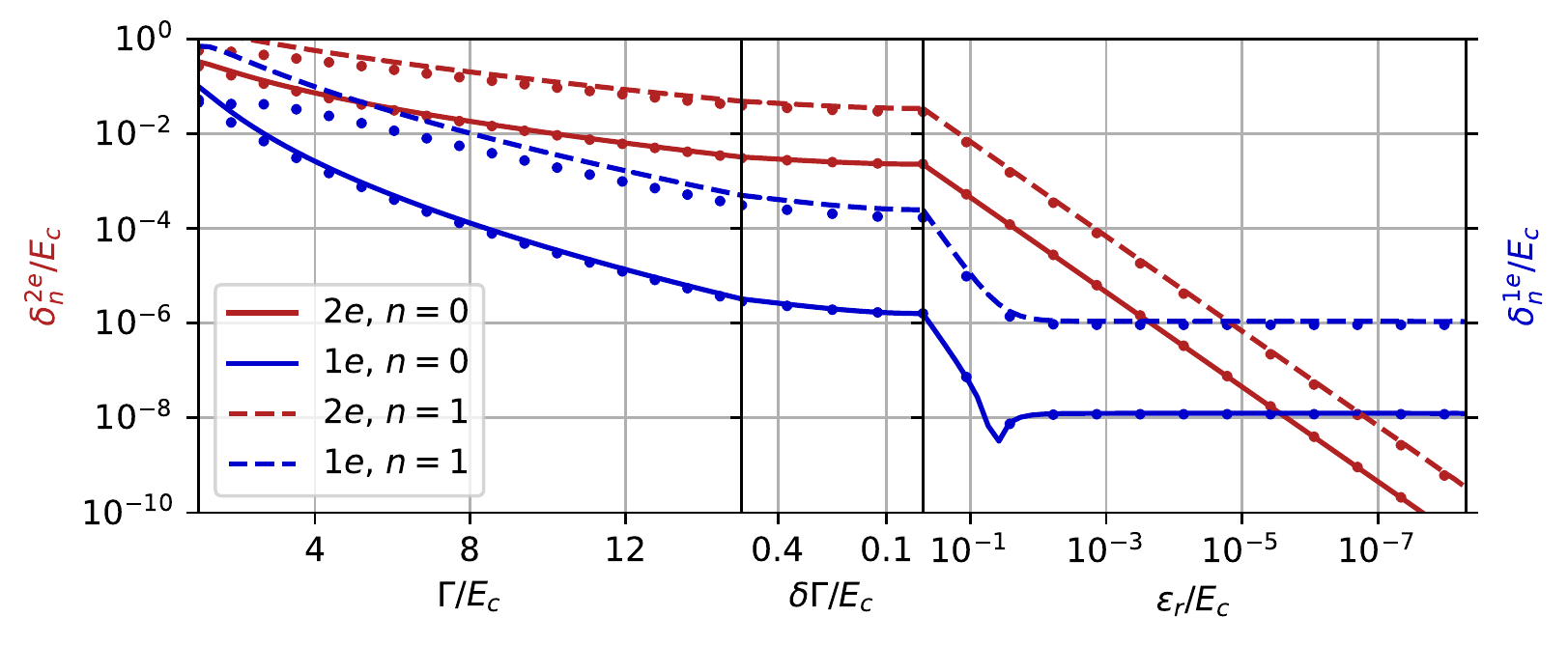}
    \caption{Dispersion of the energy levels of the resonant model of Eq.~\eqref{eq:hamiltonian} versus the model parameters, as the system is tuned from the de-tuned weak-tunneling regime (left end of the plot) to the resonant strong-tunneling regime (right end of the plot). The quantities shown are the $2e$- and $1e$-periodic components of the charge dispersion $\delta_{n}(n_g)$ of the $n$-th energy level, for $n=0$ and $n=1$. For each quantity we show both the WKB prediction (solid or dashed line) as well as numerical prediction via the diagonalization of the Hamiltonian (dots). In the left panel, $\Gamma/E_c$ is varied at fixed $\epsilon_r/E_c=0.5$ and $\delta\Gamma/E_c=0.5$. In the middle panel $\delta\Gamma/E_c$ is varied at fixed $\Gamma/E_c=15$ and $\epsilon_r/E_c=0.5$. In the third panel $\epsilon_r/E_c$ is varied at fixed $\Gamma/E_c=15$ and $\delta\Gamma=0$. Note that in the right panel the horizontal axis is also on a log scale.}
    \label{fig:comparison}
\end{figure*}

Equation~\eqref{eq:delta_n} is the central result of our paper: it describes the oscillations of the energy levels of the S-R-S transmon circuit as a function of the induced charge, including the effects of $2\pi$ and $4\pi$ quantum phase slips on equal footing.
The first term of Eq.~\eqref{eq:delta_n} gives the contribution to the charge dispersion coming from $2\pi$ phase slips, which coincides with the one computed in Ref.~\cite{averin1999,averin1999b}.
This term yields a charge dispersion with a period of $2e$ and it vanishes as $r\to 0$, since in this limit $w\to 0$.
The second term gives the contribution coming from $4\pi$ phase slips, which is finite in the limit $r\to 0$.
The last term is a $4\pi$-periodic correction to the first term, of higher order in the tunneling integral $\tau_n$.
We retain it here since, as $w$ increases, it becomes as large as the second term in the crossover between $2\pi$- and $4\pi$-dominated regimes, and eventually larger when $w\approx 1$.

Our next goal is to compare these analytical results with numerical results. To do so, we provide approximate expressions for the quantities appearing in Eq.~\eqref{eq:delta_n} in terms of the model parameters. To begin with, in the limit $T\Gamma_A\gg E_c$ in which it is appropriate to approximate the potential as a parabola, the Bohr-Sommerfeld condition gives the harmonic spectrum
\begin{equation}\label{eq:harmonic_spectrum}
E_{n}=\sqrt{2T\,\Gamma_A E_c}\,\left(n+\tfrac{1}{2}\right)\,\equiv \omega_p\,\left(n+\tfrac{1}{2}\right).
\end{equation}
We introduced the Josephson plasma frequency $\omega_p$ for later convenience. The anharmonic corrections to $E_{n}$ are of order $\sqrt{E_c/T\Gamma_A}$ and will be neglected. 

Evaluating the tunneling integrals at these energies we obtain
\begin{align}\label{eq:tau_n}
e^{-\tau_n} &= \frac{\sqrt{2\pi}}{n!}\left(\frac{b^2\omega_p}{4E_c}\right)^{n+\tfrac{1}{2}}\,e^{-a\, \omega_p/E_c}\\
e^{-\rho_n} &=e^{-(c/\sqrt{T})\,\omega_p/E_c+d \sqrt{T}\,(n+1/2)}\label{eq:rho_n}
\end{align}
where $a,b,c,d$ are positive numerical coefficients that depend weakly on $T$, and whose explicit expressions are given in appendix~\ref{app:integrals}. Finally, we also find
\begin{align}\label{eq:sigmap_n}
\sigma'_n &= \frac{\pi}{\omega_p}\,,\\\label{eq:taup_n}
\tau'_n &= \frac{1}{\omega_p}\,\log\frac{4E_c(n+\tfrac{1}{2})}{b^2 \omega_p}\,.
\end{align}
By simple replacement of Eqs.~\eqref{eq:tau_n}-\eqref{eq:taup_n} into Eq.~\eqref{eq:delta_n}, it is possible to obtain explicit asymptotic expressions for the different contributions to the charge dispersion as a function of the model parameters.

\section{Results}

Armed with these expressions, we can compare the energy levels obtained from the WKB ansatz with those obtained from a numerical diagonalization of the Hamiltonian~\eqref{eq:hamiltonian} in the charge basis.
The comparison serves both as a verification of the results obtained analytically and as a way to illustrate the behavior of the quantum phase slips amplitude versus the model parameters.
To do so, it is convenient to extract the $2e$- and $1e$-periodic components of the charge dispersion $\delta_n(n_g)$:
\begin{equation}
\delta_n(n_g) = \delta_n^{2e} \cos(2\pi n_g + \beta_n^{2e}) + \delta_n^{1e} \,\cos(4\pi n_g + \beta_n^{1e})\,
\end{equation}
This equation is just a re-writing of the right hand side of Eq.~\eqref{eq:delta_n} as a Fourier series.
In particular, $\delta_n^{2e}$ tracks the amplitude of the first term in Eq.~\eqref{eq:delta_n}, originating from $2\pi$ phase slips, while $\delta_n^{1e}$ tracks the amplitude of the second and third term in Eq.~\eqref{eq:delta_n}, originating from $4\pi$ phase slips; $\beta^{2e}_n$ and $\beta_n^{1e}$ are the corresponding total phase shifts.

In figure~\ref{fig:comparison}, we show the evolution of $\delta_n^{2e}$ and $\delta_n^{1e}$ for both the ground ($n=0$) and first excited ($n=1$) states, as the three model parameters $\Gamma, \delta\Gamma$ and $\epsilon_r$ are swept at fixed $E_c$.
The parameter sweep is such that the left end of the figure corresponds to the weak tunneling regime ($\Gamma=E_c$), finite asymmetry ($\delta\Gamma/E_c = 0.5$), \emph{and} finite detuning from resonance ($\epsilon_r/E_c = 0.5$).
On the other hand, the right end of the figure corresponds to the strong tunneling regime ($\Gamma/E_c = 15$) \emph{and} the resonant condition $\delta\Gamma=\epsilon_r=0$.

The first panel shows the exponential suppression of the charge dispersion as the tunneling rate $\Gamma$ is increased at fixed $\delta\Gamma$ and $\epsilon_r$.
This behavior is familiar from conventional transmon model \cite{koch2007} and it originates from the increase in the Josephson potential barrier height due to the increase of $\Gamma$.
The second panel shows that the trend continues as the asymmetry $\delta\Gamma$ is tuned to zero at fixed $\Gamma$ and $\epsilon_r$.
This is because the effect of decreasing $\delta\Gamma$ at fixed detuning is to increase $T\Gamma_A$ and, thus, the Josephson potential barrier height.
Up to now, both $\delta_n^{2e}$ and $\delta_n^{1e}$ exhibit a similar trend, because in these parameter ranges their magnitudes are both controlled by the exponent $\tau_n$.

\begin{figure}[t!]
    \centering
    \includegraphics[width=\columnwidth]{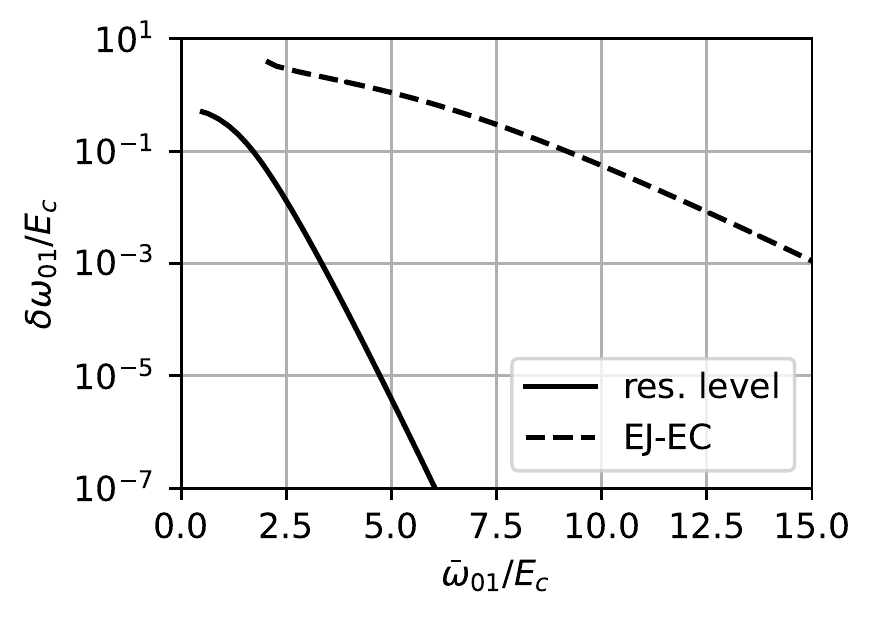}
    \caption{Comparison of the charge dispersion scaling in the resonant model (solid line) versus the traditional Cooper-pair box (transmon) model. We plot the peak-to-peak amplitude of the charge dispersion of the fundamental frequency $\omega_{01}=E_1-E_0$ versus the averaged (over $n_g$) value of $\omega_{01}$. For the resonant model, the curve shown is obtained varying the ratio $\Gamma/E_c$ with $\delta\Gamma=\epsilon_r=0$, while for the transmon model of Eq.~\eqref{eq:h_transmon} it is obtained varying $E_J/E_c$. In the first case, $\delta_{01}$ is dictated by $4\pi$ phase slips under a $-\Gamma \cos(\phi/2)$ barrier, while in the second case by $2\pi$ phase slips under a $-E_J \cos\phi$ barrier.}
    \label{fig:resonance_scaling}
\end{figure}

The third panel of figure~\ref{fig:comparison} shows the effect of tuning the level to resonance.
The $2\pi$ phase slip amplitude $\delta_n^{2e}$ drops to zero linearly towards resonance, because as the reflection coefficient $r$ approaches zero, non-adiabatic effects related to the narrowly avoided crossing at $\phi=\pi$ start to kick-in, and the Landau-Zener parameter $w$ vanishes.
On the other hand, the $4\pi$ phase slip amplitude $\delta_n^{1e}$ saturates to a finite value determined by the exponent $\rho_n$, which is not sensitive to the closing of the avoided crossing.
Eventually, the $4\pi$-periodic component overcomes the $2\pi$-periodic component of the charge dispersion at a value of $\epsilon_r$ determined by the condition $w\approx e^{-\rho_n}$, which depends slightly on $n$, as the figure shows.
This crossover is well captured by the WKB solutions.
In fact, figure~\ref{fig:comparison} shows that the agreement between the asymptotic WKB results and the numerically determined eigenvalues is reasonable even at values of $\Gamma/E_c$ not much larger than one, especially for the ground state $n=0$.

The right panel of figure~\ref{fig:comparison} also shows that if $\Gamma\gg E_c$, the crossover to the $4\pi$-dominated regime only happens very close to resonance and at charge dispersion levels so small to be practically unobservable.
For instance, in figure~\ref{fig:comparison}, $\delta_n^{1e}$ saturates at a value of order $10^{-6}\, E_c$ for $n=1$, reached when $\epsilon_r \approx 10^{-5}\, E_c$.
However, the effect becomes more striking, and experimentally detectable, when the ratio $\Gamma/E_c$ is reduced.

To highlight this, in figure~\ref{fig:resonance_scaling} we show the scaling of the charge dispersion when the tunneling strength $\Gamma$ is varied while maintaining the resonant condition.
Here we focus on the average energy difference $\bar{\omega}_{01}~=~\int_0^1 dn_g \,(E_1-E_0)$, where $E_1$ and $E_0$ are the numerically determined eigenvalues of the Hamiltonian, and on the peak-to-peak amplitude $\delta\omega_{01}$ of its charge dispersion $\delta_1-\delta_0$.
These are the quantities that can be more easily measured in a typical microwave spectroscopy experiment such as those in Refs.~\cite{kringhoj2020,bargerbos2020}, which we have in mind as a feasible way to test our predictions.
We note that, in principle, the charge dispersion of energy levels is also accessible in the I-V characteristic of the junction~\cite{likharev1985,corlevi2006,doucot2007}.

Furthermore, we compare the behavior predicted by the resonant level model with that of a conventional transmon device described by the Hamiltonian
\begin{equation}\label{eq:h_transmon}
H = 4 E_c (i\partial_\phi+n_g)^2- E_J\cos\phi
\end{equation}
with periodic boundary conditions on a $2\pi$ interval.
In the resonant level model, $\bar{\omega}_{01}$ and $\delta\omega_{01}$ were both computed numerically for increasing $\Gamma/E_c$ at fixed $\delta\Gamma=0$ and $\epsilon_r=0$.
For the transmon model, the same quantities were instead computed increasing $E_J/E_c$, and they reproduce the well-known curve for the charge dispersion of a transmon \cite{koch2007}.
Via the parametric plot of the observable quantities $\bar{\omega}_{01}$ and $\delta\omega_{01}$, computable for both models despite the different set of parameters, a direct comparison can be made.

The comparison shows that, while the charge dispersion decays exponentially in both models, the effect is much stronger in the presence of a resonant level.
This is because we are essentially comparing the tunneling amplitude under a $-\Gamma \cos(\phi/2)$ barrier and that under a $-E_J \cos\phi$ barrier: the former corresponds to a higher potential and a longer tunneling path, and is therefore exponentially smaller than the latter.
Thus, as Refs.~\cite{bargerbos2020,kringhoj2020} pointed out, resonant tunneling provides a way to reach a target charge dispersion while keeping the superconducting island closer to the Cooper-pair box limit of weak tunneling ($\Gamma \gtrsim E_c$ rather than $\Gamma\gg E_c$).
For instance, in order to achieve $\delta_{01}/E_c \approx 10^{-3}$ it is necessary to reach a ratio $\omega_{01}/E_c\approx 15$ (that is, $E_J/E_c\approx 32$) in the model of Eq.~\eqref{eq:h_transmon}, but it may be enough to reach the ratio $\omega_{01}/E_c \approx 3$ (that is, $\Gamma/E_c\approx 5$) using the resonant level model of Eq.~\eqref{eq:hamiltonian}.

\begin{figure}
    \centering\includegraphics[width=\columnwidth]
{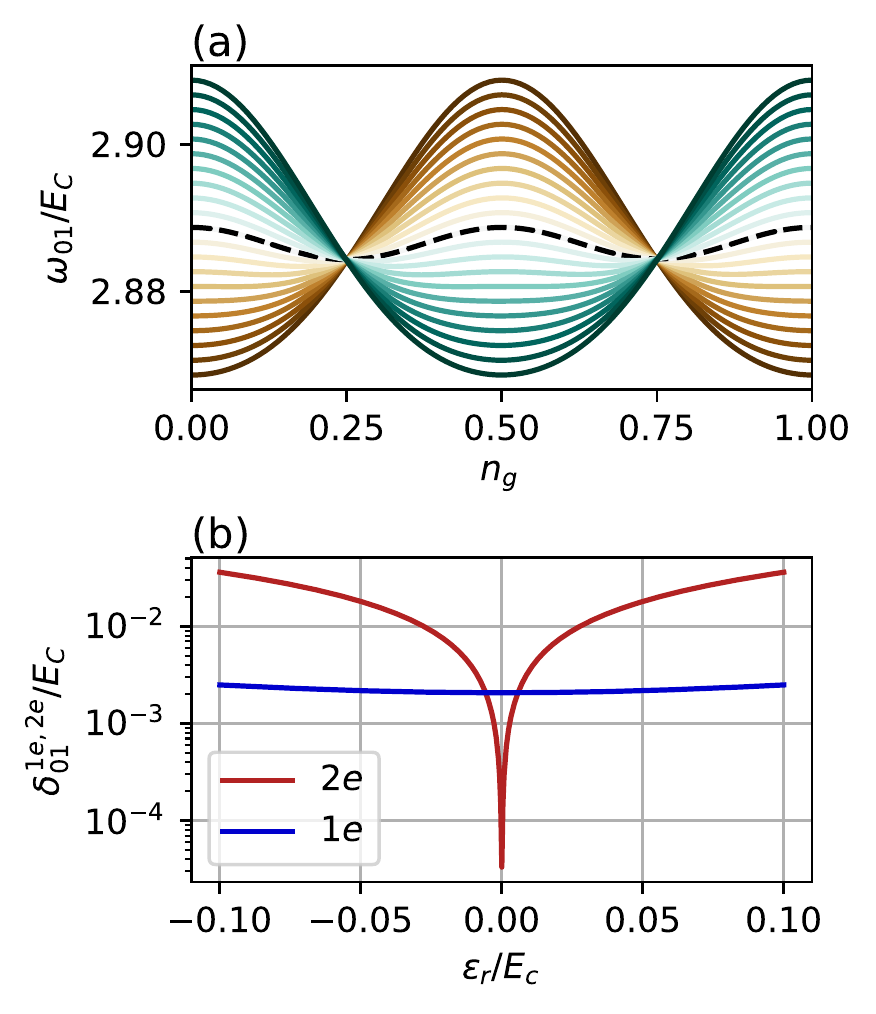}
    \caption{(\emph{a}): Evolution of the energy difference $\omega_{01} = E_1-E_0$, where $E_1$ and $E_0$ are the two lowest eigenvalues of Eq.~\eqref{eq:hamiltonian}, determined numerically, as a function of $n_g$, for different values of the detuning $\epsilon_r$ varying between $\epsilon_r/E_c=0.04$ (dark green) to $\epsilon_r/E_c=-0.04$ (dark brown). The black dashed line emphasizes the doubling of the periodicity at $\epsilon_r=0$. Other parameters are $\Gamma/E_c=5$, $\delta\Gamma / E_c = 10^{-4}$. (\emph{b}): Amplitudes of the $2e$- and $1e$-periodic components of the charge dispersion as the resonant level is swept through resonance.}\label{fig:epsilon_scaling}
\end{figure}

This fact is convenient for qubit design, since it mitigates a practical trade-off at play in the transmon: reducing the charge dispersion increases the dephasing time, but at the cost of an increase of device footprint and capacitive losses, due to the need for a large capacitor.
However, the suppression of $2\pi$ phase slips, which is at the basis of the advantageous scaling of figure~\ref{fig:resonance_scaling}, requires fine-tuning the junction to a resonance.
Thus, the effect will be very sensitive with respect to noise, especially to noise in the detuning parameter $\epsilon_r$, which would originate from charge noise in the gates required to tune the resonant level.

To illustrate this important point, in figure~\ref{fig:epsilon_scaling} we show the evolution of $\omega_{01}(n_g)$ as $\epsilon_r$ is varied from positive to negative through zero, in the case of a rather weak tunneling $\Gamma/E_c = 5$.
In the top panel, we see how the charge dispersion evolves from a conventional $2e$-periodic oscillation with a maximum at $n_g=0$ ($\epsilon_r>0$), to a $1e$-periodic curve at resonance ($\epsilon_r=0$, black dashed line), to a shifted $2e$-periodic curve with a maximum at $n_g=1/2$ ($\epsilon_r<0$).
The plot illustrates how the suppression of the charge dispersion occurs because the charge dispersion changes sign as $\epsilon_r$ passes through zero, signaling the ground state occupation of the resonant level by a Cooper pair when $\epsilon_r<0$. 
Neglecting $4\pi$ phase slips, the dashed line at $\epsilon_r=0$ would be flat.

In the bottom panel of figure~\ref{fig:epsilon_scaling} we show the $2e$- and $1e$-periodic amplitudes $\delta_{01}^{1e}\equiv \delta_1^{1e}-\delta_0^{1e}$ and $\delta_{01}^{2e}\equiv \delta_1^{2e}-\delta_0^{2e}$, extracted from the curves in the top panel (computed in a wider $\epsilon_r$ range).
The $4\pi$ phase slip amplitude stays approximately constant, while the $2\pi$ phase slip amplitude goes through a dip at resonance, with its minimum value at $\epsilon_r=0$ determined by the presence of a small, residual asymmetry ($\delta\Gamma \approx 10^{-4} E_c$ in figure~\ref{fig:epsilon_scaling})
While the region dominated by $4\pi$ phase slips has widened with respect to the right panel of figure~\ref{fig:comparison} due to the smaller ratio $\Gamma/E_c$, it still occurs in a relatively narrow interval, $\abs{\epsilon_r}/E_c \lesssim 0.01$.
The dephasing time of the plasma oscillation would be dictated by $4\pi$ phase slips only if time-dependent noise in the detuning parameter $\epsilon_r$ were to be contained in this interval.
Nevertheless, the plot also shows that in this parameter regime it would be feasible, with reasonable experimental resolution, to detect the occurrence of $4\pi$ phase slips at resonance via a spectroscopic measurement of the $\omega_{01}(n_g)$ curve.
Indeed, the residual charge dispersion at resonance is $\approx 2\times 10^{-3} E_c$ in figure~\ref{fig:epsilon_scaling}, and thus it would fall in the MHz frequency range for realistic values of $E_c/h \sim 1$~GHz.

\section{Conclusions}

We have studied in detail the quantum phase slips occurring in a Josephson junction in the presence of a resonant level mediating the tunneling of Cooper pairs.
It was known since Ref.~\cite{averin1999} that $2\pi$ phase slips are fully suppressed in the presence of a level crossing in the Andreev spectrum.
Here, we have extended this result by computing the amplitude of $4\pi$ phase slips, which remain finite in the presence of a level crossing and provide the mechanism by which the charge dispersion of the superconducting island remains finite, albeit possibly very small.
Our central result is Eq.~\eqref{eq:delta_n}: obtained within the WKB approximation, it provides asymptotic expressions for the energy levels of a Cooper-pair box in the transmon limit, including the effect of both $2\pi$ and $4\pi$ quantum phase slips, and yielding results in good agreement with numerical simulations.
To conclude our paper, we discuss several implications of our results.

\subsection{Experimental observability of $4\pi$ phase slips in transmon circuits}

The suppression of $2\pi$ phase slips occurs in a fairly narrow parameter range near resonance ($\epsilon_r=0$) and symmetric barriers ($\delta\Gamma=0$).
Within this parameter range, a crossover to a regime dominated by $4\pi$ phase slips occurs (see figure~\ref{fig:epsilon_scaling}).
The width of the crossover region around resonance, as well as the residual level of charge dispersion at resonance given by $4\pi$ phase slips, both increase with decreasing $\Gamma/E_c$.

Although the suppression of $2\pi$ phase slips at resonance has been observed in Refs.~\cite{kringhoj2020,bargerbos2020}, coherent $4\pi$ quantum phase slips were not observed.
We attribute this fact to the large ratio $\Gamma/E_c$ of those measurements.
Our calculations predict that coherent $4\pi$ quantum phase slips should be observable with the same technology of existing experiments, only in devices with larger charging energy.
For instance, let us consider a situation in which $E_c/h=1$~GHz, $\epsilon_r=\delta\Gamma=0$ and $\Gamma/h=3$~GHz.
Then, our model predicts that $\omega_{01}\approx 2.16$~GHz while $\delta\omega_{01}\approx 33$~MHz, easily in the range of detectable frequency shifts.

The direct comparison with a transmon qubit based on a conventional tunnel junction with Josephson energy $-E_J \cos\phi$ shows that the resonant level provides a much lower charge dispersion at a fixed ratio of the qubit frequency to the charging energy (see figure~\ref{fig:resonance_scaling}).
We have discussed critically the possible implications of this fact for qubit design, emphasizing that the circuit is likely to remain sensitive to charge noise modulating the energy of the resonant level.

\subsection{Connection to novel qubit designs}

Our results are relevant for the recently introduced bi-fluxon qubit~\cite{kalashnikov2020}, which uses a superconducting island tuned to the charge degeneracy point as a way to implement resonant Cooper pair tunneling with a $4\pi$-periodic effective Josephson energy.
Indeed, the model of Eqs.~\eqref{eq:hamiltonian} and \eqref{eq:potential_energy} also applies to such a case:
the two degenerate charge states of the island, with charge differing by $2e$, map to the resonant level in our model being empty or occupied.
In this mapping, the parameters $\epsilon_r$ and $\delta\Gamma$ indicate the detuning from the charge degeneracy point of the island and the asymmetry between two tunnel junctions.
For noise protection, the bi-fluxon qubit relies on the suppression of $2\pi$ quantum phase slips and ideally operates in a regime where only $4\pi$ quantum phase slips are present.
Our detailed results on the competition of $2\pi$ and $4\pi$ quantum phase slips, especially at finite detuning or junction asymmetry, are therefore relevant for its design.

A difference between the S-R-S transmon model studied in this paper and the bi-fluxon is that the circuit of the latter features an inductive shunt, similar to the fluxonium circuit~\cite{manucharyan2009}.
In the presence of an inductive loop, quantum phase slips couple coherently persistent current states characterized by a differing number of fluxons trapped in the loop \cite{koch2009}.
By tuning the applied flux, it is therefore possible to measure separately the amplitude for $2\pi$ and $4\pi$ phase slips, making such a device ideal to observe the crossover between $2\pi$ and $4\pi$-dominated regimes.
In fact, a fluxonium circuit with a weak link of the S-R-S type could be a competitive version of the bi-fluxon qubit.
We leave the analysis of this topic to future work.

\subsection{Connection to Majorana zero modes}

Our calculations also have a close connection with models of superconducting islands with Majorana zero modes (MZMs)~\cite{fu2010,pikulin2019}.
It is known that the $4\pi$ Josephson effect occurring in a junction between topological superconductors (due to the presence of a pair of coupled MZMs) \cite{kitaev2001,fu2009} suppresses the occurrence of $2\pi$ phase slips, leaving only the occurrence of $4\pi$ phase slips~\cite{pekker2013,rodriguez-mota2019,svetogorov2020}.
Even the boundary condition of Eq.~\eqref{eq:boundary_conditions} has a precise counterpart in models with topological superconducting islands, where it arises due to a fermion parity constraint on the BCS wave function \cite{fu2010,heck2011}.
In fact, the model of Eq.~\eqref{eq:hamiltonian}, together with the boundary conditions, can be mapped exactly to a model of four MZMs, two per superconducting side, coupled across a weak link.
Such a model of four coupled MZMs could arise, for instance, because of finite-size effects in a topological nanowire~\cite{pikulin2012}.

\subsection{Generality of our results}

Finally, let us discuss the generality of our results.
The regime with dominating $4\pi$ phase slips should persist even outside of the strict domain of validity of the model in Eq.~\eqref{eq:potential_energy}, because it is a consequence of the presence of a level crossing in the Andreev spectrum rather than of the precise form taken by the Josephson potential energy.
For instance, the assumption $\Gamma \ll \Delta$ could be relaxed; doing so would modify the phase dependence of the Andreev spectrum and the precise values of the WKB integrals, but not the essential feature that $2\pi$ phase slips are suppressed at resonance.

Similar conclusions can be drawn about multi-channel extensions of the single-channel model of Eq.~\eqref{eq:potential_energy}.
If the additional channels are not resonant, they simply provide a $2\pi$-periodic contribution to the Josephson energy (a similar contribution is also provided by the above-gap, continuous part of the spectrum).
This contribution will increase the height of the Josephson potential barrier, and thus lower all the quantum phase slips amplitudes, but it will not affect the resonant suppression of $2\pi$ phase slips illustrated in figure~\ref{fig:epsilon_scaling}.
The resonant suppression is controlled by the parameter $w$ of Eq.~\eqref{eq:w}, and thus by the most transparent channel only.
Qualitative deviations from our central result, Eq.~\eqref{eq:delta_n} are therefore only expected in the fine-tuned case where more than one transport channel achieves near-perfect transparency ($\abs{r}^2\ll\sqrt{E_c/\Gamma_A}$).

Our results also remain valid in the presence of a finite interaction energy $U$ for the double-occupancy of the resonant level, a term neglected in this paper.
This is true at least as long as $U \ll \Gamma$, since such a weak interaction would only renormalize the couplings in the effective Hamiltonian of Eq.~\eqref{eq:hamiltonian} \cite{kurilovich2021}.
For larger $U$, a transition to an odd-parity doublet ground state occurs close to resonance, diminishing the relevance of Eq.~\eqref{eq:hamiltonian}, which applies to an even-parity singlet ground state.
The study of quantum phase slips when the junction is in the doublet ground state is an interesting problem left to future research.

\acknowledgements{
We acknowledge helpful discussions with Marta Pita Vidal, Arno Bargerbos, Dmitry Pikulin, Vadim Cheianov, Carlo Beenakker and Valla Fatemi. This project has received funding from European Research Council (ERC) under the European Union’s Horizon 2020 research and innovation programme. BvH thanks Tjerk Oosterkamp for support through the Dutch Research Council (NWO).
}

The code and notebooks used to generate the numerical results in this work are available on Zenodo~\cite{zenodo}.

\appendix

\section{Derivation of the low-energy Hamiltonian}
\label{app:effective_hamiltonian}

In this appendix, we derive Eq.~\eqref{eq:hamiltonian} starting from the model of a level tunnel-coupled to two superconductors.
Similar derivations have appeared in the literature before, e.g. in Refs.~\cite{meng2009,kurilovich2021}.
Here we propose a simple derivation that motivates and clarifies the use of the boundary conditions of Eq.~\eqref{eq:boundary_conditions}.
The starting point is the following Hamiltonian:
\begin{equation}
H = H_\textrm{sc} + H_\textrm{dot} + H_\textrm{tunn}+ H_\textrm{c}\,.
\end{equation}
The first term $H_\textrm{sc}$ is the Hamiltonian of the two superconductors,
\begin{equation}
H_\textrm{sc} = \sum_{\alpha n \sigma} \xi_n\,c^\dagger_{\alpha n \sigma} c_{\alpha n \sigma} - \Delta\,\sum_{\alpha n}\,\left(e^{-i\phi_\alpha} c^\dagger_{\alpha \uparrow} c^\dagger_{\alpha \downarrow} + \textrm{h.c.}\right)
\end{equation}
where $\alpha=1,2$ denotes the two leads, $n$ enumerates their spin-degenerate single-particle states with energy $\xi_n$, $\sigma=\uparrow,\downarrow$ is the spin quantum number, $\Delta$ is the pairing gap, and $\phi_\alpha$ is the superconducting phase in the two leads.

The second term is the Hamiltonian of the resonant level:
\begin{equation}
H_\textrm{dot} = \epsilon_r \sum_\sigma \left(d^\dagger_\sigma d_\sigma-\tfrac{1}{2}\right)\,,
\end{equation}
where the operator $d^\dagger_\sigma, d_\sigma$ create and annihilate an electron with spin $\sigma$ on the resonant level. For simplicity, we omit an Anderson $U$. The limitations of this choice are discussed in the main text and are not crucial for what follows.
The third term is the tunneling Hamiltonian between the leads and the energy level in the dot:
\begin{equation}
H_\textrm{tunn} = \sum_{\alpha n \sigma} t_\alpha \left(d^\dagger_\sigma c_{\alpha n \sigma} + \textrm{h.c.}\right)\,.
\end{equation}
Again for simplicity, we only consider spin-conserving tunneling. In the presence of both time-reversal symmetry and spin-rotation symmetry, the couplings $t_\alpha$ can be chosen to be real. 

Finally, the last term is the charging energy between the two leads:
\begin{equation}
H_\textrm{c} = 4E_c (N - n_g)^2
\end{equation}
where $E_c=e^2/2C$ is the charging energy and $n_g$ the dimensionless charge induced by gates, and $N$ is the charge transferred between the two leads. Both $N$ and $n_g$ are expressed in units of the Cooper pair charge $2e$. Explicit expressions for $E_c$ and $n_g$ in terms of the capacitances and gate voltages of a capacitive network of two islands are given in Ref.~\cite{nazarov2009quantum}.
In writing the charging energy, we have neglected the capacitance between the superconductors and the quantum dot hosting the energy levels, as well as the capacitance between the superconductors and any gates which may control the quantum dot.

At the mean-field level description of superconductivity, $N$ is an operator which includes separate contributions from both the paired and unpaired electrons:
\begin{equation}
N = \frac{1}{2}(N_1 - N_2) + \frac{1}{4}\sum_{n\sigma} \left(c^\dagger_{1n\sigma} c_{1n\sigma} - c^\dagger_{2n\sigma} c_{2n\sigma}\right)\,.
\end{equation}
Here, we denoted with $N_1, N_2$ the number of Cooper pairs in each superconductor. They are operators with integer spectrum obeying the following commutation rules:
\begin{equation}
[N_\alpha, e^{\pm i\phi_\beta}] = \pm \delta_{\alpha\beta}\,e^{\pm i\phi_\beta}\,.
\end{equation}
We stress the fact that the operator $N$ keeps count of the charge \emph{transferred} between the superconductors in units of $2e$.
Thus, a transfer of a Cooper pair between superconductors ($N_1\to N_1\pm 1, N_2\to N_2 \mp 1$) changes $N$ by one unit (e.g. $N\to N\pm 1$). On the other hand, a transfer of a single electron changes $N$ by $\pm(1/2)$. Simply, yet amusingly, the transfer of a Cooper pair from either superconductor to the quantum dot also changes $N$ by $\pm (1/2)$.

It is convenient to use a gauge transformation that removes the operators $e^{i\phi_\alpha}$ from $H_\textrm{sc}$ and which also simplifies the form of the charging energy~\cite{keselman2019}. The gauge transformation is $H\to U H U^\dagger$, with:
\begin{equation}
U = U_1 U_2\,,\; U_\alpha = \exp\left(\frac{i \phi_\alpha}{2} \sum_{n \sigma} c^\dagger_{\alpha n \sigma} c_{\alpha n \sigma }\right)
\end{equation}
In this new gauge, we have the following changes:
\begin{align}\nonumber
H_\textrm{sc} &\to \sum_{\alpha n \sigma} \xi_n\,c^\dagger_{\alpha n \sigma} c_{\alpha n \sigma} - \Delta\,\sum_{\alpha n}\,\left(c^\dagger_{\alpha \uparrow} c^\dagger_{\alpha \downarrow} + \textrm{h.c.}\right)\,,\\\nonumber
H_\textrm{c} &\to 4E_c (N-n_g)^2\,,\\
H_\textrm{tunn} &\to \sum_{\alpha n \sigma} t_\alpha \left(e^{-i\phi_\alpha/2}\,d^\dagger_\sigma c_{\alpha n \sigma} + \textrm{h.c.}\right)\,,
\end{align}
and $H_\textrm{dot} \to H_\textrm{dot}$. Note how the tunneling terms now contain operators $e^{\pm i \phi_\alpha/2}$, which shift $N_\alpha$ by one half.

The next step is to diagonalize $H_\textrm{sc}$ and rewrite the tunneling Hamiltonian in terms of Bogoliubov quasiparticle operators:
\begin{align}
c_{\alpha n\uparrow} &= u_{\alpha n} \Gamma_{\alpha n \uparrow} + v_{\alpha n}\Gamma^\dagger_{\alpha n \downarrow}\\
c_{\alpha n\downarrow} &= u_{\alpha n} \Gamma_{\alpha n \downarrow} - v_{\alpha n}\Gamma^\dagger_{\alpha n \uparrow}
\end{align}
with $u^2_{n} = \tfrac{1}{2}(1+ \xi_{n}/\epsilon_{n})$, $v^2_{n} = \tfrac{1}{2}(1  - \xi_{n}/\epsilon_{n})$, and $\epsilon^2_{n} = \xi_{n}^2 + \Delta^2$. After the Bogoliubov rotation, the Hamiltonian changes as follows:
\begin{align*}
H_\textrm{sc} &\to \sum_{\alpha n \sigma} \epsilon_n\,\Gamma^\dagger_{\alpha n \sigma} \Gamma_{\alpha n \sigma}\\
H_\textrm{tunn} &\to \sum_{\alpha n \sigma} t_\alpha \left[e^{-i\phi_\alpha/2}\,d^\dagger_\sigma  \left(u_{n} \Gamma_{\alpha n \sigma} + \sigma v_{n}\Gamma^\dagger_{\alpha n \bar\sigma}\right)\right.\\
&\qquad\qquad\left.+ e^{i\phi_\alpha/2}\, \left(u_{n} \Gamma^\dagger_{\alpha n \sigma} + \sigma v_{n}\Gamma_{\alpha n \bar\sigma}\right)d_\sigma\right]\,,
\end{align*}
with the other terms left untouched.

At this point, assuming that $\Delta$ is the largest energy scale in the problem, we would like to integrate out the quasi-particles in the leads and derive an effective Hamiltonian describing the low-energy coupled dynamics of the condensate and of the quantum dot.
Assuming that the total number of electrons in the system is even, a generic wave function in the even-parity low-energy space can be written as
\begin{align}
\ket{\Psi} &= \sum_{n\in\mathbb{Z}} \Psi_0(n) \ket{n}\ket{0} + \sum_{n\in\mathbb{Z}+\tfrac{1}{2}} \,\Psi_2(n) \ket{n}\ket{2}\,,
\end{align}
where $\ket{n}$ are states with a given number of Cooper pairs transferred: $N\ket{n} = n\ket{n}$, $\ket{0}$ denotes the empty dot state, and $\ket{2} = d^\dagger_\uparrow d^\dagger_\downarrow \ket{0}$ denotes the state in which the dot is occupied by a pair.

Using old-fashioned perturbation theory to the second order in the tunneling term, and integrating out states with unpaired quasiparticles, we obtain the following eigenvalue problem, written in terms of the wave function amplitudes $\Psi_0(n)$ and $\Psi_2(n)$:
\begin{widetext}
\begin{align}\label{eq:simplified_equations}
\left[E - 4E_c(n-n_g)^2 + \epsilon_r\right] \,\Psi_0(n) &= -\Gamma_1\,\Psi_2(n-\tfrac{1}{2}) - \Gamma_2\,\Psi_2(n+\tfrac{1}{2})\,, \\
\left[E - 4E_c(n-n_g)^2 - \epsilon_r\right] \,\Psi_2(n) &= -\Gamma_2\,\Psi_0(n-\tfrac{1}{2}) - \Gamma_1\,\Psi_0(n+\tfrac{1}{2})\,,
\end{align}
\end{widetext}
where $\Gamma_\alpha = \sum_n (2t_\alpha^2 v_n u_n)/\epsilon_n = \pi t_\alpha^2/\delta_\alpha$, with $\delta_\alpha$ the level spacing in the superconductor. A Fourier series,
\begin{align}\label{eqs:phase_basis}
\Psi_0(\phi) &= \sum_{n\in\mathbb{Z}} e^{i\phi n}\, \Psi_0(n)\,,\\
\Psi_2(\phi) &= \sum_{n\in\mathbb{Z}+\tfrac{1}{2}} e^{i\phi n}\, \Psi_2(n)\,,
\end{align}
yields the effective Hamiltonian of the main text, acting on the spinor wave function
\begin{equation}
\Psi(\phi) = \begin{bmatrix}
\Psi_0(\phi) \\ \Psi_2(\phi)
\end{bmatrix}
\end{equation}
The boundary condition of Eq.~\eqref{eq:boundary_conditions} follows from the fact that $\Psi_0(\phi+2\pi) = \Psi_0(\phi)$ while $\Psi_2(\phi+2\pi)=-\Psi_2(\phi)$.

\section{WKB solution}\label{app:WKB}

In this appendix we derive the bound state equation~\eqref{eq:bound_state_equation} of the  main text, applying the WKB approach to the Schr\"{o}dinger equation $H\Psi = (-\Gamma_A +E)\Psi$ for the Hamiltonian in Eq.~\eqref{eq:hamiltonian}.

We find it convenient to rotate the Hamiltonian such that the $\cos(\phi/2)$ term in the potential appears on the diagonal: the basis of the eigenstates of $V(\phi)$ at $\epsilon_r=0$.
The transformation consists of a rotation of $\Psi$ by $-\pi/2$ around the $y$-axis.
Simultaneously, as already mentioned in the main text, we also multiply the wave function by a phase that gets rid of $n_g$ in the Hamiltonian, so that the transformation is
\begin{align}
\Psi &\to e^{i\phi n_g}\,e^{i(\pi/4)\tau_y} \Psi\,\\
H& \to e^{i\phi n_g}e^{i(\pi/4)\tau_y} H\, e^{-i\phi n_g}e^{-i(\pi/4)\tau_y}\,.
\end{align}
Since
\begin{equation}
e^{i(\pi/4)\tau_y} =\frac{1}{\sqrt{2}} \begin{bmatrix} 1 & 1 \\ -1 & 1 \end{bmatrix}\,,
\end{equation}
the transformation amounts to sending
\begin{align}\label{eq:final_basis}
H&\to -4E_c\partial^2_\phi + \epsilon_r \tau_x - \Gamma \cos(\phi/2) \tau_z - \delta\Gamma \sin(\phi/2)\,\tau_y\,.
\end{align}
In this new basis, the boundary condition is also different:
\begin{equation}\label{eq:boundary_condition_after_rotation}
\Psi(\phi+2\pi) = -\tau_x e^{i2\pi n_g} \Psi(\phi)
\end{equation}
In the calculation that follows we will make use of the adiabatic eigenstates of the potential $V(\phi)$ after the transformation, which in matrix form is given by 
\begin{equation}
V(\phi) =\begin{bmatrix}
-\Gamma \cos(\phi/2) & \epsilon_r + i \delta\Gamma \sin(\phi/2) \\
\epsilon_r - i \delta\Gamma \sin(\phi/2) & \Gamma \cos(\phi/2)
\end{bmatrix}
\end{equation}
The two eigenvectors $V(\phi) \chi_\pm  = \pm E_A(\phi) \chi_\pm$ are:
\begin{subequations}\label{eq:adiabatic_spinors}
\begin{align}
\chi_+ &= \mathcal{N}^{-1/2}(\phi)
\begin{bmatrix}
E_A - \Gamma \cos(\phi/2)\\
\epsilon_r - i\delta\Gamma\sin(\phi/2)
\end{bmatrix}\\
\chi_- &= \mathcal{N}^{-1/2}(\phi)
\begin{bmatrix}
-\epsilon_r - i\delta\Gamma\sin(\phi/2)\\
E_A - \Gamma \cos(\phi/2)
\end{bmatrix}
\end{align}
\end{subequations}
with a normalization factor given by
\begin{equation}
\mathcal{N}(\phi) = 2E_A(E_A-\Gamma\cos(\phi/2))\,.
\end{equation}
For later use we note the following property of these spinors:
\begin{align}\label{eq:eigenstates_2pi_shift}
\chi_+(2\pi+\phi) &= e^{i\delta(\phi)}\,\tau_x\,\chi_+(\phi)\,,\\
\chi_-(2\pi+\phi) &= -e^{-i\delta(\phi)}\,\tau_x\,\chi_-(\phi)\,.
\end{align}
where $\delta(\phi)$ is the phase of $\epsilon_r+i\delta\Gamma\sin(\phi/2)$.

To solve the Sch\"{o}dinger equation, we split the interval $[-\pi,\pi]$ into four regions as follows:
\begin{itemize}
    \item Region I: $\phi \,\in (-\pi, -\phi_c)$, where $\phi_c$ is the classical turning point such that $\Gamma_A-E_A(\phi_c) = E$.
    \item Region II: $\phi\,\in\,(-\phi_c, \phi_c)$.
    \item Region III: $\phi\,\in\,(\phi_c, \pi)$.
    \item Region IV: $\phi\,\in\,(\pi, 2\pi - \phi_c)$.
\end{itemize}
Within each region we can write the solution using the WKB ansatz, with either oscillatory or decaying/growing solutions. In detail:
\begin{widetext}
\begin{align}\label{eq:WKB_ansatz}
\Psi_\textrm{I} &= \frac{A_1\,\chi_-}{\sqrt{\kappa_1}}\,e^{-\int_{-\pi}^\phi \kappa_1 d\phi'} + \frac{A_2\,\chi_-}{\sqrt{\kappa_1}}\,e^{+\int_{-\pi}^\phi \kappa_1 d\phi'} + \frac{A_3\,\chi_+}{\sqrt{\kappa_2}}\,e^{-\int_{-\pi}^\phi \kappa_2 d\phi'} + \frac{A_4\,\chi_+}{\sqrt{\kappa_2}}\,e^{+\int_{-\pi}^\phi \kappa_2 d\phi'}\,,\\
\Psi_\textrm{II} &=\frac{B_1\,\chi_-}{\sqrt{k_1}}\,\cos\,(\tfrac{\pi}{4}+\smallint_{-\phi_c}^{\phi} k_1\,d\phi')+\frac{2B_2\,\chi_-}{\sqrt{k_1}}\,\sin\,(\tfrac{\pi}{4}+\smallint_{-\phi_c}^{\phi} k_1\,d\phi')+\frac{B_3\,\chi_+}{\sqrt{\kappa_2}}\,e^{-\int_{-\phi_c}^\phi \kappa_2 d\phi'} + \frac{B_4\,\chi_+}{\sqrt{\kappa_2}}\,e^{+\int_{-\phi_c}^\phi \kappa_2 d\phi'}\,,\\\label{eq:PsiIII}
\Psi_\textrm{III} &=\frac{C_1\,\chi_-}{\sqrt{\kappa_1}}\,e^{-\int_{\phi_c}^\phi \kappa_1 d\phi'} + \frac{C_2\,\chi_-}{\sqrt{\kappa_1}}\,e^{+\int_{\phi_c}^\phi \kappa_1 d\phi'} + \frac{C_3\,\chi_+}{\sqrt{\kappa_2}}\,e^{-\int_{\phi_c}^\phi \kappa_2 d\phi'} + \frac{C_4\,\chi_+}{\sqrt{\kappa_2}}\,e^{+\int_{\phi_c}^\phi \kappa_2 d\phi'}\,,\\\label{eq:PsiIV}
\Psi_\textrm{IV} &=\frac{D_1\,\chi_-}{\sqrt{\kappa_1}}\,e^{-\int_{\pi}^\phi \kappa_1 d\phi'} + \frac{D_2\,\chi_-}{\sqrt{\kappa_1}}\,e^{+\int_{\pi}^\phi \kappa_1 d\phi'} + \frac{D_3\,\chi_+}{\sqrt{\kappa_2}}\,e^{-\int_{\pi}^\phi \kappa_2 d\phi'} + \frac{D_4\,\chi_+}{\sqrt{\kappa_2}}\,e^{+\int_{\pi}^\phi \kappa_2 d\phi'}\,.
\end{align}
\end{widetext}
For brevity, we have introduced the following wave vectors (note that the notation differs slightly with that of Eq.~\eqref{eq:kWKB} in the main text):
\begin{align}
k_1 &= \sqrt{\frac{E-(\Gamma_A-E_A)}{4E_c}}\,,\\
\kappa_1 &=\sqrt{\frac{(\Gamma_A-E_A) - E}{4E_c}}\,,\\
\kappa_2 &= \sqrt{\frac{(\Gamma_A+E_A) - E}{4E_c}}\,.
\end{align}
The sixteen complex coefficients $A_1, \dots D_4$ must be determined via approriate matching conditions at the boundaries between the different regions.
The matching condition between regions IV and I will be determined via the boundary condition~\eqref{eq:boundary_condition_after_rotation}.
The boundaries between regions I-II, II-III and III-IV are meant to be fuzzy, and one must make use of appropriate connection formulas for the WKB solutions by obtaining approximate solutions that are valid across the boundaries. This is what we do next.

To connect solutions at the boundary between region I and II, we can use the standard WKB connection formulas that originate from linearizing the potential around the classical turning point, and then solving the Airy equations. One obtains:
\begin{equation}\label{eq:A_from_B}
\begin{pmatrix} A_1 \\ A_2 \\ A_3 \\ A_4 \end{pmatrix} =
\begin{pmatrix}
e^{+\tau_1} & 0 & 0 & 0 \\
0 & e^{-\tau_1} & 0 & 0 \\
0 & 0 & e^{+\tau_2} & 0 \\
0 & 0 & 0 & e^{-\tau_2}
\end{pmatrix}
\begin{pmatrix} B_1 \\ B_2 \\ B_3 \\ B_4 \end{pmatrix}
\end{equation}
with $\tau_1 = \int_{\phi_c}^\pi \kappa_1 \,d\phi'$ and $\tau_2 = \int_{\phi_c}^\pi \kappa_2 \,d\phi'$.

For the boundary between region II and III we can also use the standard WKB connection formulas based on the Airy equation, except that we must first take some care to rewrite the wave function in region II so that it is expressed in terms of integrals that have the boundary point $\phi_c$ as the upper end of the integration domain.

After some trigonometric manipulations one obtains the following connection matrix:
\begin{equation}\label{eq:B_from_C}
\begin{pmatrix} B_1 \\ B_2 \\ B_3 \\ B_4 \end{pmatrix} =
\begin{pmatrix}
2\cos\sigma & -\sin\sigma & 0 & 0 \\
\sin\sigma & \tfrac{1}{2}\cos\sigma & 0 & 0 \\
0 & 0 & e^{+\rho_1} & 0 \\
0 & 0 & 0 & e^{-\rho_1}
\end{pmatrix}
\begin{pmatrix} C_1 \\ C_2 \\ C_3 \\ C_4 \end{pmatrix}
\end{equation}
where $\rho_1 = \int_{-\phi_c}^{\phi_c} \kappa_2 d\phi^{'}$.

Notice that so far the connections matrices \eqref{eq:A_from_B} and \eqref{eq:B_from_C} leave the two branches of the Andreev spectrum decoupled.
This situation breaks down at the boundary between regions III and IV at $\phi=\pi$.
This is the position where Andreev levels cross at perfect transparency, and where they couple in the presence of a small but finite back-scattering.
When the level crossing is narrowly avoided, the adiabatic spinors~\eqref{eq:adiabatic_spinors} vary rapidly with phase and the WKB ansatz, which relies on a slow variation of the spinors with $\phi$, breaks down.
To proceed we must linearize the potential around $\phi=\pi$, giving the equation:
\begin{equation}\label{eq:linearized_around_pi}
-4 E_c \Psi'' + V_\pi \,\Psi + \Gamma_A \Psi = 0\,.
\end{equation}
where
\begin{equation}\label{eq:V_pi}
V_\pi = \epsilon_r \tau_x + \tfrac{1}{2}\Gamma (\phi-\pi) \tau_z- \delta \Gamma \tau_y
\end{equation}
In Eq.~\eqref{eq:linearized_around_pi} we set $E=0$ since the level crossing is at energies much higher than the bottom of the Josephson potential: thus, the form of the solutions around $\phi\approx \pi$ will not be sensitive to the precise position of low-lying energy levels.
Inspired by the fact that we need to connect asymptotically to the wave functions in region II, we try an ansatz of the form:
\begin{equation}\label{eq:ansatz_at_pi}
\Psi = \Psi_\pi e^{\sigma \kappa (\phi-\pi)}
\end{equation}
with $\sigma = \pm 1$ and $\kappa=\sqrt{\Gamma_A/4E_c}$. Inserting the ansatz in Eq.~\eqref{eq:linearized_around_pi} and neglecting the term $\propto \Psi''_\pi$ results in the following equation for $\Psi_\pi$:
\begin{equation}
-\sigma \omega_0 \Psi_\pi' + V_\pi \Psi_\pi = 0
\end{equation}
with $\omega_0 = 8E_c\kappa$. Adopting the spinor notation $\Psi_\pi = (u, d)^T$, we obtain the following coupled linear differential equation for $u$ and $d$:
\begin{align}
-\sigma (\omega_0/\Gamma)\, u' + \tilde{r}\,d + \tfrac{1}{2}\,(\phi-\pi)\,u &= 0\\
-\sigma (\omega_0/\Gamma)\, d' + \tilde{r}^*\,u - \tfrac{1}{2}\,(\phi-\pi)\,d &= 0
\end{align}
where we introduced a complex reflection coefficient $\tilde{r}$:
\begin{equation}\label{eq:r_coefficient}
\tilde{r} \equiv \frac{\epsilon_r + i\delta\Gamma}{\Gamma}\,.
\end{equation}
Note that this reflection coefficients differs from the one introduced in the main text in Eq.~\eqref{[eq:reflection_coefficient]} because of the presence of $\Gamma$ instead of $\Gamma_A$ in the denominator.
The difference arises because the linearized problem is not sensitive to the bandwidth $\Gamma_A$ of the potential, but only to its slope $\Gamma$ at $\phi=\pi$.
The phase of $\tilde{r}$ is the same as that for $r$, and can be gauged away from the linearized equations, by setting $d\to de^{-i\delta/2}$ and $u\to u e^{i\delta/2}$.
Furthermore, it is also convenient to shift and rescale the coordinate,
\begin{equation}\label{eq:phi_to_x}
x = \sqrt{\frac{\Gamma}{\omega_0}} (\phi-\pi)\,.
\end{equation}
After these two steps we obtain
\begin{align}\label{eq:coupled_u_d}
-\sigma u' + \sqrt{\lambda}\,d + \tfrac{1}{2}\,x\,u &= 0\\
-\sigma d' + \sqrt{\lambda}\,u - \tfrac{1}{2}\,x\,d &= 0
\end{align}
where the prime now refer to differentiation with respect to $x$ and we have introduced
\begin{equation}
\lambda \equiv \abs{\tilde{r}}^2 (\Gamma / \omega_0)\,,
\end{equation}
the same parameter introduced in Eq.~\eqref{eq:lambda} of the main text.
Proceeding by substitution we obtain the two equations (one for each value of $\sigma$):
\begin{equation}
u'' + \left(-\lambda - \frac{\sigma}{2} -\frac{x^2}{4}\right) u = 0
\end{equation}
which must be considered separately and, combined, give the four independent solutions we are looking for. They are instances of the Weber differential equation and are solved in terms of parabolic cylinder functions $D_p(x)$, which satisfy the differential equation $D''_p(z) + (p+\tfrac{1}{2} - z^2/4) D_p(z) = 0$. In our case we are dealing with $p=-\lambda$ when $\sigma=-1$ and $p=-\lambda-1$ when $\sigma=+1$.

Let us solve the two cases separately, beginning with $\sigma=-1$. The general solution for $u$ is the linear combination $u(x) = c_1\,D_{-\lambda}(x) + c_2 \sqrt{\lambda} D_{\lambda-1}(ix)$. The corresponding solution for $d(x)$ can be obtained using known recursion formulas for parabolic cylinder functions, which read:
 \begin{align}
D'_p(z) - p D_{p-1}(z) + \tfrac{1}{2}z D_p(z) &= 0\,, \\
D'_p(z) + D_{p+1}(z) - \tfrac{1}{2}z D_p(z) &=0\,.
\end{align}
Using these formulas we obtain $d(x) = c_1 \sqrt\lambda\, D_{-\lambda-1}(x) + i c_2 D_\lambda(ix)$. Due to a symmetry of the problem, the solutions for $\sigma=+1$ can be obtained from these by sending $x\to-x$ and exchanging $u$ and $d$, so that, overall, the general solution is
\begin{align}
\label{eq:general_solution}
&\Phi(x)= c_1 e^{-\tilde\kappa x} \begin{bmatrix} D_{-\lambda}(x)  \\  \sqrt\lambda\,D_{-\lambda-1}(x)\end{bmatrix} + c_2 e^{-\tilde\kappa x} \begin{bmatrix} \sqrt\lambda D_{\lambda-1}(ix) \\  iD_{\lambda}(ix)\end{bmatrix} \nonumber \\ & + c_3 e^{+\tilde\kappa x} \begin{bmatrix} \sqrt\lambda D_{-\lambda-1}(-x)  \\  D_{-\lambda}(-x)\end{bmatrix} + c_4 e^{+\tilde\kappa x}\begin{bmatrix} iD_{\lambda}(-ix)  \\  \sqrt\lambda\,D_{\lambda-1}(-ix)\end{bmatrix}\,.
\end{align}
Here, $\tilde\kappa = \kappa \sqrt{\omega_0/\Gamma}$.
This solution captures the interval around $\phi=\pi$ where diabatic effects not captured by the WKB ansatz may occur. This region has a width $\sim\sqrt{\lambda}$.
Thus, the solution has to be matched with $\Psi_\textrm{III}$ from Eq.~\eqref{eq:PsiIII} for $x\ll-\sqrt{\lambda}$ and with $\Psi_\textrm{IV}$ from Eq.~\eqref{eq:PsiIV} for $x\gg\sqrt{\lambda}$.
For the matching purposes, it's useful to derive the asymptotic behaviour of these WKB solutions. In the case of $\Psi_\textrm{III}$, to do so we must first rewrite the WKB solution such that the integrals run up to the level crossing. Thus, we rewrite Eq.~\eqref{eq:PsiIII} as 
\begin{align}\label{ccoeffwkb}
\Psi_\textrm{III} &=\frac{C_1\,\chi_-}{\sqrt{\kappa_1}}\,e^{-\tau_1}\,e^{+\int_\phi^\pi \kappa_1 d\phi'} + \frac{C_2\,\chi_-}{\sqrt{\kappa_1}}\,e^{+\tau_1}\,e^{-\int_\phi^\pi \kappa_1 d\phi'}\nonumber\\& + \frac{C_3\,\chi_+}{\sqrt{\kappa_2}}\,e^{-\tau_2}\,e^{+\int_\phi^\pi \kappa_2 d\phi'} + \frac{C_4\,\chi_+}{\sqrt{\kappa_2}}\,e^{+\tau_2}\,e^{-\int_\phi^\pi \kappa_2 d\phi'}\,.
\end{align}
Let's introduce the distance $R$ from the level crossing, $R=\abs{x}$. When $R\gg\sqrt{\lambda}$, the asymptotes for the adiabatic spinors are:
\begin{align}
\chi_-(-R) & \sim \begin{bmatrix} -1 \\  \sqrt{\lambda}/R \end{bmatrix}\,,\\
\chi_+(-R) &\sim \begin{bmatrix} \sqrt{\lambda}/R \\ 1 \end{bmatrix}\,,\\
\chi_-(R) & \sim \begin{bmatrix} -\sqrt{\lambda}/R \\ 1 \end{bmatrix}\,,\\
\chi_+(R) &\sim \begin{bmatrix} 1 \\ \sqrt{\lambda}/R \end{bmatrix}\,.
\end{align}
Note that $\chi_\pm(- R)=\pm \tau_x \chi_\pm(R)$.
Taking into account the fact that, approaching the level crossing,
\begin{equation}
\kappa_{1,2}\approx \kappa \mp \tfrac{1}{2}\sqrt{4\lambda + R^2}\ \sqrt{\frac{\Gamma}{\omega_0}},
\end{equation}
we obtain the following expressions for the WKB integrals:
\begin{align}\nonumber
\int_\phi^\pi \kappa_1 \,d\phi' &= \tilde{\kappa}R -\tfrac{1}{4}R^2 -\tfrac{1}{2}\lambda -\lambda \,\log\,R + \lambda \,\log\sqrt{\lambda}\,,\\
\int_\phi^\pi \kappa_2 \,d\phi' &= \tilde{\kappa}R +\tfrac{1}{4}R^2 +\tfrac{1}{2}\lambda +\lambda \,\log\,R - \lambda \,\log\sqrt{\lambda}\,.
\end{align}
Finally, when $\kappa \gg R \gg \sqrt{\lambda}$, one has that
\begin{equation}
\frac{1}{\sqrt{\kappa_{1,2}}}\approx \frac{1}{\sqrt{\kappa}}
\end{equation}
The condition $\kappa \gg R \gg \sqrt{\lambda}$ is the necessary condition for the existence of a range of coordinates where asymptotes can be matched. In practice, it requires the transition region around the level crossing at $\phi=\pi$ to be narrow enough to be far away from the classical turning point at $\phi=\phi_c$. Note that this condition is automatically satisfied since $\kappa \propto (\Gamma_A/E_c)^{1/2}$ while $\sqrt{\lambda}\sim (\Gamma_A/E_c)^{1/4}$.

With all that said, the expression approaching the level crossing from region III is:
\begin{align}\label{eq:wkbc}\nonumber
\Psi_\textrm{III} \sim &\left(\frac{C_1}{\sqrt{\kappa}}\,e^{-\tau_1}\,e^{-\lambda/2}\,\lambda^{\lambda/2} \right)\,e^{\tilde\kappa R}\,e^{-R^2/4}\,R^{-\lambda}\,\chi_-(-R)\\\nonumber
+&\left(\frac{C_2}{\sqrt{\kappa}}\,e^{+\tau_1}\,e^{+\lambda/2}\,\lambda^{-\lambda/2} \right)\,e^{-\tilde\kappa R}\,e^{R^2/4}\,R^{\lambda}\,\chi_-(-R)\\\nonumber
+&\left(\frac{C_3}{\sqrt{\kappa}}\,e^{-\tau_2}\,e^{+\lambda/2}\,\lambda^{-\lambda/2} \right)\,e^{\tilde\kappa R}\,e^{R^2/4}\,R^{\lambda}\,\chi_+(-R)\\
+&\left(\frac{C_4}{\sqrt{\kappa}}\,e^{+\tau_2}\,e^{-\lambda/2}\,\lambda^{\lambda/2} \right)\,e^{-\tilde\kappa R}\,e^{-R^2/4}\,R^{-\lambda}\,\chi_+(-R)
\end{align}
while the one for $\Psi_\textrm{IV}$, obtained from Eq.~\eqref{eq:WKB_ansatz}, is:
\begin{align}\label{eq:wkbd}\nonumber
\Psi_\textrm{IV} \sim &\;\left(\frac{D_1}{\sqrt{\kappa}}\,e^{\lambda/2}\,\lambda^{-\lambda/2} \right)\,e^{-\tilde\kappa R}\,e^{R^2/4}\,R^{\lambda}\,\chi_-(R)\\\nonumber
+&\left(\frac{D_2}{\sqrt{\kappa}}\,e^{-\lambda/2}\,\lambda^{\lambda/2} \right)\,e^{\tilde\kappa R}\,e^{-R^2/4}\,R^{-\lambda}\,\chi_-(R)\\\nonumber
+&\left(\frac{D_3}{\sqrt{\kappa}}\,e^{-\lambda/2}\,\lambda^{\lambda/2} \right)\,e^{-\tilde\kappa R}\,e^{-R^2/4} R^{-\lambda}\,\chi_+(R)\\
+&\left(\frac{D_4}{\sqrt{\kappa}}\,e^{\lambda/2}\,\lambda^{-\lambda/2} \right)\,e^{\tilde\kappa R}\,e^{R^2/4}\,R^{\lambda}\,\chi_+(R)
\end{align}

These two expressions must now be compared to and matched with the expansion of Eq.~\eqref{eq:general_solution}. 
The matching procedure will yield us a connection matrix between the wave function coefficients in regions III and IV. This connection matrix will take the form:
\begin{equation}\label{eq:C_from_D}
\begin{pmatrix} C_1 \\ C_2 \\ C_3 \\ C_4 \end{pmatrix} =\begin{pmatrix}
e^{\tau_1} & 0 & 0 & 0 \\
0 & e^{-\tau_1} & 0 & 0 \\
0 & 0 & e^{\tau_2} & 0 \\
0 & 0 & 0 & e^{-\tau_2}
\end{pmatrix}
M
\begin{pmatrix} D_1 \\ D_2 \\ D_3 \\ D_4 \end{pmatrix}
\end{equation}
where $M$ is a $4\times 4$ matrix whose elements must be determined via the matching procedure.
We expect half of the matrix elements of $M$ to be zero, because the exponentially decaying sector is decoupled from the exponentially growing sector, as assumed by the ansatz~\eqref{eq:ansatz_at_pi}.
More in detail, the matrix $M$ will have the following structure,
\begin{equation}
M = \begin{pmatrix}
m_{11} & 0 & m_{12} & 0 \\
0 & m'_{11} & 0 & m'_{12} \\
m_{21} & 0 & m_{22} & 0\\
0 & m'_{21} & 0 & m'_{22}
\end{pmatrix}
\end{equation}
with two interleaved $2\times 2$ sub-blocks $M^+$ and $M^-$ which separately connect exponentially decaying and growing solutions on either side of the level crossing:
\begin{align}
M^- & = \begin{pmatrix}
m_{11} & m_{12} \\
m_{21} & m_{22} \\
\end{pmatrix}\,,\\
M^+ & = \begin{pmatrix}
m'_{11} & m'_{12} \\
m'_{21} & m'_{22} \\
\end{pmatrix}\,.
\end{align}
To simplify the derivation of $M$, we will make use of two useful identities that connect $M^+$ and $M^-$ and thus allow to shorten the calculation.

The first identity is
\begin{equation}
\det M^+ = \det M^-\,.
\end{equation}
It follows from the fact that, given two spinors $\Phi_1=(u_1, d_1)^T$ and $\Phi_2=(u_2, d_2)^T$ which are solutions of Eq.~\eqref{eq:coupled_u_d}, one has
\begin{equation}
\frac{d}{dx}\,\det\,[\Phi_1 | \Phi_2] =0\,,
\end{equation}
where $[\Phi_1 | \Phi_2]$ is the matrix obtained joining the two spinors:
\begin{equation}
[\Phi_1 | \Phi_2]\equiv
\begin{pmatrix}
u_1 & u_2 \\ d_1 & d_2 
\end{pmatrix}\,.
\end{equation}
To verify this property one observes that:
\begin{align}
\frac{d}{dx}\,\det\,[\Phi_1 | \Phi_2] &=  \det [\Phi_1' | \Phi_2] + \det [\Phi_1 | \Phi_2']\\\nonumber
&= \sigma\det[O\Phi_1 | \Phi_2] + \sigma\det\,[\Phi_1 | O\Phi_2]\,,
\end{align}
where $\sigma=\pm 1$ and $O=\tfrac{1}{2}x\tau_z + \sqrt{\lambda} \tau_x$\,. The last passage in the equation above follows directly from Eq.~\eqref{eq:coupled_u_d}. To conclude the argument, one notices that
\begin{equation}
\det [O\Phi_1 | \Phi_2] = \det(O)\,\det\,[\Phi_1 | O^{-1}\Phi_2]
\end{equation}
Furthermore, in our case, $O^{-1} = -\det^{-1}(O) \,O$. Thus,
\begin{equation}
\det [O\Phi_1 | \Phi_2]=-\det [\Phi_1 | O\Phi_2]
\end{equation}
The conclusion is that
\begin{equation}
\det\,[\Phi_1 | \Phi_2] = \textrm{constant\,.}
\end{equation}
Let us apply it to the case in which $\Phi_1$ and $\Phi_2$ are the two exponentially decaying solutions ($\sigma=-1$) of Eq.~\eqref{eq:coupled_u_d} that enter Eq.~\eqref{eq:general_solution} with coefficients $c_1$ and $c_3$. We observe that the $\det[\Phi_1,\Phi_2]$ must remain constant also for the matched asymptotic expansions of $\Phi_1$ and $\Phi_2$ on either side of the crossing. A direct calculation gives
\begin{equation}
\det[\Phi_1 | \Phi_2] = -\frac{D_1D_3}{\tilde\kappa}
\end{equation}
for $x\gg\sqrt{\lambda}$, and, using~\eqref{eq:C_from_D}
\begin{equation}
\det[\Phi_1 | \Phi_2] = -\frac{D_1D_3}{\tilde\kappa}\,\det M^-
\end{equation}
for $x\ll-\sqrt{\lambda}$. It follows that $\det M^-=1$.  The reasoning is analogous for $\sigma=1$, so $\det M^{+}=1$ too.

The second identity we will make use of is a pseudo-inverse identity which relates $M^+$ and $M^-$:
\begin{equation}\label{eq:pseudoinverse}
M^{+}=\tau_z (M^{-})^{-1}\tau_z
\end{equation}
The idea behind this identity is that, as noticed earlier, there is a reflection symmetry around the level crossing: namely, if $[u(x),d(x)]^T$ is a solution of Eq.~\eqref{eq:coupled_u_d}, then $[d(-x), u(-x)]^T$ is also a solution.
This symmetry maps decaying solutions to growing ones and thus it suggests that there must be a relation between $M^+$ and $M^-$.
Applying this symmetry argument to the asymptotic solutions and observing that their spinors obey $\chi_\pm(-x) = \pm \tau_x\chi_\pm(x)$, one arrives at the identity \eqref{eq:pseudoinverse}.

At this point we have to find the elements of $M^-$ by looking at the asymptotic expansion of the parabolic cylinder functions~\cite{gradshteyn2014}, which can be applied term by term to \eqref{eq:general_solution} and then compared to the WKB asymptotes in Eq.~\eqref{eq:wkbc} and \eqref{eq:wkbd}. For instance, the last term in \eqref{eq:general_solution} has the following asymptotic behaviour (recall that $R=\abs{x}$):
\begin{equation}
\begin{bmatrix} \sqrt\lambda D_{\lambda-1}(ix) \\ iD_{\lambda}(ix) \end{bmatrix} \sim ie^{-i\pi\lambda/2}\,e^{R^2/4}\,R^{\lambda}\,\chi_+(-R)
\end{equation}
for $x\ll-\sqrt{\lambda}$ and
\begin{equation}
\begin{bmatrix} \sqrt\lambda D_{\lambda-1}(ix) \\ iD_{\lambda}(ix) \end{bmatrix} \sim i e^{i\pi\lambda/2}\,e^{R^2/4}\,R^{\lambda}\,\chi_-(R)
\end{equation}
for $x\gg\sqrt{\lambda}$. Matching these asymptotes with Eq.~\eqref{eq:wkbc} and \eqref{eq:wkbd} yields the matrix elements
\begin{align}
m_{11} &= 0\,,\\
m_{21} &= e^{-i\pi\lambda}\,.
\end{align}
The third term in \eqref{eq:general_solution} has the asymptotic expansion
\begin{align}\nonumber
\begin{bmatrix} D_{-\lambda}(x)  \\  \sqrt\lambda D_{-\lambda-1}(x)\end{bmatrix} & \sim -e^{i\pi \lambda} e^{-R^2/4}\,R^{-\lambda}\,\chi_-(-R) \\
&+ \frac{\sqrt{2\pi}}{\sqrt{\lambda}\Gamma(\lambda)}\,e^{R^2/4}\,R^{\lambda}\,\chi_+(-R)
\end{align}
for $x\ll\sqrt{\lambda}$ and
\begin{align}
\begin{bmatrix} D_{-\lambda}(x)  \\  \sqrt\lambda D_{-\lambda-1}(x)\end{bmatrix}  \sim e^{-R^2/4}\,R^{-\lambda}\,\chi_+(R)
\end{align}
for $x\gg\sqrt{\lambda}$. Again by comparison with~\eqref{eq:wkbc} and \eqref{eq:wkbd}, we derive
\begin{equation}
m_{22} = w
\end{equation}
where $w$ is the same as defined in the main text Eq.~\eqref{eq:w}. The determinant identity for $M^-$ then yields
\begin{equation}
m_{12}=-e^{i\pi\lambda}
\end{equation}
This completes the matrix $M^-$. The matrix $M^+$ can the be derived using the pseudo-inverse identity, and both can be combined into the final form for the connection matrix $M$ entering Eq.~\eqref{eq:C_from_D}:
\begin{equation}\label{eq:connection_III_to_IV}
M=\begin{pmatrix}
0 & 0 & -e^{i\pi\lambda} & 0 \\
0 & w & 0 & -e^{i\pi\lambda}  \\
e^{-i\pi\lambda} & 0 & w & 0 \\
0 & e^{-i\pi\lambda} & 0 & 0
\end{pmatrix}
\end{equation}

The final step is to find the connection matrix at the boundary between region IV and I. In order to do so, we impose the twisted boundary conditions~\eqref{eq:boundary_condition_after_rotation} evaluated at the point at $\phi=\pi + \epsilon$:
\begin{equation}
\Psi_\textrm{IV}(\pi+\epsilon) = -\tau_x\,e^{2\pi i n_g}\Psi_\textrm{I}(-\pi+\epsilon)
\end{equation}
Using Eq.~\eqref{eq:eigenstates_2pi_shift}, this leads to two equations:
\begin{align}
e^{-i\delta}\,(D_1+D_2) &= e^{2\pi i n_g}\,(A_3+A_4)\\
e^{+i\delta}\,(D_1+D_2) &= -e^{2\pi i n_g}\,(D_3+D_4)
\end{align}
where $\delta$ is the phase of $\epsilon_r-i\delta\Gamma$.
We need two more equations, which we can get from taking the derivative of Eq.~\eqref{eq:boundary_condition_after_rotation} at $\phi=\pi+\epsilon$:
\begin{equation}
\Psi'_\textrm{IV}(\pi+\epsilon) = -\tau_x\,e^{2\pi i n_g}\Psi'_\textrm{I}(-\pi+\epsilon)\,,
\end{equation}
to be computed neglecting the change in the slow components of the WKB wave functions.
This leads to the following connection matrix:
\begin{equation}\label{eq:D_from_A}
\begin{pmatrix} D_1 \\ D_2 \\ D_3 \\ D_4 \end{pmatrix} =
e^{2\pi i n_g}\,\begin{pmatrix}
e^{i\delta} & 0 & 0 & 0 \\
0 & e^{i\delta} & 0 & 0 \\
0 & 0 & -e^{-i\delta} & 0 \\
0 & 0 & 0 & -e^{-i\delta}
\end{pmatrix}\,
\begin{pmatrix} A_1 \\ A_2 \\ A_3 \\ A_4 \end{pmatrix}
\end{equation}

Putting together Eqs.~\eqref{eq:A_from_B}, \eqref{eq:B_from_C}, \eqref{eq:connection_III_to_IV} and \eqref{eq:D_from_A}, we obtain a linear system of equation that must be satisfied by the coefficients in region I.
After some matrix multiplication this linear system takes the form:
\begin{equation}
\vec{A}=
e^{2\pi i n_g}\,M_1
\,
M_2
\,M_3\,\vec{A}
\end{equation}
with $\vec{A}=(A_1, A_2, A_3, A_4)^T$ and
\begin{align}
M_1 &=\begin{pmatrix}
2e^{\tau}\,\cos\sigma  & -\sin\sigma & 0 & 0 \\
\sin\sigma & \tfrac{1}{2}e^{-\tau}\,\cos\sigma & 0 & 0 \\
0 & 0 & e^{\rho} & 0 \\
0 & 0 & 0 & e^{-\rho}
\end{pmatrix}\\
M_2 &= \begin{pmatrix}
0 & 0 & -e^{i \pi\lambda} & 0 \\
0 &  w & 0 & -e^{i\pi\lambda}  \\
e^{-i\pi\lambda} & 0 & w & 0 \\
0 & e^{-i\pi\lambda} & 0 & 0
\end{pmatrix}\\
M_3 &=\begin{pmatrix}
e^{i\delta} & 0 & 0 & 0 \\
0 & e^{i\delta} & 0 & 0 \\
0 & 0 & -e^{-i\delta} & 0 \\
0 & 0 & 0 & -e^{-i\delta}
\end{pmatrix}\,.
\end{align}
The WKB integrals that appear in these matrices are those defined in the main text Eq.~\eqref{eq:sigma}, \eqref{eq:tau}, and \eqref{eq:rho}.
A non-trivial solution occurs only if 
\begin{equation}
\det\left(1 - e^{2\pi i n_g} M_1 M_2 M_3\right) = 0
\end{equation}
This condition yields a transcendental equation for the energy $E$, taking the form:
\begin{equation}
\cos \sigma = \frac{4 e^\rho e^\tau\,\left[\cos(4\pi n_g) + e^\rho w \cos(2\pi n_g + \delta)\right]}{1 + e^{2\rho}\left(4 e^{2\tau}+ w^2\right)+2 e^\rho w \cos(2\pi n_g - \delta)}
\end{equation}
Using the fact that $e^{-\rho}\ll 1$ and $w e^{-\tau}\ll 1$, we can simplify the denominator on the right hand side as follows:
\begin{align}\nonumber
1 + e^{2\rho}&\left(4 e^{2\tau}+ w^2\right)+2 e^\rho w \cos(2\pi n_g - \delta) \approx 4 e^{2\rho}e^{2\tau}\,.
\end{align}
Thus, the transcendental equation takes the simpler form reported as Eq.~\eqref{eq:bound_state_equation} in the main text:
\begin{equation}\label{eq:transc}
\cos \sigma = e^{-\rho} e^{-\tau} \cos(4\pi n_g) + w\, e^{-\tau}\cos(2\pi n_g + \delta)
\end{equation}
Note that the energy enters the bound state equation via the WKB integrals $\sigma, \rho$ and $\tau$, where it appears in both the integrand and the limits of integration.

As observed in the main text, to solve this equation a good starting point is to set the right hand side to zero, since it contains only exponentially small terms. The zeros of the left hand side occur if 
\begin{equation}
\sigma(E_n) = \pi (n+\tfrac{1}{2})\,.
\end{equation}

When taking into account the right hand side, some corrections will come from the $4\pi-$phase slip term  $e^{-\rho} e^{-\tau} \cos(4\pi n_g)$ and others will come from the $2\pi$ phase slip term $w\, e^{-\tau}\cos(2\pi n_g + \delta)$. We are not interested in the corrections smaller than the corrections from $4\pi$ phase slips, so the cross-terms are neglected. For the rest, we can distinguish the following three situations:
\begin{enumerate}
    \item $we^{-\tau}\gg e^{-\rho-\tau}$: it only makes sense to keep the leading order corrections in $we^{-\tau}$ to each of the harmonics in the dispersion relation 
    \item $we^{-\tau} \approx e^{-\rho-\tau}$: we keep the leading order $we^{-\tau}$ corrections and the first order $e^{-\rho-\tau}$ -- corrections
    \item $we^{-\tau} \ll e^{-\rho-\tau}$: enough to keep only the first order in $e^{-\rho-\tau}$.
\end{enumerate}
We can conclude that in any situation it is enough to keep the leading order in  $we^{-\tau} $ and the first order in $e^{-\rho-\tau} $ for the second harmonic, although having something of the order of $e^{-\rho-\tau} $ and ignoring higher order corrections in $we^{-\tau} $ may look inconsistent when $we^{-\tau} \gg e^{-\rho-\tau} $.

Let's introduce the following notation:
\begin{align}
E=E_{n}+\delta E^{(1)}+\delta E^{(2)}+\Delta E_n+...
\end{align}
Where $\delta E^{(m)}$  stand for $m$-th order corrections in $we^{-\tau} $ ($0$-th in $e^{-\rho-\tau} $) and $\Delta E_n$ for the first order corrections in $e^{-\rho-\tau} $. By solving ~Eq.~\eqref{eq:transc} with iterative expansions, we find:
\begin{align}
\Delta E_n = \frac{(-1)^{n+1}}{\sigma'_n} e^{-\rho_n} e^{-\tau_n} \cos(4\pi n_g)\,,
\end{align}
\begin{align}
\delta E^{(1)}_n = \frac{(-1)^{n+1}}{\sigma'_n}\,w e^{-\tau_n} \cos(2\pi n_g+\delta)\,.
\end{align}
\begin{equation}
\delta E_{n}^{(2)}=-\frac{w^{2} e^{-2\tau_n}\cos^{2}(2\pi n_g+\delta)}{(\sigma'_n)^2}\left( \tau'_n+\frac{\sigma''_n}{2 \sigma'_n}\right)
\end{equation}
The corrections have quite intuitive meaning. The term with $\tau^{'}$ comes from the fact that after we consider the first order in $we^{-\tau}$ contribution, different energies see different heights of the tunneling barrier. The term proportional to $\sigma_n''/\sigma_n'$ is due to second order corrections to $\sigma_{n}$ when the splitting $ \delta E^{(1)}_{n}$ is included, and it vanishes in the harmonic limit. On the other hand, as will be shown in the next appendix, $\tau'_n$ is logarithmically large when $T\Gamma_A \gg E_c$ and thus cannot be neglected. This leads to the solution presented in the main text, Eq.~\eqref{eq:delta_n}.
Note that in the main text we have omitted the $n_g$-independent part of $\delta E_n^{(2)}$, which does not affect the charge dispersion.

\section{Evaluation of the WKB integrals}
\label{app:integrals}

In this appendix we derive expressions~\eqref{eq:tau_n},~\eqref{eq:rho_n},~\eqref{eq:sigmap_n}, and~\eqref{eq:taup_n} from the main text.
In doing so we assume that $\Gamma_A T \gg E_c$ and thus only look at leading contributions in the ratio $T\Gamma_A/E_c$ to the WKB integrals.
In this limit, the Bohr-Sommerfeld condition $\sigma(E_n)=\pi(n+\tfrac{1}{2})$ can be evaluated by expanding the integrand of $\sigma(E)$ around $\phi=0$, and adjusting the position of the classical turning point accordingly. The result is:
\begin{align}
\sigma(E)=\frac{\pi E}{\omega_p}\,,
\end{align}
where $\omega_p$ is the plasma frequency introduced in the main text. The result above immediately yields Eq.~\eqref{eq:harmonic_spectrum} of the main text as well as Eq.~\eqref{eq:sigmap_n}, $\sigma'(E)=\pi/\omega_p$.

With respect to the integral $\rho(E)$, one can see that the coefficients $c$ and $d$ in Eq.~\eqref{eq:rho_n} are given by the integrals
\begin{align}
c(T)&=\frac{1}{\sqrt{8}}\int^{\pi}_{-\pi} \sqrt{1+u(\phi)}\,d\phi\,, \\ 
d(T)&=\frac{1}{\sqrt{8}}\int^{\pi}_{-\pi} \frac{d\phi}{\sqrt{1+u(\phi)}}
\end{align}
where $u(\phi)=E_A(\phi)/\Gamma_A$. The only WKB integral which is relatively non-trivial to calculate is $\tau_n$:
\begin{equation}
\tau_n=\sqrt{\frac{\Gamma_A}{E_c}} \int_{\phi_n}^\pi \sqrt{1-y_n-u(\phi)}\,d \phi,
\end{equation}
where $y_n=E_n/\Gamma_A$ and $\pm \phi_n$ are the classical turning points for $E_n$.
It is convenient to split $\tau_n$ into three parts:
\begin{widetext}
\begin{align}
\tau_n \sqrt{\frac{E_{C}}{\Gamma_A}} \approx \int^{\pi}_{\epsilon} \sqrt{1-u(\phi)}\, d\phi - \frac{y_{n}}{2} \int^{\pi}_{\epsilon} \frac{1}{\sqrt{1-u(\phi)}}\,d\phi + \int^\epsilon_{\phi_{n}} \sqrt{1- y_{n}-u(\phi)} d \phi   
\end{align}
Here, $\epsilon$ is small enough so that $\sin^2{\epsilon/2}\ll1$ but big enough such that $\sqrt{1-y_n-u} $ can be expanded in $y_n$. By splitting these terms further, we may arrive at a representation in terms of elliptic functions:
\begin{align}
I &\approx  \int^{\pi}_{0} \sqrt{1-u(\phi)}\,d\phi - \lim_{\psi \rightarrow 0}\frac{y_{n}}{2} \int^{\pi}_{\psi} \frac{1}{\sqrt{1-u(\phi)}} d \phi- \int^{\epsilon}_{0} \sqrt{1-u(\phi)}\,d\phi\\
&+\lim_{\psi \rightarrow 0}\, \frac{y_{n}}{2} \int^{\epsilon}_{\psi} \frac{1}{\sqrt{1-u(\phi)}}\,d\phi + \int^{\epsilon}_{\phi_{n}} \sqrt{1- y_{n}-u(\phi)}\,d\phi = i_{1}-i_{2}+i_{3}+i_{4}+i_{5}
\end{align}
Since $\phi_n,\epsilon\ll1$, $ i_{3}+i_{4}+i_{5}$ is quite straightforward to calculate and is equal to \label{singularity}:
\begin{align}
    i_{3}+i_{4}+i_{5}=-\sqrt{\frac{T}{2}} \frac{\sin^2\frac{\phi_{n}}{2}}{2} +\frac{\sin^2\frac{\phi_{n}}{2}}{2}\sqrt{\frac{T}{2}} \ln \frac{\sin^{2}\frac{\phi_{n}}{2}}{\psi^2}, \ \ \ \psi \rightarrow 0
\end{align}
For $i_1$ we obtain the representation
\begin{equation}
i_1 = \frac{-4\abs{r}}{\sqrt{1+\abs{r}}}F(\mu(0),k) +\frac{8\abs{r}}{\sqrt{1+\abs{r}}}\Pi(\mu(0),1,k)
\end{equation}
where $F,\Pi$ are elliptic integrals of the first and second kind, and
\begin{align}
\mu(\phi)&=\arcsin\sqrt{\frac{u(\varphi)-\abs{r}}{u(\varphi)+\abs{r}}}\,,\\
k &= \sqrt{\frac{1-\abs{r}}{1+\abs{r}}}
\end{align}
Similarly, for $i_2$ we obtain:
\begin{align}
\sqrt{\frac{\Gamma_A}{E_{C}}}i_2 = 
(2n+1) \frac{\sqrt{2}\abs{r}}{\sqrt{1-\abs{r}}(1+\abs{r})} \lim_{\psi \rightarrow 0}  \left(2 \Pi(\mu(\psi),\frac{1}{k^2},k)- (1-\abs{r})F(\mu(0),k) \right)
\end{align}
Putting all the pieces together, we obtain Eq.~\eqref{eq:tau_n} of the main text with the coefficients
\begin{align}
b=\lim_{\psi \rightarrow 0}\psi \ e^{\frac{\sqrt{2}\abs{r}}{\sqrt{1-\abs{r}}(1+\abs{r})}  \left(2 \Pi(\mu(\psi),\frac{1}{k^2},k)- (1-\abs{r})F(\mu(0),k) \right)}\,,\\
a=\frac{\sqrt{8} \abs{r}}{(1+\abs{r})\sqrt{1-\abs{r}}} \left( -F(\mu(0),k) +2\Pi(\mu(0),1,k)\right)\,.
\end{align}
These coefficients were already reported in Ref.~\cite{averin1999b}. In a similar way, for $\tau^{'}(E_n) $ we find:
\begin{align}
\tau^{'}_{n}=\frac{1}{\omega_p}\ln \frac{2 E_n}{\Gamma_A T b^2}
\end{align}
\end{widetext}

\bibliography{references}

%apsrev4-2.bst 2019-01-14 (MD) hand-edited version of apsrev4-1.bst
%Control: key (0)
%Control: author (8) initials jnrlst
%Control: editor formatted (1) identically to author
%Control: production of article title (0) allowed
%Control: page (0) single
%Control: year (1) truncated
%Control: production of eprint (0) enabled
\begin{thebibliography}{74}%
\makeatletter
\providecommand \@ifxundefined [1]{%
 \@ifx{#1\undefined}
}%
\providecommand \@ifnum [1]{%
 \ifnum #1\expandafter \@firstoftwo
 \else \expandafter \@secondoftwo
 \fi
}%
\providecommand \@ifx [1]{%
 \ifx #1\expandafter \@firstoftwo
 \else \expandafter \@secondoftwo
 \fi
}%
\providecommand \natexlab [1]{#1}%
\providecommand \enquote  [1]{``#1''}%
\providecommand \bibnamefont  [1]{#1}%
\providecommand \bibfnamefont [1]{#1}%
\providecommand \citenamefont [1]{#1}%
\providecommand \href@noop [0]{\@secondoftwo}%
\providecommand \href [0]{\begingroup \@sanitize@url \@href}%
\providecommand \@href[1]{\@@startlink{#1}\@@href}%
\providecommand \@@href[1]{\endgroup#1\@@endlink}%
\providecommand \@sanitize@url [0]{\catcode `\\12\catcode `\$12\catcode
  `\&12\catcode `\#12\catcode `\^12\catcode `\_12\catcode `\%12\relax}%
\providecommand \@@startlink[1]{}%
\providecommand \@@endlink[0]{}%
\providecommand \url  [0]{\begingroup\@sanitize@url \@url }%
\providecommand \@url [1]{\endgroup\@href {#1}{\urlprefix }}%
\providecommand \urlprefix  [0]{URL }%
\providecommand \Eprint [0]{\href }%
\providecommand \doibase [0]{https://doi.org/}%
\providecommand \selectlanguage [0]{\@gobble}%
\providecommand \bibinfo  [0]{\@secondoftwo}%
\providecommand \bibfield  [0]{\@secondoftwo}%
\providecommand \translation [1]{[#1]}%
\providecommand \BibitemOpen [0]{}%
\providecommand \bibitemStop [0]{}%
\providecommand \bibitemNoStop [0]{.\EOS\space}%
\providecommand \EOS [0]{\spacefactor3000\relax}%
\providecommand \BibitemShut  [1]{\csname bibitem#1\endcsname}%
\let\auto@bib@innerbib\@empty
%</preamble>
\bibitem [{\citenamefont {Haviland}(2010)}]{haviland2010}%
  \BibitemOpen
  \bibfield  {author} {\bibinfo {author} {\bibfnamefont {D.}~\bibnamefont
  {Haviland}},\ }\bibfield  {title} {\bibinfo {title} {Quantum phase slips},\
  }\href {https://doi.org/10.1038/nphys1747} {\bibfield  {journal} {\bibinfo
  {journal} {Nature Physics}\ }\textbf {\bibinfo {volume} {6}},\ \bibinfo
  {pages} {565} (\bibinfo {year} {2010})}\BibitemShut {NoStop}%
\bibitem [{\citenamefont {Chow}\ \emph {et~al.}(1998)\citenamefont {Chow},
  \citenamefont {Delsing},\ and\ \citenamefont {Haviland}}]{chow1998}%
  \BibitemOpen
  \bibfield  {author} {\bibinfo {author} {\bibfnamefont {E.}~\bibnamefont
  {Chow}}, \bibinfo {author} {\bibfnamefont {P.}~\bibnamefont {Delsing}},\ and\
  \bibinfo {author} {\bibfnamefont {D.~B.}\ \bibnamefont {Haviland}},\
  }\bibfield  {title} {\bibinfo {title} {{Length}-{Scale} {Dependence} of the
  {Superconductor}-to-{Insulator} {Quantum} {Phase} {Transition} in {One}
  {Dimension}},\ }\href {https://doi.org/10.1103/PhysRevLett.81.204} {\bibfield
   {journal} {\bibinfo  {journal} {Phys. Rev. Lett.}\ }\textbf {\bibinfo
  {volume} {81}},\ \bibinfo {pages} {204} (\bibinfo {year} {1998})}\BibitemShut
  {NoStop}%
\bibitem [{\citenamefont {Lau}\ \emph {et~al.}(2001)\citenamefont {Lau},
  \citenamefont {Markovic}, \citenamefont {Bockrath}, \citenamefont
  {Bezryadin},\ and\ \citenamefont {Tinkham}}]{lau2001}%
  \BibitemOpen
  \bibfield  {author} {\bibinfo {author} {\bibfnamefont {C.~N.}\ \bibnamefont
  {Lau}}, \bibinfo {author} {\bibfnamefont {N.}~\bibnamefont {Markovic}},
  \bibinfo {author} {\bibfnamefont {M.}~\bibnamefont {Bockrath}}, \bibinfo
  {author} {\bibfnamefont {A.}~\bibnamefont {Bezryadin}},\ and\ \bibinfo
  {author} {\bibfnamefont {M.}~\bibnamefont {Tinkham}},\ }\bibfield  {title}
  {\bibinfo {title} {Quantum phase slips in superconducting nanowires},\ }\href
  {https://doi.org/10.1103/PhysRevLett.87.217003} {\bibfield  {journal}
  {\bibinfo  {journal} {Phys. Rev. Lett.}\ }\textbf {\bibinfo {volume} {87}},\
  \bibinfo {pages} {217003} (\bibinfo {year} {2001})}\BibitemShut {NoStop}%
\bibitem [{\citenamefont {Mooij}\ and\ \citenamefont
  {Nazarov}(2006)}]{mooij2006}%
  \BibitemOpen
  \bibfield  {author} {\bibinfo {author} {\bibfnamefont {J.}~\bibnamefont
  {Mooij}}\ and\ \bibinfo {author} {\bibfnamefont {Y.~V.}\ \bibnamefont
  {Nazarov}},\ }\bibfield  {title} {\bibinfo {title} {Superconducting nanowires
  as quantum phase-slip junctions},\ }\href
  {https://www.nature.com/articles/nphys234} {\bibfield  {journal} {\bibinfo
  {journal} {Nature Physics}\ }\textbf {\bibinfo {volume} {2}} (\bibinfo {year}
  {2006})}\BibitemShut {NoStop}%
\bibitem [{\citenamefont {Pop}\ \emph {et~al.}(2010)\citenamefont {Pop},
  \citenamefont {Protopopov}, \citenamefont {Lecocq}, \citenamefont {Peng},
  \citenamefont {Pannetier}, \citenamefont {Buisson},\ and\ \citenamefont
  {Guichard}}]{pop2010}%
  \BibitemOpen
  \bibfield  {author} {\bibinfo {author} {\bibfnamefont {I.~M.}\ \bibnamefont
  {Pop}}, \bibinfo {author} {\bibfnamefont {I.}~\bibnamefont {Protopopov}},
  \bibinfo {author} {\bibfnamefont {F.}~\bibnamefont {Lecocq}}, \bibinfo
  {author} {\bibfnamefont {Z.}~\bibnamefont {Peng}}, \bibinfo {author}
  {\bibfnamefont {B.}~\bibnamefont {Pannetier}}, \bibinfo {author}
  {\bibfnamefont {O.}~\bibnamefont {Buisson}},\ and\ \bibinfo {author}
  {\bibfnamefont {W.}~\bibnamefont {Guichard}},\ }\bibfield  {title} {\bibinfo
  {title} {Measurement of the effect of quantum phase slips in a {Josephson}
  junction chain},\ }\href {https://www.nature.com/articles/nphys1697}
  {\bibfield  {journal} {\bibinfo  {journal} {Nature Physics}\ }\textbf
  {\bibinfo {volume} {6}},\ \bibinfo {pages} {589} (\bibinfo {year}
  {2010})}\BibitemShut {NoStop}%
\bibitem [{\citenamefont {Astafiev}\ \emph {et~al.}(2012)\citenamefont
  {Astafiev}, \citenamefont {Ioffe}, \citenamefont {Kafanov}, \citenamefont
  {Pashkin}, \citenamefont {Arutyunov}, \citenamefont {Shahar}, \citenamefont
  {Cohen},\ and\ \citenamefont {Tsai}}]{astafiev2012}%
  \BibitemOpen
  \bibfield  {author} {\bibinfo {author} {\bibfnamefont {O.}~\bibnamefont
  {Astafiev}}, \bibinfo {author} {\bibfnamefont {L.}~\bibnamefont {Ioffe}},
  \bibinfo {author} {\bibfnamefont {S.}~\bibnamefont {Kafanov}}, \bibinfo
  {author} {\bibfnamefont {Y.~A.}\ \bibnamefont {Pashkin}}, \bibinfo {author}
  {\bibfnamefont {K.~Y.}\ \bibnamefont {Arutyunov}}, \bibinfo {author}
  {\bibfnamefont {D.}~\bibnamefont {Shahar}}, \bibinfo {author} {\bibfnamefont
  {O.}~\bibnamefont {Cohen}},\ and\ \bibinfo {author} {\bibfnamefont {J.~S.}\
  \bibnamefont {Tsai}},\ }\bibfield  {title} {\bibinfo {title} {Coherent
  quantum phase slip},\ }\href {https://doi.org/10.1038/nature10930} {\bibfield
   {journal} {\bibinfo  {journal} {Nature}\ }\textbf {\bibinfo {volume}
  {484}},\ \bibinfo {pages} {355} (\bibinfo {year} {2012})}\BibitemShut
  {NoStop}%
\bibitem [{\citenamefont {Manucharyan}\ \emph {et~al.}(2012)\citenamefont
  {Manucharyan}, \citenamefont {Masluk}, \citenamefont {Kamal}, \citenamefont
  {Koch}, \citenamefont {Glazman},\ and\ \citenamefont
  {Devoret}}]{manucharyan2012}%
  \BibitemOpen
  \bibfield  {author} {\bibinfo {author} {\bibfnamefont {V.~E.}\ \bibnamefont
  {Manucharyan}}, \bibinfo {author} {\bibfnamefont {N.~A.}\ \bibnamefont
  {Masluk}}, \bibinfo {author} {\bibfnamefont {A.}~\bibnamefont {Kamal}},
  \bibinfo {author} {\bibfnamefont {J.}~\bibnamefont {Koch}}, \bibinfo {author}
  {\bibfnamefont {L.~I.}\ \bibnamefont {Glazman}},\ and\ \bibinfo {author}
  {\bibfnamefont {M.~H.}\ \bibnamefont {Devoret}},\ }\bibfield  {title}
  {\bibinfo {title} {Evidence for coherent quantum phase slips across a
  {Josephson} junction array},\ }\href
  {https://doi.org/10.1103/PhysRevB.85.024521} {\bibfield  {journal} {\bibinfo
  {journal} {Phys. Rev. B}\ }\textbf {\bibinfo {volume} {85}},\ \bibinfo
  {pages} {024521} (\bibinfo {year} {2012})}\BibitemShut {NoStop}%
\bibitem [{\citenamefont {Zaikin}\ \emph {et~al.}(1997)\citenamefont {Zaikin},
  \citenamefont {Golubev}, \citenamefont {van Otterlo},\ and\ \citenamefont
  {Zim\'anyi}}]{zaikin1997}%
  \BibitemOpen
  \bibfield  {author} {\bibinfo {author} {\bibfnamefont {A.~D.}\ \bibnamefont
  {Zaikin}}, \bibinfo {author} {\bibfnamefont {D.~S.}\ \bibnamefont {Golubev}},
  \bibinfo {author} {\bibfnamefont {A.}~\bibnamefont {van Otterlo}},\ and\
  \bibinfo {author} {\bibfnamefont {G.~T.}\ \bibnamefont {Zim\'anyi}},\
  }\bibfield  {title} {\bibinfo {title} {Quantum phase slips and transport in
  ultrathin superconducting wires},\ }\href
  {https://doi.org/10.1103/PhysRevLett.78.1552} {\bibfield  {journal} {\bibinfo
   {journal} {Phys. Rev. Lett.}\ }\textbf {\bibinfo {volume} {78}},\ \bibinfo
  {pages} {1552} (\bibinfo {year} {1997})}\BibitemShut {NoStop}%
\bibitem [{\citenamefont {Hekking}\ and\ \citenamefont
  {Glazman}(1997)}]{hekking1997}%
  \BibitemOpen
  \bibfield  {author} {\bibinfo {author} {\bibfnamefont {F.~W.~J.}\
  \bibnamefont {Hekking}}\ and\ \bibinfo {author} {\bibfnamefont {L.~I.}\
  \bibnamefont {Glazman}},\ }\bibfield  {title} {\bibinfo {title} {Quantum
  fluctuations in the equilibrium state of a thin superconducting loop},\
  }\href {https://doi.org/10.1103/PhysRevB.55.6551} {\bibfield  {journal}
  {\bibinfo  {journal} {Phys. Rev. B}\ }\textbf {\bibinfo {volume} {55}},\
  \bibinfo {pages} {6551} (\bibinfo {year} {1997})}\BibitemShut {NoStop}%
\bibitem [{\citenamefont {Fazio}\ and\ \citenamefont {{van der
  Zant}}(2001)}]{fazio2001}%
  \BibitemOpen
  \bibfield  {author} {\bibinfo {author} {\bibfnamefont {R.}~\bibnamefont
  {Fazio}}\ and\ \bibinfo {author} {\bibfnamefont {H.}~\bibnamefont {{van der
  Zant}}},\ }\bibfield  {title} {\bibinfo {title} {Quantum phase transitions
  and vortex dynamics in superconducting networks},\ }\href
  {https://doi.org/https://doi.org/10.1016/S0370-1573(01)00022-9} {\bibfield
  {journal} {\bibinfo  {journal} {Physics Reports}\ }\textbf {\bibinfo {volume}
  {355}},\ \bibinfo {pages} {235} (\bibinfo {year} {2001})}\BibitemShut
  {NoStop}%
\bibitem [{\citenamefont {Golubev}\ and\ \citenamefont
  {Zaikin}(2001)}]{golubev2001}%
  \BibitemOpen
  \bibfield  {author} {\bibinfo {author} {\bibfnamefont {D.~S.}\ \bibnamefont
  {Golubev}}\ and\ \bibinfo {author} {\bibfnamefont {A.~D.}\ \bibnamefont
  {Zaikin}},\ }\bibfield  {title} {\bibinfo {title} {Quantum tunneling of the
  order parameter in superconducting nanowires},\ }\href
  {https://doi.org/10.1103/PhysRevB.64.014504} {\bibfield  {journal} {\bibinfo
  {journal} {Phys. Rev. B}\ }\textbf {\bibinfo {volume} {64}},\ \bibinfo
  {pages} {014504} (\bibinfo {year} {2001})}\BibitemShut {NoStop}%
\bibitem [{\citenamefont {Matveev}\ \emph {et~al.}(2002)\citenamefont
  {Matveev}, \citenamefont {Larkin},\ and\ \citenamefont
  {Glazman}}]{matveev2002}%
  \BibitemOpen
  \bibfield  {author} {\bibinfo {author} {\bibfnamefont {K.~A.}\ \bibnamefont
  {Matveev}}, \bibinfo {author} {\bibfnamefont {A.~I.}\ \bibnamefont
  {Larkin}},\ and\ \bibinfo {author} {\bibfnamefont {L.~I.}\ \bibnamefont
  {Glazman}},\ }\bibfield  {title} {\bibinfo {title} {Persistent current in
  superconducting nanorings},\ }\href
  {https://doi.org/10.1103/PhysRevLett.89.096802} {\bibfield  {journal}
  {\bibinfo  {journal} {Phys. Rev. Lett.}\ }\textbf {\bibinfo {volume} {89}},\
  \bibinfo {pages} {096802} (\bibinfo {year} {2002})}\BibitemShut {NoStop}%
\bibitem [{\citenamefont {B\"uchler}\ \emph {et~al.}(2004)\citenamefont
  {B\"uchler}, \citenamefont {Geshkenbein},\ and\ \citenamefont
  {Blatter}}]{buchler2004}%
  \BibitemOpen
  \bibfield  {author} {\bibinfo {author} {\bibfnamefont {H.~P.}\ \bibnamefont
  {B\"uchler}}, \bibinfo {author} {\bibfnamefont {V.~B.}\ \bibnamefont
  {Geshkenbein}},\ and\ \bibinfo {author} {\bibfnamefont {G.}~\bibnamefont
  {Blatter}},\ }\bibfield  {title} {\bibinfo {title} {Quantum fluctuations in
  thin superconducting wires of finite length},\ }\href
  {https://doi.org/10.1103/PhysRevLett.92.067007} {\bibfield  {journal}
  {\bibinfo  {journal} {Phys. Rev. Lett.}\ }\textbf {\bibinfo {volume} {92}},\
  \bibinfo {pages} {067007} (\bibinfo {year} {2004})}\BibitemShut {NoStop}%
\bibitem [{\citenamefont {Refael}\ \emph {et~al.}(2007)\citenamefont {Refael},
  \citenamefont {Demler}, \citenamefont {Oreg},\ and\ \citenamefont
  {Fisher}}]{refael2007}%
  \BibitemOpen
  \bibfield  {author} {\bibinfo {author} {\bibfnamefont {G.}~\bibnamefont
  {Refael}}, \bibinfo {author} {\bibfnamefont {E.}~\bibnamefont {Demler}},
  \bibinfo {author} {\bibfnamefont {Y.}~\bibnamefont {Oreg}},\ and\ \bibinfo
  {author} {\bibfnamefont {D.~S.}\ \bibnamefont {Fisher}},\ }\bibfield  {title}
  {\bibinfo {title} {Superconductor-to-normal transitions in dissipative chains
  of mesoscopic grains and nanowires},\ }\href
  {https://doi.org/10.1103/PhysRevB.75.014522} {\bibfield  {journal} {\bibinfo
  {journal} {Phys. Rev. B}\ }\textbf {\bibinfo {volume} {75}},\ \bibinfo
  {pages} {014522} (\bibinfo {year} {2007})}\BibitemShut {NoStop}%
\bibitem [{\citenamefont {Halperin}\ \emph {et~al.}(2010)\citenamefont
  {Halperin}, \citenamefont {Refael},\ and\ \citenamefont
  {Demler}}]{halperin2010}%
  \BibitemOpen
  \bibfield  {author} {\bibinfo {author} {\bibfnamefont {B.~I.}\ \bibnamefont
  {Halperin}}, \bibinfo {author} {\bibfnamefont {G.}~\bibnamefont {Refael}},\
  and\ \bibinfo {author} {\bibfnamefont {E.}~\bibnamefont {Demler}},\
  }\bibfield  {title} {\bibinfo {title} {Resistance in superconductors},\
  }\href {https://arxiv.org/pdf/1005.3347.pdf} {\bibfield  {journal} {\bibinfo
  {journal} {International Journal of Modern Physics B}\ }\textbf {\bibinfo
  {volume} {24}},\ \bibinfo {pages} {4039} (\bibinfo {year}
  {2010})}\BibitemShut {NoStop}%
\bibitem [{\citenamefont {Averin}\ \emph {et~al.}(1985)\citenamefont {Averin},
  \citenamefont {Zorin},\ and\ \citenamefont {Likharev}}]{averin1985}%
  \BibitemOpen
  \bibfield  {author} {\bibinfo {author} {\bibfnamefont {D.~V.}\ \bibnamefont
  {Averin}}, \bibinfo {author} {\bibfnamefont {A.~B.}\ \bibnamefont {Zorin}},\
  and\ \bibinfo {author} {\bibfnamefont {K.~K.}\ \bibnamefont {Likharev}},\
  }\bibfield  {title} {\bibinfo {title} {Bloch oscillations in small
  {Josephson} junctions},\ }\href
  {http://www.jetp.ras.ru/cgi-bin/dn/e_061_02_0407} {\bibfield  {journal}
  {\bibinfo  {journal} {Sov. Phys. JETP}\ }\textbf {\bibinfo {volume} {61}},\
  \bibinfo {pages} {407} (\bibinfo {year} {1985})}\BibitemShut {NoStop}%
\bibitem [{\citenamefont {Bouchiat}\ \emph {et~al.}(1998)\citenamefont
  {Bouchiat}, \citenamefont {Vion}, \citenamefont {Joyez}, \citenamefont
  {Esteve},\ and\ \citenamefont {Devoret}}]{bouchiat1998}%
  \BibitemOpen
  \bibfield  {author} {\bibinfo {author} {\bibfnamefont {V.}~\bibnamefont
  {Bouchiat}}, \bibinfo {author} {\bibfnamefont {D.}~\bibnamefont {Vion}},
  \bibinfo {author} {\bibfnamefont {P.}~\bibnamefont {Joyez}}, \bibinfo
  {author} {\bibfnamefont {D.}~\bibnamefont {Esteve}},\ and\ \bibinfo {author}
  {\bibfnamefont {M.}~\bibnamefont {Devoret}},\ }\bibfield  {title} {\bibinfo
  {title} {Quantum coherence with a single {Cooper} pair},\ }\href
  {https://doi.org/10.1238/Physica.Topical.076a00165} {\bibfield  {journal}
  {\bibinfo  {journal} {Physica Scripta}\ }\textbf {\bibinfo {volume} {1998}},\
  \bibinfo {pages} {165} (\bibinfo {year} {1998})}\BibitemShut {NoStop}%
\bibitem [{\citenamefont {Nakamura}\ \emph {et~al.}(1999)\citenamefont
  {Nakamura}, \citenamefont {Pashkin},\ and\ \citenamefont
  {Tsai}}]{nakamura1999}%
  \BibitemOpen
  \bibfield  {author} {\bibinfo {author} {\bibfnamefont {Y.}~\bibnamefont
  {Nakamura}}, \bibinfo {author} {\bibfnamefont {Y.~A.}\ \bibnamefont
  {Pashkin}},\ and\ \bibinfo {author} {\bibfnamefont {J.}~\bibnamefont
  {Tsai}},\ }\bibfield  {title} {\bibinfo {title} {Coherent control of
  macroscopic quantum states in a {single-Cooper-pair} box},\ }\href
  {https://doi.org/10.1038/19718} {\bibfield  {journal} {\bibinfo  {journal}
  {nature}\ }\textbf {\bibinfo {volume} {398}},\ \bibinfo {pages} {786}
  (\bibinfo {year} {1999})}\BibitemShut {NoStop}%
\bibitem [{\citenamefont {Vion}\ \emph {et~al.}(2002)\citenamefont {Vion},
  \citenamefont {Aassime}, \citenamefont {Cottet}, \citenamefont {Joyez},
  \citenamefont {Pothier}, \citenamefont {Urbina}, \citenamefont {Esteve},\
  and\ \citenamefont {Devoret}}]{vion2002}%
  \BibitemOpen
  \bibfield  {author} {\bibinfo {author} {\bibfnamefont {D.}~\bibnamefont
  {Vion}}, \bibinfo {author} {\bibfnamefont {A.}~\bibnamefont {Aassime}},
  \bibinfo {author} {\bibfnamefont {A.}~\bibnamefont {Cottet}}, \bibinfo
  {author} {\bibfnamefont {P.}~\bibnamefont {Joyez}}, \bibinfo {author}
  {\bibfnamefont {H.}~\bibnamefont {Pothier}}, \bibinfo {author} {\bibfnamefont
  {C.}~\bibnamefont {Urbina}}, \bibinfo {author} {\bibfnamefont
  {D.}~\bibnamefont {Esteve}},\ and\ \bibinfo {author} {\bibfnamefont {M.~H.}\
  \bibnamefont {Devoret}},\ }\bibfield  {title} {\bibinfo {title} {Manipulating
  the quantum state of an electrical circuit},\ }\href
  {https://doi.org/10.1126/science.1069372} {\bibfield  {journal} {\bibinfo
  {journal} {Science}\ }\textbf {\bibinfo {volume} {296}},\ \bibinfo {pages}
  {886} (\bibinfo {year} {2002})}\BibitemShut {NoStop}%
\bibitem [{\citenamefont {Koch}\ \emph {et~al.}(2007)\citenamefont {Koch},
  \citenamefont {Yu}, \citenamefont {Gambetta}, \citenamefont {Houck},
  \citenamefont {Schuster}, \citenamefont {Majer}, \citenamefont {Blais},
  \citenamefont {Devoret}, \citenamefont {Girvin},\ and\ \citenamefont
  {Schoelkopf}}]{koch2007}%
  \BibitemOpen
  \bibfield  {author} {\bibinfo {author} {\bibfnamefont {J.}~\bibnamefont
  {Koch}}, \bibinfo {author} {\bibfnamefont {T.~M.}\ \bibnamefont {Yu}},
  \bibinfo {author} {\bibfnamefont {J.}~\bibnamefont {Gambetta}}, \bibinfo
  {author} {\bibfnamefont {A.~A.}\ \bibnamefont {Houck}}, \bibinfo {author}
  {\bibfnamefont {D.~I.}\ \bibnamefont {Schuster}}, \bibinfo {author}
  {\bibfnamefont {J.}~\bibnamefont {Majer}}, \bibinfo {author} {\bibfnamefont
  {A.}~\bibnamefont {Blais}}, \bibinfo {author} {\bibfnamefont {M.~H.}\
  \bibnamefont {Devoret}}, \bibinfo {author} {\bibfnamefont {S.~M.}\
  \bibnamefont {Girvin}},\ and\ \bibinfo {author} {\bibfnamefont {R.~J.}\
  \bibnamefont {Schoelkopf}},\ }\bibfield  {title} {\bibinfo {title}
  {Charge-insensitive qubit design derived from the {Cooper} pair box},\ }\href
  {https://doi.org/10.1103/PhysRevA.76.042319} {\bibfield  {journal} {\bibinfo
  {journal} {Phys. Rev. A}\ }\textbf {\bibinfo {volume} {76}},\ \bibinfo
  {pages} {042319} (\bibinfo {year} {2007})}\BibitemShut {NoStop}%
\bibitem [{\citenamefont {Schreier}\ \emph {et~al.}(2008)\citenamefont
  {Schreier}, \citenamefont {Houck}, \citenamefont {Koch}, \citenamefont
  {Schuster}, \citenamefont {Johnson}, \citenamefont {Chow}, \citenamefont
  {Gambetta}, \citenamefont {Majer}, \citenamefont {Frunzio}, \citenamefont
  {Devoret}, \citenamefont {Girvin},\ and\ \citenamefont
  {Schoelkopf}}]{schreier2008}%
  \BibitemOpen
  \bibfield  {author} {\bibinfo {author} {\bibfnamefont {J.~A.}\ \bibnamefont
  {Schreier}}, \bibinfo {author} {\bibfnamefont {A.~A.}\ \bibnamefont {Houck}},
  \bibinfo {author} {\bibfnamefont {J.}~\bibnamefont {Koch}}, \bibinfo {author}
  {\bibfnamefont {D.~I.}\ \bibnamefont {Schuster}}, \bibinfo {author}
  {\bibfnamefont {B.~R.}\ \bibnamefont {Johnson}}, \bibinfo {author}
  {\bibfnamefont {J.~M.}\ \bibnamefont {Chow}}, \bibinfo {author}
  {\bibfnamefont {J.~M.}\ \bibnamefont {Gambetta}}, \bibinfo {author}
  {\bibfnamefont {J.}~\bibnamefont {Majer}}, \bibinfo {author} {\bibfnamefont
  {L.}~\bibnamefont {Frunzio}}, \bibinfo {author} {\bibfnamefont {M.~H.}\
  \bibnamefont {Devoret}}, \bibinfo {author} {\bibfnamefont {S.~M.}\
  \bibnamefont {Girvin}},\ and\ \bibinfo {author} {\bibfnamefont {R.~J.}\
  \bibnamefont {Schoelkopf}},\ }\bibfield  {title} {\bibinfo {title}
  {Suppressing charge noise decoherence in superconducting charge qubits},\
  }\href {https://doi.org/10.1103/PhysRevB.77.180502} {\bibfield  {journal}
  {\bibinfo  {journal} {Phys. Rev. B}\ }\textbf {\bibinfo {volume} {77}},\
  \bibinfo {pages} {180502} (\bibinfo {year} {2008})}\BibitemShut {NoStop}%
\bibitem [{\citenamefont {Rist{\`e}}\ \emph {et~al.}(2013)\citenamefont
  {Rist{\`e}}, \citenamefont {Bultink}, \citenamefont {Tiggelman},
  \citenamefont {Schouten}, \citenamefont {Lehnert},\ and\ \citenamefont
  {DiCarlo}}]{riste2013}%
  \BibitemOpen
  \bibfield  {author} {\bibinfo {author} {\bibfnamefont {D.}~\bibnamefont
  {Rist{\`e}}}, \bibinfo {author} {\bibfnamefont {C.}~\bibnamefont {Bultink}},
  \bibinfo {author} {\bibfnamefont {M.~J.}\ \bibnamefont {Tiggelman}}, \bibinfo
  {author} {\bibfnamefont {R.~N.}\ \bibnamefont {Schouten}}, \bibinfo {author}
  {\bibfnamefont {K.~W.}\ \bibnamefont {Lehnert}},\ and\ \bibinfo {author}
  {\bibfnamefont {L.}~\bibnamefont {DiCarlo}},\ }\bibfield  {title} {\bibinfo
  {title} {Millisecond charge-parity fluctuations and induced decoherence in a
  superconducting transmon qubit},\ }\href {https://doi.org/10.1038/ncomms2936}
  {\bibfield  {journal} {\bibinfo  {journal} {Nature communications}\ }\textbf
  {\bibinfo {volume} {4}},\ \bibinfo {pages} {1913} (\bibinfo {year}
  {2013})}\BibitemShut {NoStop}%
\bibitem [{\citenamefont {Serniak}\ \emph {et~al.}(2018)\citenamefont
  {Serniak}, \citenamefont {Hays}, \citenamefont {de~Lange}, \citenamefont
  {Diamond}, \citenamefont {Shankar}, \citenamefont {Burkhart}, \citenamefont
  {Frunzio}, \citenamefont {Houzet},\ and\ \citenamefont
  {Devoret}}]{serniak2018}%
  \BibitemOpen
  \bibfield  {author} {\bibinfo {author} {\bibfnamefont {K.}~\bibnamefont
  {Serniak}}, \bibinfo {author} {\bibfnamefont {M.}~\bibnamefont {Hays}},
  \bibinfo {author} {\bibfnamefont {G.}~\bibnamefont {de~Lange}}, \bibinfo
  {author} {\bibfnamefont {S.}~\bibnamefont {Diamond}}, \bibinfo {author}
  {\bibfnamefont {S.}~\bibnamefont {Shankar}}, \bibinfo {author} {\bibfnamefont
  {L.~D.}\ \bibnamefont {Burkhart}}, \bibinfo {author} {\bibfnamefont
  {L.}~\bibnamefont {Frunzio}}, \bibinfo {author} {\bibfnamefont
  {M.}~\bibnamefont {Houzet}},\ and\ \bibinfo {author} {\bibfnamefont {M.~H.}\
  \bibnamefont {Devoret}},\ }\bibfield  {title} {\bibinfo {title} {Hot
  nonequilibrium quasiparticles in transmon qubits},\ }\href
  {https://doi.org/10.1103/PhysRevLett.121.157701} {\bibfield  {journal}
  {\bibinfo  {journal} {Phys. Rev. Lett.}\ }\textbf {\bibinfo {volume} {121}},\
  \bibinfo {pages} {157701} (\bibinfo {year} {2018})}\BibitemShut {NoStop}%
\bibitem [{\citenamefont {Serniak}\ \emph {et~al.}(2019)\citenamefont
  {Serniak}, \citenamefont {Diamond}, \citenamefont {Hays}, \citenamefont
  {Fatemi}, \citenamefont {Shankar}, \citenamefont {Frunzio}, \citenamefont
  {Schoelkopf},\ and\ \citenamefont {Devoret}}]{serniak2019}%
  \BibitemOpen
  \bibfield  {author} {\bibinfo {author} {\bibfnamefont {K.}~\bibnamefont
  {Serniak}}, \bibinfo {author} {\bibfnamefont {S.}~\bibnamefont {Diamond}},
  \bibinfo {author} {\bibfnamefont {M.}~\bibnamefont {Hays}}, \bibinfo {author}
  {\bibfnamefont {V.}~\bibnamefont {Fatemi}}, \bibinfo {author} {\bibfnamefont
  {S.}~\bibnamefont {Shankar}}, \bibinfo {author} {\bibfnamefont
  {L.}~\bibnamefont {Frunzio}}, \bibinfo {author} {\bibfnamefont
  {R.}~\bibnamefont {Schoelkopf}},\ and\ \bibinfo {author} {\bibfnamefont
  {M.}~\bibnamefont {Devoret}},\ }\bibfield  {title} {\bibinfo {title} {Direct
  dispersive monitoring of charge parity in offset-charge-sensitive
  transmons},\ }\href {https://doi.org/10.1103/PhysRevApplied.12.014052}
  {\bibfield  {journal} {\bibinfo  {journal} {Phys. Rev. Appl.}\ }\textbf
  {\bibinfo {volume} {12}},\ \bibinfo {pages} {014052} (\bibinfo {year}
  {2019})}\BibitemShut {NoStop}%
\bibitem [{\citenamefont {Uilhoorn}\ \emph {et~al.}(2021)\citenamefont
  {Uilhoorn}, \citenamefont {Kroll}, \citenamefont {Bargerbos}, \citenamefont
  {Nabi}, \citenamefont {Yang}, \citenamefont {Krogstrup}, \citenamefont
  {Kouwenhoven}, \citenamefont {Kou},\ and\ \citenamefont
  {de~Lange}}]{uilhoorn2021}%
  \BibitemOpen
  \bibfield  {author} {\bibinfo {author} {\bibfnamefont {W.}~\bibnamefont
  {Uilhoorn}}, \bibinfo {author} {\bibfnamefont {J.~G.}\ \bibnamefont {Kroll}},
  \bibinfo {author} {\bibfnamefont {A.}~\bibnamefont {Bargerbos}}, \bibinfo
  {author} {\bibfnamefont {S.~D.}\ \bibnamefont {Nabi}}, \bibinfo {author}
  {\bibfnamefont {C.-K.}\ \bibnamefont {Yang}}, \bibinfo {author}
  {\bibfnamefont {P.}~\bibnamefont {Krogstrup}}, \bibinfo {author}
  {\bibfnamefont {L.~P.}\ \bibnamefont {Kouwenhoven}}, \bibinfo {author}
  {\bibfnamefont {A.}~\bibnamefont {Kou}},\ and\ \bibinfo {author}
  {\bibfnamefont {G.}~\bibnamefont {de~Lange}},\ }\href
  {https://doi.org/10.48550/ARXIV.2105.11038} {\bibinfo {title} {Quasiparticle
  trapping by orbital effect in a hybrid superconducting-semiconducting
  circuit}} (\bibinfo {year} {2021})\BibitemShut {NoStop}%
\bibitem [{\citenamefont {Kurter}\ \emph {et~al.}(2022)\citenamefont {Kurter},
  \citenamefont {Murray}, \citenamefont {Gordon}, \citenamefont {Wymore},
  \citenamefont {Sandberg}, \citenamefont {Shelby}, \citenamefont {Eddins},
  \citenamefont {Adiga}, \citenamefont {Finck}, \citenamefont {Rivera} \emph
  {et~al.}}]{kurter2022}%
  \BibitemOpen
  \bibfield  {author} {\bibinfo {author} {\bibfnamefont {C.}~\bibnamefont
  {Kurter}}, \bibinfo {author} {\bibfnamefont {C.}~\bibnamefont {Murray}},
  \bibinfo {author} {\bibfnamefont {R.}~\bibnamefont {Gordon}}, \bibinfo
  {author} {\bibfnamefont {B.}~\bibnamefont {Wymore}}, \bibinfo {author}
  {\bibfnamefont {M.}~\bibnamefont {Sandberg}}, \bibinfo {author}
  {\bibfnamefont {R.}~\bibnamefont {Shelby}}, \bibinfo {author} {\bibfnamefont
  {A.}~\bibnamefont {Eddins}}, \bibinfo {author} {\bibfnamefont
  {V.}~\bibnamefont {Adiga}}, \bibinfo {author} {\bibfnamefont
  {A.}~\bibnamefont {Finck}}, \bibinfo {author} {\bibfnamefont
  {E.}~\bibnamefont {Rivera}}, \emph {et~al.},\ }\bibfield  {title} {\bibinfo
  {title} {Quasiparticle tunneling as a probe of josephson junction barrier and
  capacitor material in superconducting qubits},\ }\href
  {https://doi.org/10.1038/s41534-022-00542-2} {\bibfield  {journal} {\bibinfo
  {journal} {npj Quantum Information}\ }\textbf {\bibinfo {volume} {8}},\
  \bibinfo {pages} {31} (\bibinfo {year} {2022})}\BibitemShut {NoStop}%
\bibitem [{\citenamefont {Erlandsson}\ \emph {et~al.}(2022)\citenamefont
  {Erlandsson}, \citenamefont {Sabonis}, \citenamefont {Kringhøj},
  \citenamefont {Larsen}, \citenamefont {Krogstrup}, \citenamefont
  {Petersson},\ and\ \citenamefont {Marcus}}]{erlandsson2022}%
  \BibitemOpen
  \bibfield  {author} {\bibinfo {author} {\bibfnamefont {O.}~\bibnamefont
  {Erlandsson}}, \bibinfo {author} {\bibfnamefont {D.}~\bibnamefont {Sabonis}},
  \bibinfo {author} {\bibfnamefont {A.}~\bibnamefont {Kringhøj}}, \bibinfo
  {author} {\bibfnamefont {T.~W.}\ \bibnamefont {Larsen}}, \bibinfo {author}
  {\bibfnamefont {P.}~\bibnamefont {Krogstrup}}, \bibinfo {author}
  {\bibfnamefont {K.~D.}\ \bibnamefont {Petersson}},\ and\ \bibinfo {author}
  {\bibfnamefont {C.~M.}\ \bibnamefont {Marcus}},\ }\href
  {https://doi.org/10.48550/ARXIV.2202.05974} {\bibinfo {title} {Parity
  switching in a full-shell superconductor-semiconductor nanowire qubit}}
  (\bibinfo {year} {2022})\BibitemShut {NoStop}%
\bibitem [{\citenamefont {Hassler}\ \emph {et~al.}(2011)\citenamefont
  {Hassler}, \citenamefont {Akhmerov},\ and\ \citenamefont
  {Beenakker}}]{hassler2011}%
  \BibitemOpen
  \bibfield  {author} {\bibinfo {author} {\bibfnamefont {F.}~\bibnamefont
  {Hassler}}, \bibinfo {author} {\bibfnamefont {A.}~\bibnamefont {Akhmerov}},\
  and\ \bibinfo {author} {\bibfnamefont {C.}~\bibnamefont {Beenakker}},\
  }\bibfield  {title} {\bibinfo {title} {The top-transmon: a hybrid
  superconducting qubit for parity-protected quantum computation},\ }\href
  {https://doi.org/10.1088/1367-2630/13/9/095004} {\bibfield  {journal}
  {\bibinfo  {journal} {New Journal of Physics}\ }\textbf {\bibinfo {volume}
  {13}},\ \bibinfo {pages} {095004} (\bibinfo {year} {2011})}\BibitemShut
  {NoStop}%
\bibitem [{\citenamefont {Aasen}\ \emph {et~al.}(2016)\citenamefont {Aasen},
  \citenamefont {Hell}, \citenamefont {Mishmash}, \citenamefont {Higginbotham},
  \citenamefont {Danon}, \citenamefont {Leijnse}, \citenamefont {Jespersen},
  \citenamefont {Folk}, \citenamefont {Marcus}, \citenamefont {Flensberg},\
  and\ \citenamefont {Alicea}}]{aasen2016}%
  \BibitemOpen
  \bibfield  {author} {\bibinfo {author} {\bibfnamefont {D.}~\bibnamefont
  {Aasen}}, \bibinfo {author} {\bibfnamefont {M.}~\bibnamefont {Hell}},
  \bibinfo {author} {\bibfnamefont {R.~V.}\ \bibnamefont {Mishmash}}, \bibinfo
  {author} {\bibfnamefont {A.}~\bibnamefont {Higginbotham}}, \bibinfo {author}
  {\bibfnamefont {J.}~\bibnamefont {Danon}}, \bibinfo {author} {\bibfnamefont
  {M.}~\bibnamefont {Leijnse}}, \bibinfo {author} {\bibfnamefont {T.~S.}\
  \bibnamefont {Jespersen}}, \bibinfo {author} {\bibfnamefont {J.~A.}\
  \bibnamefont {Folk}}, \bibinfo {author} {\bibfnamefont {C.~M.}\ \bibnamefont
  {Marcus}}, \bibinfo {author} {\bibfnamefont {K.}~\bibnamefont {Flensberg}},\
  and\ \bibinfo {author} {\bibfnamefont {J.}~\bibnamefont {Alicea}},\
  }\bibfield  {title} {\bibinfo {title} {Milestones toward {Majorana}-based
  quantum computing},\ }\href {https://doi.org/10.1103/PhysRevX.6.031016}
  {\bibfield  {journal} {\bibinfo  {journal} {Phys. Rev. X}\ }\textbf {\bibinfo
  {volume} {6}},\ \bibinfo {pages} {031016} (\bibinfo {year}
  {2016})}\BibitemShut {NoStop}%
\bibitem [{\citenamefont {Aguado}(2020)}]{aguado2020}%
  \BibitemOpen
  \bibfield  {author} {\bibinfo {author} {\bibfnamefont {R.}~\bibnamefont
  {Aguado}},\ }\bibfield  {title} {\bibinfo {title} {A perspective on
  semiconductor-based superconducting qubits},\ }\href
  {https://doi.org/10.1063/5.0024124} {\bibfield  {journal} {\bibinfo
  {journal} {Applied Physics Letters}\ }\textbf {\bibinfo {volume} {117}},\
  \bibinfo {pages} {240501} (\bibinfo {year} {2020})}\BibitemShut {NoStop}%
\bibitem [{\citenamefont {Gyenis}\ \emph {et~al.}(2021)\citenamefont {Gyenis},
  \citenamefont {Di~Paolo}, \citenamefont {Koch}, \citenamefont {Blais},
  \citenamefont {Houck},\ and\ \citenamefont {Schuster}}]{gyenis2021}%
  \BibitemOpen
  \bibfield  {author} {\bibinfo {author} {\bibfnamefont {A.}~\bibnamefont
  {Gyenis}}, \bibinfo {author} {\bibfnamefont {A.}~\bibnamefont {Di~Paolo}},
  \bibinfo {author} {\bibfnamefont {J.}~\bibnamefont {Koch}}, \bibinfo {author}
  {\bibfnamefont {A.}~\bibnamefont {Blais}}, \bibinfo {author} {\bibfnamefont
  {A.~A.}\ \bibnamefont {Houck}},\ and\ \bibinfo {author} {\bibfnamefont
  {D.~I.}\ \bibnamefont {Schuster}},\ }\bibfield  {title} {\bibinfo {title}
  {Moving beyond the transmon: {Noise}-protected superconducting quantum
  circuits},\ }\href {https://doi.org/10.1103/PRXQuantum.2.030101} {\bibfield
  {journal} {\bibinfo  {journal} {PRX Quantum}\ }\textbf {\bibinfo {volume}
  {2}},\ \bibinfo {pages} {030101} (\bibinfo {year} {2021})}\BibitemShut
  {NoStop}%
\bibitem [{\citenamefont {Kringh\o{}j}\ \emph {et~al.}(2020)\citenamefont
  {Kringh\o{}j}, \citenamefont {van Heck}, \citenamefont {Larsen},
  \citenamefont {Erlandsson}, \citenamefont {Sabonis}, \citenamefont
  {Krogstrup}, \citenamefont {Casparis}, \citenamefont {Petersson},\ and\
  \citenamefont {Marcus}}]{kringhoj2020}%
  \BibitemOpen
  \bibfield  {author} {\bibinfo {author} {\bibfnamefont {A.}~\bibnamefont
  {Kringh\o{}j}}, \bibinfo {author} {\bibfnamefont {B.}~\bibnamefont {van
  Heck}}, \bibinfo {author} {\bibfnamefont {T.~W.}\ \bibnamefont {Larsen}},
  \bibinfo {author} {\bibfnamefont {O.}~\bibnamefont {Erlandsson}}, \bibinfo
  {author} {\bibfnamefont {D.}~\bibnamefont {Sabonis}}, \bibinfo {author}
  {\bibfnamefont {P.}~\bibnamefont {Krogstrup}}, \bibinfo {author}
  {\bibfnamefont {L.}~\bibnamefont {Casparis}}, \bibinfo {author}
  {\bibfnamefont {K.~D.}\ \bibnamefont {Petersson}},\ and\ \bibinfo {author}
  {\bibfnamefont {C.~M.}\ \bibnamefont {Marcus}},\ }\bibfield  {title}
  {\bibinfo {title} {Suppressed charge dispersion via resonant tunneling in a
  single-channel transmon},\ }\href
  {https://doi.org/10.1103/PhysRevLett.124.246803} {\bibfield  {journal}
  {\bibinfo  {journal} {Phys. Rev. Lett.}\ }\textbf {\bibinfo {volume} {124}},\
  \bibinfo {pages} {246803} (\bibinfo {year} {2020})}\BibitemShut {NoStop}%
\bibitem [{\citenamefont {Bargerbos}\ \emph {et~al.}(2020)\citenamefont
  {Bargerbos}, \citenamefont {Uilhoorn}, \citenamefont {Yang}, \citenamefont
  {Krogstrup}, \citenamefont {Kouwenhoven}, \citenamefont {de~Lange},
  \citenamefont {van Heck},\ and\ \citenamefont {Kou}}]{bargerbos2020}%
  \BibitemOpen
  \bibfield  {author} {\bibinfo {author} {\bibfnamefont {A.}~\bibnamefont
  {Bargerbos}}, \bibinfo {author} {\bibfnamefont {W.}~\bibnamefont {Uilhoorn}},
  \bibinfo {author} {\bibfnamefont {C.-K.}\ \bibnamefont {Yang}}, \bibinfo
  {author} {\bibfnamefont {P.}~\bibnamefont {Krogstrup}}, \bibinfo {author}
  {\bibfnamefont {L.~P.}\ \bibnamefont {Kouwenhoven}}, \bibinfo {author}
  {\bibfnamefont {G.}~\bibnamefont {de~Lange}}, \bibinfo {author}
  {\bibfnamefont {B.}~\bibnamefont {van Heck}},\ and\ \bibinfo {author}
  {\bibfnamefont {A.}~\bibnamefont {Kou}},\ }\bibfield  {title} {\bibinfo
  {title} {Observation of vanishing charge dispersion of a nearly open
  superconducting island},\ }\href
  {https://doi.org/10.1103/PhysRevLett.124.246802} {\bibfield  {journal}
  {\bibinfo  {journal} {Phys. Rev. Lett.}\ }\textbf {\bibinfo {volume} {124}},\
  \bibinfo {pages} {246802} (\bibinfo {year} {2020})}\BibitemShut {NoStop}%
\bibitem [{Note1()}]{Note1}%
  \BibitemOpen
  \bibinfo {note} {A low-transparency QPC differs from an oxide junction
  because in the former the entire phase dispersion of the ground state
  originates from a single transport channel, and thus a single Andreev bound
  state, while in the latter from hundreds or even thousands of transport
  channels. The two junctions have equivalent ground state properties, but
  different densities of states close to the gap edge; the sketch in
  figure~\ref {fig:intro}c schematically depicts the first case.}\BibitemShut
  {Stop}%
\bibitem [{\citenamefont {Holstein}(1988)}]{holstein1988}%
  \BibitemOpen
  \bibfield  {author} {\bibinfo {author} {\bibfnamefont {B.~R.}\ \bibnamefont
  {Holstein}},\ }\bibfield  {title} {\bibinfo {title} {Semiclassical treatment
  of the periodic potential},\ }\href {https://doi.org/10.1119/1.15405}
  {\bibfield  {journal} {\bibinfo  {journal} {American Journal of Physics}\
  }\textbf {\bibinfo {volume} {56}},\ \bibinfo {pages} {894} (\bibinfo {year}
  {1988})}\BibitemShut {NoStop}%
\bibitem [{\citenamefont {Altland}\ and\ \citenamefont
  {Simons}(2010)}]{altland}%
  \BibitemOpen
  \bibfield  {author} {\bibinfo {author} {\bibfnamefont {A.}~\bibnamefont
  {Altland}}\ and\ \bibinfo {author} {\bibfnamefont {B.~D.}\ \bibnamefont
  {Simons}},\ }\href
  {https://login.ezproxy.leidenuniv.nl:2443/login?URL=https://search.ebscohost.com/login.aspx?direct=true&db=e000xww&AN=329335&site=ehost-live}
  {\emph {\bibinfo {title} {Condensed Matter Field Theory.}}},\ \bibinfo
  {edition} {2nd}\ ed.\ (\bibinfo  {publisher} {Cambridge University Press},\
  \bibinfo {year} {2010})\ pp.\ \bibinfo {pages} {154--155}\BibitemShut
  {NoStop}%
\bibitem [{\citenamefont {Averin}(1999{\natexlab{a}})}]{averin1999}%
  \BibitemOpen
  \bibfield  {author} {\bibinfo {author} {\bibfnamefont {D.~V.}\ \bibnamefont
  {Averin}},\ }\bibfield  {title} {\bibinfo {title} {Coulomb blockade in
  superconducting quantum point contacts},\ }\href
  {https://doi.org/10.1103/PhysRevLett.82.3685} {\bibfield  {journal} {\bibinfo
   {journal} {Phys. Rev. Lett.}\ }\textbf {\bibinfo {volume} {82}},\ \bibinfo
  {pages} {3685} (\bibinfo {year} {1999}{\natexlab{a}})}\BibitemShut {NoStop}%
\bibitem [{\citenamefont {Ivanov}\ and\ \citenamefont
  {Feigel’man}(1998)}]{ivanov1998}%
  \BibitemOpen
  \bibfield  {author} {\bibinfo {author} {\bibfnamefont {D.}~\bibnamefont
  {Ivanov}}\ and\ \bibinfo {author} {\bibfnamefont {M.}~\bibnamefont
  {Feigel’man}},\ }\bibfield  {title} {\bibinfo {title} {Coulomb effects in a
  ballistic one-channel {S-S-S} device},\ }\href
  {https://doi.org/https://doi.org/10.1134/1.558666} {\bibfield  {journal}
  {\bibinfo  {journal} {Journal of Experimental and Theoretical Physics}\
  }\textbf {\bibinfo {volume} {87}},\ \bibinfo {pages} {349} (\bibinfo {year}
  {1998})}\BibitemShut {NoStop}%
\bibitem [{\citenamefont {Averin}(1999{\natexlab{b}})}]{averin1999b}%
  \BibitemOpen
  \bibfield  {author} {\bibinfo {author} {\bibfnamefont {D.}~\bibnamefont
  {Averin}},\ }\bibfield  {title} {\bibinfo {title} {Quantum dynamics of
  superconducting point contacts: chiral anomaly, {Landau}--{Zener}
  transitions, and all that},\ }\href {https://doi.org/10.1006/spmi.1999.0748}
  {\bibfield  {journal} {\bibinfo  {journal} {Superlattices and
  microstructures}\ }\textbf {\bibinfo {volume} {25}},\ \bibinfo {pages} {891}
  (\bibinfo {year} {1999}{\natexlab{b}})}\BibitemShut {NoStop}%
\bibitem [{\citenamefont {Beenakker}\ and\ \citenamefont {van
  Houten}(1992)}]{carlo1992}%
  \BibitemOpen
  \bibfield  {author} {\bibinfo {author} {\bibfnamefont {C.~W.~J.}\
  \bibnamefont {Beenakker}}\ and\ \bibinfo {author} {\bibfnamefont
  {H.}~\bibnamefont {van Houten}},\ }\bibfield  {title} {\bibinfo {title}
  {Resonant {Josephson} current through a quantum dot},\ }in\ \href
  {https://arxiv.org/pdf/cond-mat/0111505.pdf} {\emph {\bibinfo {booktitle}
  {Single-Electron Tunneling and Mesoscopic Devices}}},\ \bibinfo {editor}
  {edited by\ \bibinfo {editor} {\bibfnamefont {H.}~\bibnamefont {Koch}}\ and\
  \bibinfo {editor} {\bibfnamefont {H.}~\bibnamefont {L{\"u}bbig}}}\ (\bibinfo
  {publisher} {Springer Berlin Heidelberg},\ \bibinfo {address} {Berlin,
  Heidelberg},\ \bibinfo {year} {1992})\ pp.\ \bibinfo {pages}
  {175--179}\BibitemShut {NoStop}%
\bibitem [{\citenamefont {Devyatov}\ and\ \citenamefont
  {Kupriyanov}(1997)}]{devyatov1997}%
  \BibitemOpen
  \bibfield  {author} {\bibinfo {author} {\bibfnamefont {I.~A.}\ \bibnamefont
  {Devyatov}}\ and\ \bibinfo {author} {\bibfnamefont {M.~Y.}\ \bibnamefont
  {Kupriyanov}},\ }\bibfield  {title} {\bibinfo {title} {Resonant {Josephson}
  tunneling through {S}-{I}-{S} junctions of arbitrary size},\ }\href
  {https://doi.org/10.1134/1.558305} {\bibfield  {journal} {\bibinfo  {journal}
  {Journal of Experimental and Theoretical Physics}\ }\textbf {\bibinfo
  {volume} {85}},\ \bibinfo {pages} {189} (\bibinfo {year} {1997})}\BibitemShut
  {NoStop}%
\bibitem [{\citenamefont {Beenakker}(1991)}]{carlo1991}%
  \BibitemOpen
  \bibfield  {author} {\bibinfo {author} {\bibfnamefont {C.~W.~J.}\
  \bibnamefont {Beenakker}},\ }\bibfield  {title} {\bibinfo {title} {Universal
  limit of critical-current fluctuations in mesoscopic {Josephson} junctions},\
  }\href {https://doi.org/10.1103/PhysRevLett.67.3836} {\bibfield  {journal}
  {\bibinfo  {journal} {Phys. Rev. Lett.}\ }\textbf {\bibinfo {volume} {67}},\
  \bibinfo {pages} {3836} (\bibinfo {year} {1991})}\BibitemShut {NoStop}%
\bibitem [{\citenamefont {Kurilovich}\ \emph {et~al.}(2021)\citenamefont
  {Kurilovich}, \citenamefont {Kurilovich}, \citenamefont {Fatemi},
  \citenamefont {Devoret},\ and\ \citenamefont {Glazman}}]{kurilovich2021}%
  \BibitemOpen
  \bibfield  {author} {\bibinfo {author} {\bibfnamefont {P.~D.}\ \bibnamefont
  {Kurilovich}}, \bibinfo {author} {\bibfnamefont {V.~D.}\ \bibnamefont
  {Kurilovich}}, \bibinfo {author} {\bibfnamefont {V.}~\bibnamefont {Fatemi}},
  \bibinfo {author} {\bibfnamefont {M.~H.}\ \bibnamefont {Devoret}},\ and\
  \bibinfo {author} {\bibfnamefont {L.~I.}\ \bibnamefont {Glazman}},\
  }\bibfield  {title} {\bibinfo {title} {Microwave response of an {Andreev}
  bound state},\ }\href {https://doi.org/10.1103/PhysRevB.104.174517}
  {\bibfield  {journal} {\bibinfo  {journal} {Phys. Rev. B}\ }\textbf {\bibinfo
  {volume} {104}},\ \bibinfo {pages} {174517} (\bibinfo {year}
  {2021})}\BibitemShut {NoStop}%
\bibitem [{\citenamefont {Kitaev}(2001)}]{kitaev2001}%
  \BibitemOpen
  \bibfield  {author} {\bibinfo {author} {\bibfnamefont {A.~Y.}\ \bibnamefont
  {Kitaev}},\ }\bibfield  {title} {\bibinfo {title} {Unpaired {Majorana}
  fermions in quantum wires},\ }\href
  {https://doi.org/10.1070/1063-7869/44/10s/s29} {\bibfield  {journal}
  {\bibinfo  {journal} {Physics-Uspekhi}\ }\textbf {\bibinfo {volume} {44}},\
  \bibinfo {pages} {131} (\bibinfo {year} {2001})}\BibitemShut {NoStop}%
\bibitem [{\citenamefont {Fu}\ and\ \citenamefont {Kane}(2009)}]{fu2009}%
  \BibitemOpen
  \bibfield  {author} {\bibinfo {author} {\bibfnamefont {L.}~\bibnamefont
  {Fu}}\ and\ \bibinfo {author} {\bibfnamefont {C.~L.}\ \bibnamefont {Kane}},\
  }\bibfield  {title} {\bibinfo {title} {{Josephson} current and noise at a
  superconductor/quantum - spin - {Hall}-insulator/superconductor junction},\
  }\href {https://doi.org/10.1103/PhysRevB.79.161408} {\bibfield  {journal}
  {\bibinfo  {journal} {Phys. Rev. B}\ }\textbf {\bibinfo {volume} {79}},\
  \bibinfo {pages} {161408} (\bibinfo {year} {2009})}\BibitemShut {NoStop}%
\bibitem [{\citenamefont {Beenakker}(1992)}]{beenakker1992three}%
  \BibitemOpen
  \bibfield  {author} {\bibinfo {author} {\bibfnamefont {C.}~\bibnamefont
  {Beenakker}},\ }\bibfield  {title} {\bibinfo {title} {Three “universal”
  mesoscopic {Josephson} effects},\ }in\ \href
  {https://arxiv.org/pdf/cond-mat/0406127.pdf} {\emph {\bibinfo {booktitle}
  {Transport Phenomena in Mesoscopic Systems}}}\ (\bibinfo  {publisher}
  {Springer},\ \bibinfo {year} {1992})\ pp.\ \bibinfo {pages}
  {235--253}\BibitemShut {NoStop}%
\bibitem [{\citenamefont {Mart{\'\i}n-Rodero}\ and\ \citenamefont
  {Levy~Yeyati}(2011)}]{martinrodero2011}%
  \BibitemOpen
  \bibfield  {author} {\bibinfo {author} {\bibfnamefont {A.}~\bibnamefont
  {Mart{\'\i}n-Rodero}}\ and\ \bibinfo {author} {\bibfnamefont
  {A.}~\bibnamefont {Levy~Yeyati}},\ }\bibfield  {title} {\bibinfo {title}
  {{Josephson} and {Andreev }transport through quantum dots},\ }\href
  {https://doi.org/10.1080/00018732.2011.624266} {\bibfield  {journal}
  {\bibinfo  {journal} {Advances in Physics}\ }\textbf {\bibinfo {volume}
  {60}},\ \bibinfo {pages} {899} (\bibinfo {year} {2011})}\BibitemShut
  {NoStop}%
\bibitem [{\citenamefont {Janvier}\ \emph {et~al.}(2015)\citenamefont
  {Janvier}, \citenamefont {Tosi}, \citenamefont {Bretheau}, \citenamefont
  {Girit}, \citenamefont {Stern}, \citenamefont {Bertet}, \citenamefont
  {Joyez}, \citenamefont {Vion}, \citenamefont {Esteve}, \citenamefont
  {Goffman} \emph {et~al.}}]{janvier2015}%
  \BibitemOpen
  \bibfield  {author} {\bibinfo {author} {\bibfnamefont {C.}~\bibnamefont
  {Janvier}}, \bibinfo {author} {\bibfnamefont {L.}~\bibnamefont {Tosi}},
  \bibinfo {author} {\bibfnamefont {L.}~\bibnamefont {Bretheau}}, \bibinfo
  {author} {\bibfnamefont {{\c{C}}.}~\bibnamefont {Girit}}, \bibinfo {author}
  {\bibfnamefont {M.}~\bibnamefont {Stern}}, \bibinfo {author} {\bibfnamefont
  {P.}~\bibnamefont {Bertet}}, \bibinfo {author} {\bibfnamefont
  {P.}~\bibnamefont {Joyez}}, \bibinfo {author} {\bibfnamefont
  {D.}~\bibnamefont {Vion}}, \bibinfo {author} {\bibfnamefont {D.}~\bibnamefont
  {Esteve}}, \bibinfo {author} {\bibfnamefont {M.}~\bibnamefont {Goffman}},
  \emph {et~al.},\ }\bibfield  {title} {\bibinfo {title} {Coherent manipulation
  of {Andreev} states in superconducting atomic contacts},\ }\href
  {10.1126/science.aab2179} {\bibfield  {journal} {\bibinfo  {journal}
  {Science}\ }\textbf {\bibinfo {volume} {349}},\ \bibinfo {pages} {1199}
  (\bibinfo {year} {2015})}\BibitemShut {NoStop}%
\bibitem [{\citenamefont {Hays}\ \emph {et~al.}(2018)\citenamefont {Hays},
  \citenamefont {de~Lange}, \citenamefont {Serniak}, \citenamefont {van
  Woerkom}, \citenamefont {Bouman}, \citenamefont {Krogstrup}, \citenamefont
  {Nyg\aa{}rd}, \citenamefont {Geresdi},\ and\ \citenamefont
  {Devoret}}]{hays2018}%
  \BibitemOpen
  \bibfield  {author} {\bibinfo {author} {\bibfnamefont {M.}~\bibnamefont
  {Hays}}, \bibinfo {author} {\bibfnamefont {G.}~\bibnamefont {de~Lange}},
  \bibinfo {author} {\bibfnamefont {K.}~\bibnamefont {Serniak}}, \bibinfo
  {author} {\bibfnamefont {D.~J.}\ \bibnamefont {van Woerkom}}, \bibinfo
  {author} {\bibfnamefont {D.}~\bibnamefont {Bouman}}, \bibinfo {author}
  {\bibfnamefont {P.}~\bibnamefont {Krogstrup}}, \bibinfo {author}
  {\bibfnamefont {J.}~\bibnamefont {Nyg\aa{}rd}}, \bibinfo {author}
  {\bibfnamefont {A.}~\bibnamefont {Geresdi}},\ and\ \bibinfo {author}
  {\bibfnamefont {M.~H.}\ \bibnamefont {Devoret}},\ }\bibfield  {title}
  {\bibinfo {title} {Direct microwave measurement of {Andreev}-bound-state
  dynamics in a semiconductor-nanowire {Josephson} junction},\ }\href
  {https://doi.org/10.1103/PhysRevLett.121.047001} {\bibfield  {journal}
  {\bibinfo  {journal} {Phys. Rev. Lett.}\ }\textbf {\bibinfo {volume} {121}},\
  \bibinfo {pages} {047001} (\bibinfo {year} {2018})}\BibitemShut {NoStop}%
\bibitem [{\citenamefont {Hays}\ \emph {et~al.}(2021)\citenamefont {Hays},
  \citenamefont {Fatemi}, \citenamefont {Bouman}, \citenamefont {Cerrillo},
  \citenamefont {Diamond}, \citenamefont {Serniak}, \citenamefont {Connolly},
  \citenamefont {Krogstrup}, \citenamefont {Nygård}, \citenamefont {Yeyati},
  \citenamefont {Geresdi},\ and\ \citenamefont {Devoret}}]{hays2021}%
  \BibitemOpen
  \bibfield  {author} {\bibinfo {author} {\bibfnamefont {M.}~\bibnamefont
  {Hays}}, \bibinfo {author} {\bibfnamefont {V.}~\bibnamefont {Fatemi}},
  \bibinfo {author} {\bibfnamefont {D.}~\bibnamefont {Bouman}}, \bibinfo
  {author} {\bibfnamefont {J.}~\bibnamefont {Cerrillo}}, \bibinfo {author}
  {\bibfnamefont {S.}~\bibnamefont {Diamond}}, \bibinfo {author} {\bibfnamefont
  {K.}~\bibnamefont {Serniak}}, \bibinfo {author} {\bibfnamefont
  {T.}~\bibnamefont {Connolly}}, \bibinfo {author} {\bibfnamefont
  {P.}~\bibnamefont {Krogstrup}}, \bibinfo {author} {\bibfnamefont
  {J.}~\bibnamefont {Nygård}}, \bibinfo {author} {\bibfnamefont {A.~L.}\
  \bibnamefont {Yeyati}}, \bibinfo {author} {\bibfnamefont {A.}~\bibnamefont
  {Geresdi}},\ and\ \bibinfo {author} {\bibfnamefont {M.~H.}\ \bibnamefont
  {Devoret}},\ }\bibfield  {title} {\bibinfo {title} {Coherent manipulation of
  an {Andreev} spin qubit},\ }\href {https://doi.org/10.1126/science.abf0345}
  {\bibfield  {journal} {\bibinfo  {journal} {Science}\ }\textbf {\bibinfo
  {volume} {373}},\ \bibinfo {pages} {430} (\bibinfo {year}
  {2021})}\BibitemShut {NoStop}%
\bibitem [{\citenamefont {Bargerbos}\ \emph {et~al.}(2022)\citenamefont
  {Bargerbos}, \citenamefont {Pita-Vidal}, \citenamefont
  {\ifmmode~\check{Z}\else \v{Z}\fi{}itko}, \citenamefont {\'Avila},
  \citenamefont {Splitthoff}, \citenamefont {Gr\"unhaupt}, \citenamefont
  {Wesdorp}, \citenamefont {Andersen}, \citenamefont {Liu}, \citenamefont
  {Kouwenhoven}, \citenamefont {Aguado}, \citenamefont {Kou},\ and\
  \citenamefont {van Heck}}]{arno2022}%
  \BibitemOpen
  \bibfield  {author} {\bibinfo {author} {\bibfnamefont {A.}~\bibnamefont
  {Bargerbos}}, \bibinfo {author} {\bibfnamefont {M.}~\bibnamefont
  {Pita-Vidal}}, \bibinfo {author} {\bibfnamefont {R.}~\bibnamefont
  {\ifmmode~\check{Z}\else \v{Z}\fi{}itko}}, \bibinfo {author} {\bibfnamefont
  {J.}~\bibnamefont {\'Avila}}, \bibinfo {author} {\bibfnamefont {L.~J.}\
  \bibnamefont {Splitthoff}}, \bibinfo {author} {\bibfnamefont
  {L.}~\bibnamefont {Gr\"unhaupt}}, \bibinfo {author} {\bibfnamefont {J.~J.}\
  \bibnamefont {Wesdorp}}, \bibinfo {author} {\bibfnamefont {C.~K.}\
  \bibnamefont {Andersen}}, \bibinfo {author} {\bibfnamefont {Y.}~\bibnamefont
  {Liu}}, \bibinfo {author} {\bibfnamefont {L.~P.}\ \bibnamefont
  {Kouwenhoven}}, \bibinfo {author} {\bibfnamefont {R.}~\bibnamefont {Aguado}},
  \bibinfo {author} {\bibfnamefont {A.}~\bibnamefont {Kou}},\ and\ \bibinfo
  {author} {\bibfnamefont {B.}~\bibnamefont {van Heck}},\ }\bibfield  {title}
  {\bibinfo {title} {Singlet-doublet transitions of a quantum dot {Josephson}
  junction detected in a transmon circuit},\ }\href
  {https://doi.org/10.1103/PRXQuantum.3.030311} {\bibfield  {journal} {\bibinfo
   {journal} {PRX Quantum}\ }\textbf {\bibinfo {volume} {3}},\ \bibinfo {pages}
  {030311} (\bibinfo {year} {2022})}\BibitemShut {NoStop}%
\bibitem [{\citenamefont {Meng}\ \emph {et~al.}(2009)\citenamefont {Meng},
  \citenamefont {Florens},\ and\ \citenamefont {Simon}}]{meng2009}%
  \BibitemOpen
  \bibfield  {author} {\bibinfo {author} {\bibfnamefont {T.}~\bibnamefont
  {Meng}}, \bibinfo {author} {\bibfnamefont {S.}~\bibnamefont {Florens}},\ and\
  \bibinfo {author} {\bibfnamefont {P.}~\bibnamefont {Simon}},\ }\bibfield
  {title} {\bibinfo {title} {Self-consistent description of {Andreev} bound
  states in {Josephson} quantum dot devices},\ }\href
  {https://doi.org/10.1103/PhysRevB.79.224521} {\bibfield  {journal} {\bibinfo
  {journal} {Phys. Rev. B}\ }\textbf {\bibinfo {volume} {79}},\ \bibinfo
  {pages} {224521} (\bibinfo {year} {2009})}\BibitemShut {NoStop}%
\bibitem [{\citenamefont {Recher}\ \emph {et~al.}(2010)\citenamefont {Recher},
  \citenamefont {Nazarov},\ and\ \citenamefont {Kouwenhoven}}]{recher2010}%
  \BibitemOpen
  \bibfield  {author} {\bibinfo {author} {\bibfnamefont {P.}~\bibnamefont
  {Recher}}, \bibinfo {author} {\bibfnamefont {Y.~V.}\ \bibnamefont
  {Nazarov}},\ and\ \bibinfo {author} {\bibfnamefont {L.~P.}\ \bibnamefont
  {Kouwenhoven}},\ }\bibfield  {title} {\bibinfo {title} {{Josephson}
  light-emitting diode},\ }\href
  {https://doi.org/10.1103/PhysRevLett.104.156802} {\bibfield  {journal}
  {\bibinfo  {journal} {Phys. Rev. Lett.}\ }\textbf {\bibinfo {volume} {104}},\
  \bibinfo {pages} {156802} (\bibinfo {year} {2010})}\BibitemShut {NoStop}%
\bibitem [{\citenamefont {Oriekhov}\ \emph {et~al.}(2021)\citenamefont
  {Oriekhov}, \citenamefont {Cheipesh},\ and\ \citenamefont
  {Beenakker}}]{oriekhov2021}%
  \BibitemOpen
  \bibfield  {author} {\bibinfo {author} {\bibfnamefont {D.~O.}\ \bibnamefont
  {Oriekhov}}, \bibinfo {author} {\bibfnamefont {Y.}~\bibnamefont {Cheipesh}},\
  and\ \bibinfo {author} {\bibfnamefont {C.~W.~J.}\ \bibnamefont {Beenakker}},\
  }\bibfield  {title} {\bibinfo {title} {Voltage staircase in a current-biased
  quantum-dot {Josephson} junction},\ }\href
  {https://doi.org/10.1103/PhysRevB.103.094518} {\bibfield  {journal} {\bibinfo
   {journal} {Phys. Rev. B}\ }\textbf {\bibinfo {volume} {103}},\ \bibinfo
  {pages} {094518} (\bibinfo {year} {2021})}\BibitemShut {NoStop}%
\bibitem [{\citenamefont {Ivanov}\ and\ \citenamefont
  {Feigel'man}(1999)}]{ivanov1999}%
  \BibitemOpen
  \bibfield  {author} {\bibinfo {author} {\bibfnamefont {D.~A.}\ \bibnamefont
  {Ivanov}}\ and\ \bibinfo {author} {\bibfnamefont {M.~V.}\ \bibnamefont
  {Feigel'man}},\ }\bibfield  {title} {\bibinfo {title} {Two-level
  {H}amiltonian of a superconducting quantum point contact},\ }\href
  {https://doi.org/10.1103/PhysRevB.59.8444} {\bibfield  {journal} {\bibinfo
  {journal} {Phys. Rev. B}\ }\textbf {\bibinfo {volume} {59}},\ \bibinfo
  {pages} {8444} (\bibinfo {year} {1999})}\BibitemShut {NoStop}%
\bibitem [{\citenamefont {Zazunov}\ \emph {et~al.}(2003)\citenamefont
  {Zazunov}, \citenamefont {Shumeiko}, \citenamefont {Bratus'}, \citenamefont
  {Lantz},\ and\ \citenamefont {Wendin}}]{zazunov2003}%
  \BibitemOpen
  \bibfield  {author} {\bibinfo {author} {\bibfnamefont {A.}~\bibnamefont
  {Zazunov}}, \bibinfo {author} {\bibfnamefont {V.~S.}\ \bibnamefont
  {Shumeiko}}, \bibinfo {author} {\bibfnamefont {E.~N.}\ \bibnamefont
  {Bratus'}}, \bibinfo {author} {\bibfnamefont {J.}~\bibnamefont {Lantz}},\
  and\ \bibinfo {author} {\bibfnamefont {G.}~\bibnamefont {Wendin}},\
  }\bibfield  {title} {\bibinfo {title} {{Andreev} level qubit},\ }\href
  {https://doi.org/10.1103/PhysRevLett.90.087003} {\bibfield  {journal}
  {\bibinfo  {journal} {Phys. Rev. Lett.}\ }\textbf {\bibinfo {volume} {90}},\
  \bibinfo {pages} {087003} (\bibinfo {year} {2003})}\BibitemShut {NoStop}%
\bibitem [{\citenamefont {Landau}\ and\ \citenamefont
  {Lifshitz}(2013)}]{landau2013}%
  \BibitemOpen
  \bibfield  {author} {\bibinfo {author} {\bibfnamefont {L.~D.}\ \bibnamefont
  {Landau}}\ and\ \bibinfo {author} {\bibfnamefont {E.~M.}\ \bibnamefont
  {Lifshitz}},\ }\href@noop {} {\emph {\bibinfo {title} {Quantum mechanics:
  non-relativistic theory}}},\ Vol.~\bibinfo {volume} {3}\ (\bibinfo
  {publisher} {Elsevier},\ \bibinfo {year} {2013})\ pp.\ \bibinfo {pages}
  {167--170}\BibitemShut {NoStop}%
\bibitem [{\citenamefont {Gradshteyn}\ and\ \citenamefont
  {Ryzhik}(2014)}]{gradshteyn2014}%
  \BibitemOpen
  \bibfield  {author} {\bibinfo {author} {\bibfnamefont {I.~S.}\ \bibnamefont
  {Gradshteyn}}\ and\ \bibinfo {author} {\bibfnamefont {I.~M.}\ \bibnamefont
  {Ryzhik}},\ }\href@noop {} {\emph {\bibinfo {title} {Table of integrals,
  series, and products}}}\ (\bibinfo  {publisher} {Academic press},\ \bibinfo
  {year} {2014})\ pp.\ \bibinfo {pages} {1064--1067}\BibitemShut {NoStop}%
\bibitem [{\citenamefont {Likharev}\ and\ \citenamefont
  {Zorin}(1985)}]{likharev1985}%
  \BibitemOpen
  \bibfield  {author} {\bibinfo {author} {\bibfnamefont {K.}~\bibnamefont
  {Likharev}}\ and\ \bibinfo {author} {\bibfnamefont {A.}~\bibnamefont
  {Zorin}},\ }\bibfield  {title} {\bibinfo {title} {Theory of the {Bloch-wave}
  oscillations in small {Josephson} junctions},\ }\href
  {https://doi.org/10.1007/BF00683782} {\bibfield  {journal} {\bibinfo
  {journal} {Journal of low temperature physics}\ }\textbf {\bibinfo {volume}
  {59}},\ \bibinfo {pages} {347} (\bibinfo {year} {1985})}\BibitemShut
  {NoStop}%
\bibitem [{\citenamefont {Corlevi}\ \emph {et~al.}(2006)\citenamefont
  {Corlevi}, \citenamefont {Guichard}, \citenamefont {Hekking},\ and\
  \citenamefont {Haviland}}]{corlevi2006}%
  \BibitemOpen
  \bibfield  {author} {\bibinfo {author} {\bibfnamefont {S.}~\bibnamefont
  {Corlevi}}, \bibinfo {author} {\bibfnamefont {W.}~\bibnamefont {Guichard}},
  \bibinfo {author} {\bibfnamefont {F.~W.~J.}\ \bibnamefont {Hekking}},\ and\
  \bibinfo {author} {\bibfnamefont {D.~B.}\ \bibnamefont {Haviland}},\
  }\bibfield  {title} {\bibinfo {title} {Phase-charge duality of a {Josephson}
  junction in a fluctuating electromagnetic environment},\ }\href
  {https://doi.org/10.1103/PhysRevLett.97.096802} {\bibfield  {journal}
  {\bibinfo  {journal} {Phys. Rev. Lett.}\ }\textbf {\bibinfo {volume} {97}},\
  \bibinfo {pages} {096802} (\bibinfo {year} {2006})}\BibitemShut {NoStop}%
\bibitem [{\citenamefont {Dou\ifmmode~\mbox{\c{c}}\else \c{c}\fi{}ot}\ and\
  \citenamefont {Ioffe}(2007)}]{doucot2007}%
  \BibitemOpen
  \bibfield  {author} {\bibinfo {author} {\bibfnamefont {B.}~\bibnamefont
  {Dou\ifmmode~\mbox{\c{c}}\else \c{c}\fi{}ot}}\ and\ \bibinfo {author}
  {\bibfnamefont {L.~B.}\ \bibnamefont {Ioffe}},\ }\bibfield  {title} {\bibinfo
  {title} {Voltage-current curves for small {Josephson} junction arrays:
  {Semiclassical} treatment},\ }\href
  {https://doi.org/10.1103/PhysRevB.76.214507} {\bibfield  {journal} {\bibinfo
  {journal} {Phys. Rev. B}\ }\textbf {\bibinfo {volume} {76}},\ \bibinfo
  {pages} {214507} (\bibinfo {year} {2007})}\BibitemShut {NoStop}%
\bibitem [{\citenamefont {Kalashnikov}\ \emph {et~al.}(2020)\citenamefont
  {Kalashnikov}, \citenamefont {Hsieh}, \citenamefont {Zhang}, \citenamefont
  {Lu}, \citenamefont {Kamenov}, \citenamefont {Di~Paolo}, \citenamefont
  {Blais}, \citenamefont {Gershenson},\ and\ \citenamefont
  {Bell}}]{kalashnikov2020}%
  \BibitemOpen
  \bibfield  {author} {\bibinfo {author} {\bibfnamefont {K.}~\bibnamefont
  {Kalashnikov}}, \bibinfo {author} {\bibfnamefont {W.~T.}\ \bibnamefont
  {Hsieh}}, \bibinfo {author} {\bibfnamefont {W.}~\bibnamefont {Zhang}},
  \bibinfo {author} {\bibfnamefont {W.-S.}\ \bibnamefont {Lu}}, \bibinfo
  {author} {\bibfnamefont {P.}~\bibnamefont {Kamenov}}, \bibinfo {author}
  {\bibfnamefont {A.}~\bibnamefont {Di~Paolo}}, \bibinfo {author}
  {\bibfnamefont {A.}~\bibnamefont {Blais}}, \bibinfo {author} {\bibfnamefont
  {M.~E.}\ \bibnamefont {Gershenson}},\ and\ \bibinfo {author} {\bibfnamefont
  {M.}~\bibnamefont {Bell}},\ }\bibfield  {title} {\bibinfo {title} {Bifluxon:
  {Fluxon}-parity-protected superconducting qubit},\ }\href
  {https://doi.org/10.1103/PRXQuantum.1.010307} {\bibfield  {journal} {\bibinfo
   {journal} {PRX Quantum}\ }\textbf {\bibinfo {volume} {1}},\ \bibinfo {pages}
  {010307} (\bibinfo {year} {2020})}\BibitemShut {NoStop}%
\bibitem [{\citenamefont {Manucharyan}\ \emph {et~al.}(2009)\citenamefont
  {Manucharyan}, \citenamefont {Koch}, \citenamefont {Glazman},\ and\
  \citenamefont {Devoret}}]{manucharyan2009}%
  \BibitemOpen
  \bibfield  {author} {\bibinfo {author} {\bibfnamefont {V.~E.}\ \bibnamefont
  {Manucharyan}}, \bibinfo {author} {\bibfnamefont {J.}~\bibnamefont {Koch}},
  \bibinfo {author} {\bibfnamefont {L.~I.}\ \bibnamefont {Glazman}},\ and\
  \bibinfo {author} {\bibfnamefont {M.~H.}\ \bibnamefont {Devoret}},\
  }\bibfield  {title} {\bibinfo {title} {Fluxonium: {Single} {Cooper}-pair
  circuit free of charge offsets},\ }\href
  {https://doi.org/10.1126/science.1175552} {\bibfield  {journal} {\bibinfo
  {journal} {Science}\ }\textbf {\bibinfo {volume} {326}},\ \bibinfo {pages}
  {113} (\bibinfo {year} {2009})}\BibitemShut {NoStop}%
\bibitem [{\citenamefont {Koch}\ \emph {et~al.}(2009)\citenamefont {Koch},
  \citenamefont {Manucharyan}, \citenamefont {Devoret},\ and\ \citenamefont
  {Glazman}}]{koch2009}%
  \BibitemOpen
  \bibfield  {author} {\bibinfo {author} {\bibfnamefont {J.}~\bibnamefont
  {Koch}}, \bibinfo {author} {\bibfnamefont {V.}~\bibnamefont {Manucharyan}},
  \bibinfo {author} {\bibfnamefont {M.~H.}\ \bibnamefont {Devoret}},\ and\
  \bibinfo {author} {\bibfnamefont {L.~I.}\ \bibnamefont {Glazman}},\
  }\bibfield  {title} {\bibinfo {title} {Charging effects in the inductively
  shunted {Josephson} junction},\ }\href
  {https://doi.org/10.1103/PhysRevLett.103.217004} {\bibfield  {journal}
  {\bibinfo  {journal} {Phys. Rev. Lett.}\ }\textbf {\bibinfo {volume} {103}},\
  \bibinfo {pages} {217004} (\bibinfo {year} {2009})}\BibitemShut {NoStop}%
\bibitem [{\citenamefont {Fu}(2010)}]{fu2010}%
  \BibitemOpen
  \bibfield  {author} {\bibinfo {author} {\bibfnamefont {L.}~\bibnamefont
  {Fu}},\ }\bibfield  {title} {\bibinfo {title} {{Electron} {Teleportation} via
  {Majorana} {Bound} {States} in a {Mesoscopic} {Superconductor}},\ }\href
  {https://doi.org/10.1103/PhysRevLett.104.056402} {\bibfield  {journal}
  {\bibinfo  {journal} {Phys. Rev. Lett.}\ }\textbf {\bibinfo {volume} {104}},\
  \bibinfo {pages} {056402} (\bibinfo {year} {2010})}\BibitemShut {NoStop}%
\bibitem [{\citenamefont {Pikulin}\ \emph {et~al.}(2019)\citenamefont
  {Pikulin}, \citenamefont {Flensberg}, \citenamefont {Glazman}, \citenamefont
  {Houzet},\ and\ \citenamefont {Lutchyn}}]{pikulin2019}%
  \BibitemOpen
  \bibfield  {author} {\bibinfo {author} {\bibfnamefont {D.}~\bibnamefont
  {Pikulin}}, \bibinfo {author} {\bibfnamefont {K.}~\bibnamefont {Flensberg}},
  \bibinfo {author} {\bibfnamefont {L.~I.}\ \bibnamefont {Glazman}}, \bibinfo
  {author} {\bibfnamefont {M.}~\bibnamefont {Houzet}},\ and\ \bibinfo {author}
  {\bibfnamefont {R.~M.}\ \bibnamefont {Lutchyn}},\ }\bibfield  {title}
  {\bibinfo {title} {Coulomb {Blockade} of a {Nearly} {Open } {Majorana}
  {Island}},\ }\href {https://doi.org/10.1103/PhysRevLett.122.016801}
  {\bibfield  {journal} {\bibinfo  {journal} {Phys. Rev. Lett.}\ }\textbf
  {\bibinfo {volume} {122}},\ \bibinfo {pages} {016801} (\bibinfo {year}
  {2019})}\BibitemShut {NoStop}%
\bibitem [{\citenamefont {Pekker}\ \emph {et~al.}(2013)\citenamefont {Pekker},
  \citenamefont {Hou}, \citenamefont {Bergman}, \citenamefont {Goldberg},
  \citenamefont {Adagideli},\ and\ \citenamefont {Hassler}}]{pekker2013}%
  \BibitemOpen
  \bibfield  {author} {\bibinfo {author} {\bibfnamefont {D.}~\bibnamefont
  {Pekker}}, \bibinfo {author} {\bibfnamefont {C.-Y.}\ \bibnamefont {Hou}},
  \bibinfo {author} {\bibfnamefont {D.~L.}\ \bibnamefont {Bergman}}, \bibinfo
  {author} {\bibfnamefont {S.}~\bibnamefont {Goldberg}}, \bibinfo {author}
  {\bibfnamefont {I.}~\bibnamefont {Adagideli}},\ and\ \bibinfo {author}
  {\bibfnamefont {F.}~\bibnamefont {Hassler}},\ }\bibfield  {title} {\bibinfo
  {title} {Suppression of $2\ensuremath{\pi}$ phase slip due to hidden zero
  modes in one-dimensional topological superconductors},\ }\href
  {https://doi.org/10.1103/PhysRevB.87.064506} {\bibfield  {journal} {\bibinfo
  {journal} {Phys. Rev. B}\ }\textbf {\bibinfo {volume} {87}},\ \bibinfo
  {pages} {064506} (\bibinfo {year} {2013})}\BibitemShut {NoStop}%
\bibitem [{\citenamefont {Rodr\'{\i}guez-Mota}\ \emph
  {et~al.}(2019)\citenamefont {Rodr\'{\i}guez-Mota}, \citenamefont
  {Vishveshwara},\ and\ \citenamefont {Pereg-Barnea}}]{rodriguez-mota2019}%
  \BibitemOpen
  \bibfield  {author} {\bibinfo {author} {\bibfnamefont {R.}~\bibnamefont
  {Rodr\'{\i}guez-Mota}}, \bibinfo {author} {\bibfnamefont {S.}~\bibnamefont
  {Vishveshwara}},\ and\ \bibinfo {author} {\bibfnamefont {T.}~\bibnamefont
  {Pereg-Barnea}},\ }\bibfield  {title} {\bibinfo {title} {Revisiting
  $2\ensuremath{\pi}$ phase slip suppression in topological {Josephson}
  junctions},\ }\href {https://doi.org/10.1103/PhysRevB.99.024517} {\bibfield
  {journal} {\bibinfo  {journal} {Phys. Rev. B}\ }\textbf {\bibinfo {volume}
  {99}},\ \bibinfo {pages} {024517} (\bibinfo {year} {2019})}\BibitemShut
  {NoStop}%
\bibitem [{\citenamefont {Svetogorov}\ \emph {et~al.}(2020)\citenamefont
  {Svetogorov}, \citenamefont {Loss},\ and\ \citenamefont
  {Klinovaja}}]{svetogorov2020}%
  \BibitemOpen
  \bibfield  {author} {\bibinfo {author} {\bibfnamefont {A.~E.}\ \bibnamefont
  {Svetogorov}}, \bibinfo {author} {\bibfnamefont {D.}~\bibnamefont {Loss}},\
  and\ \bibinfo {author} {\bibfnamefont {J.}~\bibnamefont {Klinovaja}},\
  }\bibfield  {title} {\bibinfo {title} {Critical current for an insulating
  regime of an underdamped current-biased topological josephson junction},\
  }\href {https://doi.org/10.1103/PhysRevResearch.2.033448} {\bibfield
  {journal} {\bibinfo  {journal} {Phys. Rev. Research}\ }\textbf {\bibinfo
  {volume} {2}},\ \bibinfo {pages} {033448} (\bibinfo {year}
  {2020})}\BibitemShut {NoStop}%
\bibitem [{\citenamefont {van Heck}\ \emph {et~al.}(2011)\citenamefont {van
  Heck}, \citenamefont {Hassler}, \citenamefont {Akhmerov},\ and\ \citenamefont
  {Beenakker}}]{heck2011}%
  \BibitemOpen
  \bibfield  {author} {\bibinfo {author} {\bibfnamefont {B.}~\bibnamefont {van
  Heck}}, \bibinfo {author} {\bibfnamefont {F.}~\bibnamefont {Hassler}},
  \bibinfo {author} {\bibfnamefont {A.~R.}\ \bibnamefont {Akhmerov}},\ and\
  \bibinfo {author} {\bibfnamefont {C.~W.~J.}\ \bibnamefont {Beenakker}},\
  }\bibfield  {title} {\bibinfo {title} {Coulomb stability of the
  4$\ensuremath{\pi}$-periodic {Josephson} effect of {Majorana} fermions},\
  }\href {https://doi.org/10.1103/PhysRevB.84.180502} {\bibfield  {journal}
  {\bibinfo  {journal} {Phys. Rev. B}\ }\textbf {\bibinfo {volume} {84}},\
  \bibinfo {pages} {180502} (\bibinfo {year} {2011})}\BibitemShut {NoStop}%
\bibitem [{\citenamefont {Pikulin}\ and\ \citenamefont
  {Nazarov}(2012)}]{pikulin2012}%
  \BibitemOpen
  \bibfield  {author} {\bibinfo {author} {\bibfnamefont {D.~I.}\ \bibnamefont
  {Pikulin}}\ and\ \bibinfo {author} {\bibfnamefont {Y.~V.}\ \bibnamefont
  {Nazarov}},\ }\bibfield  {title} {\bibinfo {title} {Phenomenology and
  dynamics of a {Majorana Josephson} junction},\ }\href
  {https://doi.org/10.1103/PhysRevB.86.140504} {\bibfield  {journal} {\bibinfo
  {journal} {Phys. Rev. B}\ }\textbf {\bibinfo {volume} {86}},\ \bibinfo
  {pages} {140504} (\bibinfo {year} {2012})}\BibitemShut {NoStop}%
\bibitem [{\citenamefont {Vakhtel}\ and\ \citenamefont {van
  Heck}(2022)}]{zenodo}%
  \BibitemOpen
  \bibfield  {author} {\bibinfo {author} {\bibfnamefont {T.}~\bibnamefont
  {Vakhtel}}\ and\ \bibinfo {author} {\bibfnamefont {B.}~\bibnamefont {van
  Heck}},\ }\href {https://doi.org/10.5281/zenodo.7347760} {\bibinfo {title}
  {{Quantum phase slips in a resonant Josephson junction}}} (\bibinfo {year}
  {2022}),\ \bibinfo {note} {on Zenodo}\BibitemShut {NoStop}%
\bibitem [{\citenamefont {Nazarov}\ and\ \citenamefont
  {Blanter}(2009)}]{nazarov2009quantum}%
  \BibitemOpen
  \bibfield  {author} {\bibinfo {author} {\bibfnamefont {Y.~V.}\ \bibnamefont
  {Nazarov}}\ and\ \bibinfo {author} {\bibfnamefont {Y.~M.}\ \bibnamefont
  {Blanter}},\ }\href@noop {} {\emph {\bibinfo {title} {Quantum transport:
  introduction to nanoscience}}}\ (\bibinfo  {publisher} {Cambridge university
  press},\ \bibinfo {year} {2009})\BibitemShut {NoStop}%
\bibitem [{\citenamefont {Keselman}\ \emph {et~al.}(2019)\citenamefont
  {Keselman}, \citenamefont {Murthy}, \citenamefont {van Heck},\ and\
  \citenamefont {Bauer}}]{keselman2019}%
  \BibitemOpen
  \bibfield  {author} {\bibinfo {author} {\bibfnamefont {A.}~\bibnamefont
  {Keselman}}, \bibinfo {author} {\bibfnamefont {C.}~\bibnamefont {Murthy}},
  \bibinfo {author} {\bibfnamefont {B.}~\bibnamefont {van Heck}},\ and\
  \bibinfo {author} {\bibfnamefont {B.}~\bibnamefont {Bauer}},\ }\bibfield
  {title} {\bibinfo {title} {{Spectral response of Josephson junctions with
  low-energy quasiparticles}},\ }\href
  {https://doi.org/10.21468/SciPostPhys.7.4.050} {\bibfield  {journal}
  {\bibinfo  {journal} {SciPost Phys.}\ }\textbf {\bibinfo {volume} {7}},\
  \bibinfo {pages} {050} (\bibinfo {year} {2019})}\BibitemShut {NoStop}%
\end{thebibliography}%

\end{document}